\documentclass[sigconf]{acmart}  % ,nonacm for arXiv

\AtBeginDocument{%
  \providecommand\BibTeX{{%
    \normalfont B\kern-0.5em{\scshape i\kern-0.25em b}\kern-0.8em\TeX}}}

\setcopyright{acmcopyright}
\copyrightyear{2021}
\acmYear{2021}
\acmDOI{10.1145/NNNNNNN.NNNNNNN}  % leave as is for now

\acmConference[PPDP '21]{PPDP '21}{September 6–8, 2021}{Tallinn, Estonia}
\acmBooktitle{Proceedings of the 23rd Symposium on Principles and Practice of Declarative Programming, PPDP 2021, Tallinn, Estonia, September 6–8, 2021}
\acmPrice{15.00}
\acmISBN{978-1-4503-8689-0}

\usepackage{listings}
\RequirePackage[undo-recent-deprecations]{expl3} %% Fix: latex3 problem with ebproof
\usepackage{ebproof}
\usepackage{multicol}
\usepackage{xspace}

\newcommand{\polyrpc}{$\lambda_{rpc}^{\forall}$\xspace}

\newcommand{\cs}{$\lambda_{cs}$\xspace}
\newcommand{\polycs}{$\lambda_{cs}^{\forall}$\xspace}

\newcommand{\client}{\textbf{c}}
\newcommand{\server}{\textbf{s}}

\newcommand{\evalRPC}[3]{#1\Downarrow_{#2}#3}

\newcommand{\lamL}[3]{\lambda^{#1}#2.#3}

\newcommand{\subst}[2]{\{#1/#2\}}
\newcommand{\llet}[3]{\textsf{let} \ #1 = #2 \ \textsf{in} \ #3}
\newcommand{\ldokeyword}{\textsf{do}}
\newcommand{\ldo}[3]{\textsf{do} \ #1 \leftarrow #2 \ \textsf{in} \ #3}
\newcommand{\ldovoid}[2]{\textsf{do} \ #1 \ \textsf{in} \ #2}
\newcommand{\lunit}[1]{\textsf{unit} \ #1}

\newcommand{\textsfGen}{\textsf{gen}}
\newcommand{\gen}[3]{\textsfGen(#1,#2,#3)}

\newcommand{\textsfReq}{\textsf{req}}
\newcommand{\req}[2]{\textsfReq(#1,#2)}

\newcommand{\textsfCall}{\textsf{call}}
\newcommand{\call}[2]{\textsfCall(#1,#2)}

\newcommand{\textsfCase}{\textsf{case}}
\newcommand{\textsfOf}{\textsf{of}}
\newcommand{\case}[2]{\textsfCase ~ #1 ~\textsfOf ~ #2}

\newcommand{\textsfIf}{\textsf{if}}

\newcommand{\funL}[1]{\xrightarrow{#1}}

\newcommand{\tyenv}{\Gamma}
\newcommand{\tyenvExt}[2]{\Gamma,#1:#2}
\newcommand{\tyenvExtWith}[1]{\Gamma,#1}

\newcommand{\typing}[4]{#1\vdash_{#2} #3 : #4}

\newcommand{\funtyping}[5]{#1\vdash #4 : #5}     % ;#2, #3
\newcommand{\codetyping}[4]{#1\vdash_{code} #3 : #4}  % _{#2}

\newcommand{\conftyping}[2]{\vdash #1 : #2}
\newcommand{\stacktyping}[3]{\vdash_{#1} #2 : #3}
\newcommand{\fvtyping}[2]{#1\vdash #2}

\newcommand{\ccomp}[1]{\mathcal{C}\llbracket#1\rrbracket}
\newcommand{\vcomp}[1]{\mathcal{V}\llbracket#1\rrbracket}
\newcommand{\ecomp}[1]{\llbracket#1\rrbracket}

\newcommand{\FUNS}{\Phi}
\newcommand{\funstore}{\FUNS}

\newcommand{\funcode}{\funstore}

\newcommand{\clo}[2]{clo({#1},{#2})}
\newcommand{\cloty}[1]{Clo(#1)}
\newcommand{\valty}[1]{relocatable(#1)}

\newcommand{\Loc}{Loc}

\newcommand{\optrpc}[1]{\mathcal{O}[\![#1]\!]}

\newcommand{\stack}{\Delta}
\newcommand{\emptystack}{\epsilon}

\newcommand{\conf}{\Sigma}
\newcommand{\confcs}[2]{\langle #1 | #2 \rangle}
\newcommand{\stackcsWith}[2]{ #1  |  #2 }
\newcommand{\stackcs}{\stackcsWith{ \stack_\client }{ \stack_\server }}
\newcommand{\confuntyped}{\sigma}

\newcommand{\run}{\longrightarrow}
\newcommand{\runequiv}{\Longrightarrow}
\newcommand{\structeqv}{\equiv}

\newcommand{\loctyvars}{\overline{l} \ \overline{\alpha}}
\newcommand{\loctys}{\overline{\Loc}\ \overline{A}}

\newcommand{\CloCodeType}{Ty}
\newcommand{\CloCode}{Code}
\newcommand{\OpenCode}{OpenCode}

\newcommand{\anonymousurl}[1]{https://anonymized.url}

\newtheorem{lemma}{Lemma}[section]
\newtheorem{theorem}{Theorem}[section]
\newtheorem{fact}{Fact}[section]
\newtheorem{definition}{Definition}[section]

\begin{document}

\title{A Typed Slicing Compilation of the Polymorphic RPC calculus}

\author{Kwanghoon Choi}
\affiliation{%
  \institution{Chonnam National University}
  \streetaddress{77 Yongbong-ro, Buk-gu}
  \city{Gwangju}
  \country{Republic of Korea}}
\email{kwanghoon.choi@jnu.ac.kr}

\author{James Cheney}
\affiliation{%
  \institution{The University of Edinburgh}
  \streetaddress{10 Crichton St, Newington}
  \city{Edinburgh}
  \country{United Kingdom}}
\email{jcheney@inf.ed.ac.uk}

\author{Sam Lindley}
\affiliation{%
  \institution{The University of Edinburgh}
  \streetaddress{10 Crichton St, Newington}
  \city{Edinburgh}
  \country{United Kingdom}}
\email{Sam.Lindley@ed.ac.uk}

\author{Bob Reynders}
\affiliation{%
  \institution{Chonnam National University}
  \streetaddress{77 Yongbong-ro, Buk-gu}
  \city{Gwangju}
  \country{Republic of Korea}}
\email{tzbob@gmail.com}

\renewcommand{\shortauthors}{K. Choi, J. Cheney, S. Lindley, and B. Reynders}

\begin{abstract}
The polymorphic RPC calculus allows programmers to write succinct
multitier programs using polymorphic location constructs. However,
until now it lacked an implementation. We develop an experimental
programming language based on the polymorphic RPC calculus.
We introduce a polymorphic Client-Server (CS) calculus with the client
and server parts separated. In contrast to existing untyped CS
calculi, our calculus is not only able to resolve polymorphic
locations statically, but it is also able to do so dynamically.
We design a type-based slicing compilation of the polymorphic RPC
calculus into this CS calculus, proving type and semantic
correctness. We propose a method to erase types unnecessary for
execution but retaining locations at runtime by translating the
polymorphic CS calculus into an untyped CS calculus, proving semantic
correctness.
\end{abstract}


%% %%
%% %% Keywords. The author(s) should pick words that accurately describe
%% %% the work being presented. Separate the keywords with commas.

\maketitle

\section{Introduction}
\label{sec:introduction}

Multi-tier programming languages address the complexity of developing
distributed systems by providing abstractions for communication
between peers.
For instance, a web application is a basic distributed system
consisting of a client, which provides access to a user interface, and
a server, which provides access to a persistent database, where the
client and server are connected by the HTTP network protocol.
Typically, the client and server code must be developed as two
separate programs and run on two separate machines, adding to the
programmer's burden.
The two programs must be tested together, which is more complex than
testing one program on a single machine.
As the web application evolves, suitable invariants between client and
server programs must be carefully maintained.
Worse, when tasks cross the boundary of the client and server, the
programmer must split the work across the two programs, often baking
in implementation decisions that are hard to understand, revisit or
maintain.

Multi-tier programming solves this problem by allowing programmers to
write client and server expressions together in a single programming
language, and by automatically slicing the unified program into client
and server programs that are connected together with networking
libraries automatically.

An important feature of multi-tier programming languages is the
ability to specify locations where code should run. RPC calculi~\cite{Cooper:2009:RC:1599410.1599439,choijfp2019,CHOI:scp2020} offer a
promising, yet lightweight, semantic foundation for multi-tier
programming:
Firstly, programmers have only to add location annotations, for
example, $\client$ for client and $\server$ for server, to lambda
abstractions to write remote procedures such as $\lambda^{\client}x.M$
and $\lambda^{\server}x.M$.
Secondly, remote procedure calls are as simple as local calls, reusing
the standard function application syntax.
Thirdly, RPC calculi allow unrestricted composition of differently
located procedures.
These features are not provided by existing multi-tier programming
languages such as ML5~\cite{Murphy:2007:TDP:1793574.1793585,Murphy:2008:MTM:1467784}, Eliom~\cite{Radanne2017,Radanne:2018:TWP:3184558.3185953}, Hop~\cite{Serrano2006,
  Serrano:2012:MPH:2240236.2240253,Serrano:2016:GH:3022670.2951916},
Ur/Web~\cite{Chlipala2015}, ScalaLoci~\cite{Weisenburger:2018:DSD:3288538.3276499,weisenburger_et_al:LIPIcs:2019:10795},
and Gavial~\cite{Reynders2020}.

However, the simplicity of RPC calculi gives rise to a difficult
choice between convenience and efficiency: it can be hard to determine
statically whether a given call site is local or remote.
Links~\cite{Cooper:2006:LWP:1777707.1777724}, a practical multi-tier
web programming language, is based on the untyped RPC
calculus~\cite{Cooper:2009:RC:1599410.1599439}, which provides no
static location information and thus depends entirely on runtime
location-checking.
As function calls are pervasive in functional languages, even a small
overhead may be costly.
For instance, local computations that could run efficiently on the
server may see a significant slow-down as a result of having to
dynamically check whether a client call is required, even when the
check always determines that a client call is unnecessary.
Given compile-time location information at each call site, such
overheads can be avoided.

%% SL: it would be good to sort out the horrible formatting where
%% there isn't enough space between the text and the shapes in the
%% figure!

%% KC: Thanks!

%% SL: The figure uses "monomorphization", but currently the rest of
%% the text uses "monomorphisation". We should be consistent. Are we
%% using US spelling (e.g. monomorphization and behavior) or British
%% spelling (e.g. monomorphisation and behaviour)?

%% KC: Updated!

\begin{figure*}[t]
  \centering
  \includegraphics[width=0.8\textwidth]{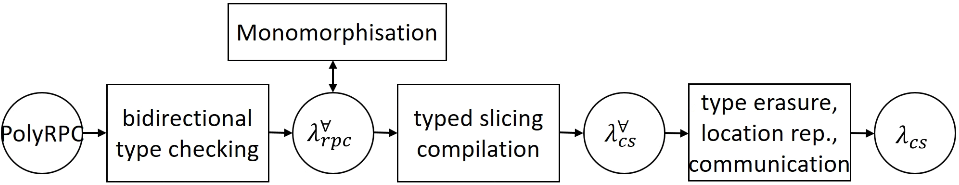}
  \caption{A PolyRPC Compiler}
  \Description[A PolyRPC Compiler]{A PolyRPC Compiler}
  \label{fig:polyrpccompiler}
\end{figure*}

The simply-typed RPC calculus~\cite{choijfp2019} is designed to offer
complete location information statically through types in order to
determine all remote procedure calls at compile-time.
Such typed location information is not only useful for avoiding the
overhead of runtime location checking but also guides the design of
simple slicing compilation methods for both stateless and stateful
client-server calculi.
We anticipate further applications of static location information to
multi-tier programming in the future.
Despite this prospect, the simply-typed RPC calculus only allows
programmers to statically specify fixed locations; it does not support
polymorphic locations, which are useful for writing succinct programs.
For instance, Eliom~\cite{radanneaplas2016,radanneifl2016,Radanne2017,Radanne:2018:TWP:3184558.3185953}
provides a macro feature to make it possible to write code for the
client and for the server at the same time; polymorphic locations
offer similar functionality.

%% SL: I'm tempted to suggest a citation for monomorphisation, but I'm
%% not sure if there is a canonical reference. The Featherweight Go
%% paper at OOPSLA 2020 purports to be the first paper to fully
%% formalise one particular form of monomorphisation, and includes a
%% reasonably comprehensive collection of references on
%% monomorphisation.
%
%% KC: I agree. I have put a citation right after the monomorphisation
%% translation.
%
%% I learned an interesting English word, purport, that fits for the
%% situation. :)

In previous work introducing the polymorphic RPC calculus, {\polyrpc},~\cite{CHOI:scp2020}, it was proposed to implement polymorphic
locations by a so called \emph{monomorphisation} translation, which
translates polymorphically located programs into monomorphic ones at
compile time, by specialising each location-polymorphic function and
compiling it once for each possible location assignment of concrete
locations to polymorphic location variables.
On top of existing monomorphically typed RPC calculi~\cite{choijfp2019}, this translation could be used as the basis for an
implementation of the polymorphic RPC calculus. Although the previous
work did not provide or evaluate an implementation, but one clear
concern about monomorphisation is that it can lead to a code explosion
problem.
%

%% SL: if there are only two locations, then does this just mean that
%% we at worst double the size of the generated code?
%
%% If yes, then this doesn't seem like a huge cost, particularly given
%% that we probably have to generate code for different client and
%% server platforms anyway.
%
%% KC: I agree. When there are only two locations and only
%% single-location abstractions are used, this might not be a huge
%% cost.
%
%% SL: Perhaps it could be worse than this, e.g., if we use curried
%% functions?
%
%% KC: By the curried functions, I presume that you mean
%% multiple-location abstractions such as {l1 l2 l3}.V?
%
%% Even if there are only two locations, multiple-location abstractions 
%% could be vulnerable to the problem.
%
%% In summary, the increment of code size (by the monomorphisation)
%% depends on the # of kinds of location (e.g., c and s as you
%% assume) and also on the # of nested location abstractions.

In this paper,
%we propose a type-based implementation for the
%polymorphic RPC calculus and design an experimental programming
%language for the Web based on the calculus.
%
%For the typed implementation,
we design a \emph{polymorphic}
Client-Server (CS) calculus, {\polycs}, and a \emph{type-based}
slicing compilation of the polymorphic RPC calculus into \polycs.
Prior CS
calculi~\cite{Cooper:2006:LWP:1777707.1777724,choijfp2019,CHOI:scp2020}
and their slicing compilers are untyped.
In (typed or untyped) CS calculi, the client part is clearly separated
from the server part, and communication between the two is inserted
automatically.

The first highlight of the polymorphic CS calculus is that the type
system guarantees that while functions may be passed to arbitrary
locations, every function must run at the declared location.
Regardless of how slicing compilation is specified, the type system
ensures that first-class functions are well-behaved in located
contexts.

The second highlight is that our polymorphic CS calculus is designed
to support the combination of the static approach relying on
monomorphisation and a complementary dynamic approach for handling
polymorphic-location programs directly.
In the dynamic approach, locations are passed and examined at runtime,
thus avoiding the code explosion problem.
%
%% resolving the code explosion problem~\cite{CHOI:scp2020}, by
%% duplicating each polymorphically located lambda abstraction into a
%% client one and a server one .

The idea of dynamically passing locations is reminiscent of
intentional polymorphism~\cite{Harper:1995:CPU:199448.199475} and
type-erasure semantics~\cite{crary_weirich_morrisett_2002}.
%
%% SL: probably unnecessary to mention the following as it goes
%% without saying.
%
%% These works do not involve any client-server distributed systems and
%% thus make no guarantees about running functions at the right location.
%
%% KC: I agree.
%
Accordingly, we propose an efficient implementation strategy for
polymorphic CS calculus by erasing all types unnecessary for
computation, but retaining those locations required at runtime. We
introduce explicit CS communication primitives.

The third highlight is a monadic abstraction for {\it trampolined
  style} RPC communication where a single ``scheduler'' loop, called
trampoline, manages all transfers of control by remote procedure call.
This allows us to treat the polymorphic CS calculus like a sequential
calculus over the client and server.

For a practical aspect, we design an experimental multi-tier
programming language for the Web, named {\it PolyRPC}, based on the
polymorphic RPC calculus and implement its compiler based on the
polymorphic CS calculus as shown in Figure \ref{fig:polyrpccompiler}.
In the language, the calculi are extended with basic programming
features, such as recursion, data types, and references.
In our PolyRPC compiler, the front-end is equipped with a simple
bidirectional type checker~\cite{10.1145/2500365.2500582}.
Monomorphisation is implemented, and it can be enabled or disabled.
When monomorphisation is disabled, polymorphic locations are resolved
dynamically.
The back-end comprises a slicing compiler for the polymorphic CS
calculus, followed by type-erasure and location representation stages
with the introduction of explicit communication primitives for the CS
based Web system.

Using this programming system, we validate the usefulness of
polymorphic locations by developing a multi-tier ToDo list program.

The contributions of this paper are summarized as follows:
\begin{itemize}
\item For a polymorphic RPC calculus, we introduce a new polymorphic
  CS calculus, \polycs, and prove type-soundness.

\item We design a typed slicing compilation of \polyrpc into \polycs
  via a static approach and a dynamic approach, and prove its type
  correctness and semantic correctness.

\item We describe an implementation of \polycs by erasing types but
  retaining locations in terms and by making client-server
  communication explicit, and we prove semantic correctness.

\item We design and implement an experimental multi-tier programming
  language for the Web, and discuss a case study with a multi-tier
  ToDoMVC program.
\end{itemize}

Section~\ref{sec:runningexample} presents a case study to help
understand the polymorphic RPC calculus in practice.
Section~\ref{sec:polyrpc} gives a formal account of the polymorphic
RPC calculus.
Section~\ref{sec:polycs} proposes a polymorphic client-server
calculus, proves type-soundness of this calculus, describes a typed
slicing compilation of RPC calculus into CS calculus, and proves type
and semantic correctness of compilation.
Section~\ref{sec:implementation} details how to implement the
polymorphic CS calculus using type-erasure.
Related work is discussed in Section \ref{sec:relatedwork} and
Section~\ref{sec:conclusion} concludes.
Proofs are available in the extended version~\cite{choi2021typedslicingcompilation}.

%%%%%

\section{Case Study: A Multi-tier todo list}
\label{sec:runningexample}

\begin{figure}[h]
  \centering
  \includegraphics[width=.4\textwidth]{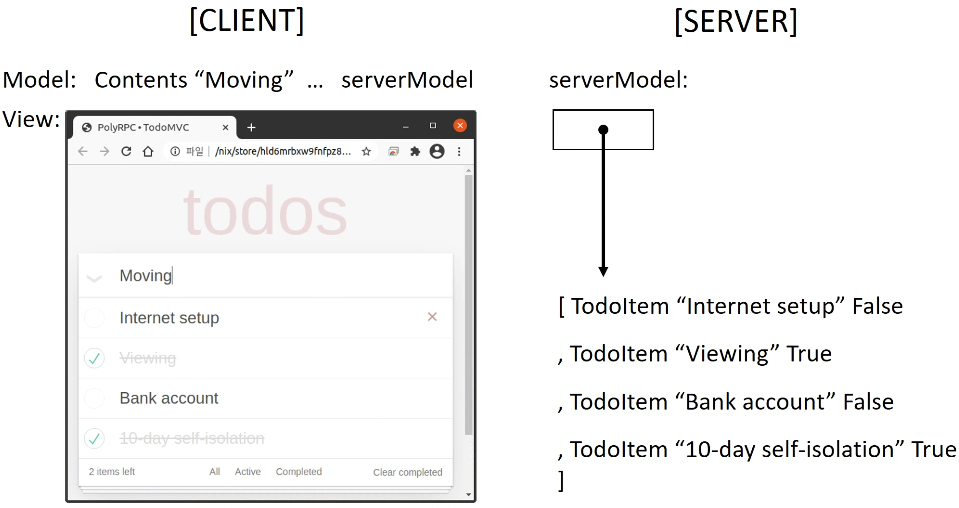}
  \caption{Running the Multi-tier TodoMVC Program}
  \Description[Running the Multi-tier TodoMVC Program]{Running the Multi-tier TodoMVC Program}
  \label{fig:todomvcrunning}
\end{figure}

\begin{figure*}[t]
  \centering
\begin{minipage}[t]{.7\textwidth}
\begin{lstlisting}
data TodoItem = TodoItem String Bool;
data Model = Content String (List [TodoItem]) (Ref {server} [List [TodoItem]]);
data Msg = Update String | Submit | Toggle Int | Delete Int ;

showItem : TodoItem -client-> Int -client-> Html [Msg]
  = \item @ client idx @ client. case item { TodoItem content done =>
        Element "li" []
         [ Element "input" [ Attribute "type" "checkbox", onClick (Toggle idx)
                  , Property "checked" (if done then "false" else "true") ] []
         , Element "label" [] [ Txt content ]
         , Element "button" [ onClick (Delete idx) ] [ Txt "X" ]   ] };
showList : List [TodoItem] -client-> Html [Msg]
  = \items@client. Element "ul" [] (mapWithCount {client} 0 showItem items);
header : String -client-> Html [Msg]
  = \str @ client. Element "input"
          [ Attribute "placeholder" "What needs to be done?"
          , Property "value" str, onInput Update, onEnter Submit ] [];
view : Model -client-> Html [Msg]
  = \m @ client. case m { Content str visibleList ref =>
          Element "div" [] [ header str, showList visibleList ] };
toggleItem: {l}. TodoItem -l-> TodoItem
  = {l}. \ti @ l. case ti { TodoItem content done =>
        TodoItem content (not {l} done) };
update : Msg -client-> Model -client-> Model
  = \msg @ client model @ client.
     case model { Content line visibleList ref =>
       case msg {
         Update str => Content str visibleList ref;
         Submit => let { u : Unit =
	     ref := {server} ( TodoItem line False :: ! {server} ref )
	   } Content "" (! {server} ref) ref end ;
         Toggle idx => let { u : Unit =
             ref := {server} (mapOnIndex {server} idx (toggleItem {server})
                               (! {server} ref))
           } Content line (! {server} ref) ref end ;
         Delete idx => let { u : Unit =
	     ref := {server} (delete {server} idx (! {server} ref))
           } Content line (! {server} ref) ref end
     }};
serverModel : Ref {server} [List [TodoItem]] = ref {server} Nil;
init : Model = Content "" Nil serverModel;
main : Page [Model Msg] = Page init view update
\end{lstlisting}
\end{minipage}
\caption{A Multi-tier TodoMVC Program}
\Description[A Multi-tier TodoMVC Program]{A Multi-tier TodoMVC Program}
\label{fig:todomvccode}
\end{figure*}

In this section we illustrate polymorphic RPC calculus with an
example multi-tier web application.
Our example is the `Hello world' of web programming: a todo list
application.
The TodoMVC program manages a list of work items, and is structured
using the Model-View-Update (MVU~\cite{fowler:LIPIcs:2020:13171})
pattern.
It is written in \emph{PolyRPC}, an experimental programming language
based on the polymorphic RPC calculus in Section \ref{sec:polyrpc}.

%% This section motivates a reader with an example of a multi-tier
%% TodoMVC program. This is a web-based program to manage a list of work
%% items, and is structured by the Model-View-Update (MVU
%% \cite{fowler:LIPIcs:2020:13171}) design pattern.  It is written in
%% PolyRPC\footnote{https://github.com/kwanghoon/polyrpc}, which is an
%% experimental programming language based on the polymorphic RPC
%% calculus in Section \ref{sec:polyrpc}.  Its running is fully supported
%% by our dynamic approach to be explained from Section \ref{sec:polycs}.
%% TodoMVC is known as a `Hello World' program in web programming, and is
%% useful for comparing different web programming languages and
%% frameworks.

The multi-tier TodoMVC program consists of a web-based UI on the
client, and a model for managing the items on the server. The UI
allows a user to ask the server to add a new item, mark an item as
completed, and delete an item. Figure \ref{fig:todomvcrunning} shows
the program running and depicts the configuration of the client and
the server.

We present source code for TodoMVC in Figure~\ref{fig:todomvccode}.
A longer version, complete with CSS styling, is available
online\footnote{https://github.com/kwanghoon/todomvc}.
Following the MVU design pattern, the main value (Line 42) declares a
page with the initial model, a view function, and an updating
function.

A model of type {\texttt {Model}} (Line 2) is a triple of a text
string that the user is typing, a list of visible items of type
{\texttt {List [TodoItem]}}, and a reference to a list of all items at
the server of type {\texttt {Ref \{server\} [List [TodoItem]]}} where
\texttt{Ref \{server\}} is the location application type of
\texttt{Ref} to \texttt{server} using a notation $\{-\}$ and
\texttt{List [TodoItem]} is the type application type of \texttt{List}
to \texttt{TodoItem} using another notation $[ - ]$ in PolyRPC.

%% SL: The type suggests that main is a value and not a function. I
%% changed the text above accordingly. Is this the correct
%% interpretation.

%% KC: Yes, this is correct. The main is a value. 

%% an abbreviated source code of the fully functional multi-tier TodoMVC
%% program\footnote{https://github.com/kwanghoon/todomvc} in Figure
%% \ref{fig:todomvccode}. With the MVU design pattern, the main declares
%% a page with the initial model, a view function, and an updating
%% function in Line 42.

%% PolyRPC allows programmers to omit writing type abstractions and type
%% arguments that can be reconstructed automatically by bidirectional
%% type checking~\cite{10.1145/2500365.2500582}.

The {\texttt {view}} function (Line 18) takes a model and returns an
HTML value.
The \texttt{client} annotation on the function type ensures that the
function is run on the client.

%% In PolyRPC, programmers can specify where to run a function. For
%% example, the {\texttt {view}} function has type {\texttt{Model
%%     -client-> Html [Msg]}} where the client location is annotated to
%% the function type.

A user interacts with the constructed HTML of type {\texttt {Html
    [Msg]}} through event handler actions that generate messages of
type {\texttt {Msg}}.
%
%% When the user inputs new characters `M', `o', `v', `i', `n', `g' in
%% sequence as in Figure \ref{fig:todomvcrunning}, the {\texttt
%%   {onInput}} event (Line 17) generates messages {\texttt {Update
%%     ``M''}}, {\texttt {Update ``Mo''}}, ... , {\texttt {Update
%%     ``Moving''}}, respectively.
%% %
%% When the user types an Enter key, the {\texttt {onEnter}} event
%% generates a {\texttt {Submit}} message to add a new item to the list
%% with the typed string.
%% %
%% When the user clicks a checkbox associated with an item, the {\texttt
%%   {onClick}} event (Line 8) generates a {\texttt {Toggle index}}
%% message with the index for the item.
%% %
%% When the user presses the X button adjacent to an item as shown in
%% Figure \ref{fig:todomvcrunning}, the {\texttt {onClick}} event (Line
%% 11) generates a {\texttt {Delete index}} message for the item.
Each message is handled by the {\texttt {update}} function (Line 24)
which runs on the client and updates the existing model using references, according to
the message.
%% %
%% For each {\texttt {Update str}} message (Line 28), the existing string
%% is replaced by a new one that the user is typing.
%% %
%% For the other messages, the list of items is first updated on the
%% server before being read back for display on the client.

%
A locative and polymorphic reference type \texttt {Ref \{Loc\} [A]} is
an abstract data type parameterized by locations {\texttt {Loc}}, as
well as types {\texttt {A}}, with three interface functions
\begin{itemize}
\item \texttt{ref : \{l\}. [a]. a -l-> Ref \{l\} [a] }
\item \texttt{(!) : \{l\}. [a]. Ref \{l\} [a] -l-> a }
\item \texttt{(:=) : \{l\}. [a]. Ref \{l\} [a] -l-> a -l-> Unit }.
\end{itemize}
where \texttt{\{l\}.A} is a location abstraction type over a location
variable {\texttt {l}} and \texttt{[a].A} is a type abstraction type
over a type variable {\texttt {a}}.
The server model is represented as a reference to a work item list
stored on the server is {\texttt {Ref \{server\} [List [TodoItem]]}}.
It is initialised to an empty list (Line 40).
Only {\texttt {ref \{server\}}}, {\texttt {! \{server\}}}, and
{\texttt {:= \{server\}}} can create, read, and modify server
references.
We write {\texttt {M \{Loc\}}} for the location application of $M$ to
$\Loc$.
%
%% Just replacing {\texttt{\{server\}}} in the types and terms by
%% {\texttt {\{client\}}} here would be enough for a fully client-side
%% version of the todo list.

A key property is that every reference of type \texttt{Ref \{Loc\}
  [A]} is dereferenced only at the right location \texttt{Loc}.
This is enforced by the type signatures of the three interface
functions.
Located references can be implemented efficiently without attaching
any location information to them at runtime in a tagless manner.

%% A new item whose name is provided by user input is added to
%% a list of items stored at the server by
%% \begin{center}
%% {\texttt {ref :=\{server\} (TodoItem line False :: !\{server\} ref)}}
%% \end{center}
%% (Line 30) where {\texttt {ref}} is a reference of type {\texttt {Ref
%%     \{server\} [List [TodoItem]]}}. In the code, {\texttt {!\{server\}
%%     ref}} retrieves the existing list, and then {\texttt {ref :=
%%     \{server\}}} updates the list by adding the new element onto the
%% head of the old list using the list constructor {\texttt {(::)}}.
%% %
%% It would be nice to avoid having to write the location parameters
%% explicitly, using a type inference. This will be discussed later.

%% Note that the code described above is intended to run on the client
%% and make a request to the server twice.  One server request can be
%% avoided by placing the code inside a server function followed by an
%% immediate application:
%% \begin{center}
%% {\texttt {(\textbackslash \_ @ server. \texttt ref :=\{server\} (TodoItem line False :: !\{server\} ref) ) ()}}
%% \end{center}

Programmers can define user-defined data types with polymorphic
locations. For example, one can define a polymorphic location model
type by abstracting the location {\texttt{server}} of the reference
type in Line 2:
\begin{center}
\texttt{data Model = \{l\}. Content String (List [TodoItem]) (Ref {l} [List [TodoItem]])}
\end{center}
Accordingly, {\texttt{init}}, {\texttt{view}}, and {\texttt{update}}
can be rewritten to use location-parametric models:
\begin{itemize}
\item \texttt{init : \{l\}. Model \{l\}}
\item \texttt{view : \{l\}. Model \{l\} -client-> Html [Msg]}
\item \texttt{update : \{l\}.Msg-client->Model\{l\}-client->Model\{l\}}
\end{itemize}
Then one can write a polymorphic page value
\begin{flushleft}
\ \ \texttt{page : \{l\}.Page [(Model \{l\}) Msg] = \\
\ \   \ \ \ \ \ Page (init \{l\}) (view \{l\}) (update \{l\})}
\end{flushleft}
where {\texttt{page\{client\}}} is a client only TodoMVC program while
{\texttt{page\{server\}}} is a multi-tier TodoMVC program that behaves like our original example.

%%%%%

\section{The Polymorphic RPC Calculus}
\label{sec:polyrpc}

This section reminds the reader of the polymorphic RPC calculus~\cite{CHOI:scp2020}. It is a polymorphically typed call-by-value
$\lambda$-calculus with location annotations on $\lambda$-abstractions
specifying where to run. The calculus offers the notion of polymorphic
location to write polymorphically located functions succinctly, which
is convenient for programmers.

\subsection{The Syntax and the Semantics}
\label{sec:polyrpc:syntax&semantics}

\begin{figure}[ht]
\begin{tabular}{ l  l  l  c  l }
\multicolumn{5}{l}{\textbf{Syntax}} \\
 \!\!\!\!\!\!          & Location 	\!\!\!	& $a,b$ 	\!\!\!	& $\!\!\! ::= \!\!\!$ 	& $\client \ \ | \  \  \server$
            \ \ \ \ \ \ \ \ \ \  $\Loc \ \ \ \ \ ::= \ \ \ \ \ a  \ \ |  \ \ l$
            	 \\
  %          &                   		& $\Loc$   	& $::=$  	& $a  \ \ |  \ \ l$   \\
            & Term 	& $L,M,N$	& $\!\!\! ::= \!\!\!$ 	& $V  \ |  \ L \ M  \ |  \ M[A]  \ |  \ M[\Loc]$			%	\\
  %          &   	&       	& $|$ 	&
                 $\ | \ (L,M)  \ \ |  \ \ \pi_i(M)$				\\
                 & Value    			& $V,W$		& $\!\!\! ::= \!\!\!$ 	& $x  \ \ |  \ \ \lambda^{Loc} x.M  \ \ |  \ \ \Lambda\alpha.V  \ \ |  \ \ \Lambda l.V \ | \ (V,W)$ \\
  %          &     			& 		& $|$ 	& $(V,W)$
\\[0.1cm]
\end{tabular}

\begin{tabular}{ l  c }
\multicolumn{2}{l}{\textbf{Semantics}} \\
&
{       \begin{prooftree}
       		\infer[left label=(Abs)]0{ \evalRPC{\lamL{b}{x}{M}}{a}{\lamL{b}{x}{M} }}
              \end{prooftree}
}                                            
\\[0.5cm]
&
{
    \begin{prooftree}
  		\hypo{ \evalRPC{L}{a}{\lamL{b}{x}{N}} }
  		\hypo{ \evalRPC{M}{a}{W} }
  		\hypo{ \evalRPC{N\subst{W}{x}}{b}{V}  }
  		\infer[left label=(App)]3{ \evalRPC{L \ M}{a}{V}  }
	\end{prooftree}
            }
  \\[0.5cm]
 &
{       \begin{prooftree}
       		\infer[left label=(Tabs)]0{ \evalRPC{\Lambda\alpha.V}{a}{\Lambda\alpha.V }}
	\end{prooftree}}
	\ \ \ \ \
{	\begin{prooftree}
  		\hypo{ \evalRPC{M}{a}{\Lambda\alpha.V} }
  		\infer[left label=(Tapp)]1{ \evalRPC{M[B]}{a}{V\subst{B}{\alpha}}  }
	\end{prooftree}}
\\[0.5cm]
 &
 	\mbox{
	\begin{prooftree}
      		\infer[left label=(Labs)]0{ \evalRPC{\Lambda l.V}{a}{\Lambda l.V} }
	\end{prooftree}
	\ \ \ \ \ \ \
	\begin{prooftree}
 		\hypo{ \evalRPC{M}{a}{\Lambda l.{V}} }
   		\infer[left label=(Lapp)]1{ \evalRPC{M[b]}{a}{V\subst{b}{l}}  }
	\end{prooftree}
 	}
\\[0.5cm]
 &
{       \begin{prooftree}
       	\hypo{ \evalRPC{L}{a}{V} }
       	\hypo{ \evalRPC{M}{a}{W} }
       	\infer[left label=(Pair)]2{ \evalRPC{(L,M)}{a}{(V,W) }}
	\end{prooftree}}
	\ \ \ \ \
{	\begin{prooftree}
  		\hypo{ \evalRPC{M}{a}{(V_1,V_2)}  \ \ \ i\in\{1,2\}}
  		\infer[left label=(Proj-i)]1{ \evalRPC{\pi_i(M)}{a}{V_i}  }
	\end{prooftree}}
\end{tabular}
\caption{The semantics for \polyrpc}
\Description[The semantics for \polyrpc]{The semantics for \polyrpc}
\label{fig:polyrpc}
\end{figure}

Figure \ref{fig:polyrpc} shows the syntax and semantics of the
polymorphic RPC calculus, {\polyrpc} that allows programmers to use
the same syntax of $\lambda$-application for both local and remote
calls, and allows them to compose differently located functions
arbitrarily.  An important feature is the notion of location variable
$l$ for which a location constant $a$ can be substituted. A syntactic
object $\Loc$ is either a location constant or a location variable.
Assuming the client-server model in the calculus, location constants
are either $\client$ denoting client or $\server$ denoting server.

In the syntax, $M$ denotes terms, and $V$ denotes values. Every
$\lambda$-abstraction $\lamL{\Loc}{x}{M}$ has a location annotation of
$\Loc$. By substituting a location $b$ for a location variable
annotation, $(\lamL{l}{x}{M})\subst{b}{l}$ becomes a monomorphic
$\lambda$-abstraction $\lamL{b}{x}{(M\subst{b}{l})}$. This location
variable is abstracted by the location abstraction construct $\Lambda
l.V$, and it is instantiated by the location application construct
$M[\Loc]$. Term applications are denoted by $L \ M$.  The rest of the
syntax are straightforward.

	The semantics of {\polyrpc} is defined in the style of a big-step operational semantics whose evaluation judgments, $\evalRPC{M}{a}{V}$, denote that a term $M$ evaluates to a value $V$ at location $a$.
	In the semantics, location annotated $\lambda$-abstractions, type abstractions, and location abstractions are all values. So, (Abs), (Tabs), and (Labs) are straightforwardly defined as an identity evaluation relation over them.
	(App) defines local calls when $a=b$ and remote calls when $a\not=b$ in the same syntax of lambda applications. The evaluation of an application $L \ M$ at location $a$ performs $\beta$-reduction at location $b$, where a $\lambda$-abstraction $\lamL{b}{x}{N}$ from $L$ has as an annotation, with a value $W$ from $M$, and it continues to evaluate the $\beta$-reduced term $N\subst{W}{x}$, which is a substitution of $W$ for $x$ in $N$, at the same location.
        The remaining semantics rules are easily understood.

        As a running example, let us consider a simple term:
        \[
        (\Lambda l. \lamL{l}{g}{g \ 1}) [\server] \ (\lamL{\client}{x}{x})
        \ \ \ \mbox{where} \ g \ \mbox{has type} \  Int \funL{\client} Int
        \]
        Evaluation starting at client goes to server by $(\lamL{\server}{g}{g \ 1}) \ (\lamL{\client}{x}{x})$ and then to the client by $(\lamL{\client}{x}{x}) \ 1$ resulting in $1$ there. The result comes back to the server and then to the client, ending the evaluation. 
        
\begin{figure}[ht]
\begin{minipage}[t]{\columnwidth}    
\begin{tabular}{l l c c l}
\multicolumn{5}{l}{\textbf{Types}} \\
   & Type & $A,B,C$ & $::=$ &
		$base \ | \  A\funL{Loc}B 	\ | \ \alpha
				 \ | \  A \times B \ | \ \forall\alpha.A 	\ | \ \forall l.A$
\\
%%    &      &         & $ |$ &
%% 		$\forall\alpha.A 	\ \ | \ \ \forall l.A $
%% \\
  & Type env. & $\Gamma$ & $::=$ &
		$\emptyset \ \ | \ \ \Gamma, x:A \ \ | \ \ \Gamma, \alpha \ \ | \ \ \Gamma, l$
\\[0.1cm]
\end{tabular}

\begin{tabular}{l l}
\multicolumn{2}{l}{\textbf{Typing Rules}} \\
 &
  	{\begin{prooftree}
		\hypo{  \tyenv(x)=A }
		\infer[left label=(T-Var)]1{ \typing{\tyenv}{\Loc}{x}{A} }
	\end{prooftree}}
	\ \ \ \ \
	{\begin{prooftree}
		\hypo{ \typing{\tyenvExt{x}{A}}{\Loc}{M}{B} }
		%\hypo{ flv(\Loc) \subseteq \Delta  }
		\infer[left label=(T-Abs)]1{ \typing{\tyenv}{\Loc'}{\lamL{\Loc}{x}{M}}{A\funL{\Loc}B} }
	\end{prooftree}}
\\[0.5cm]
&
	{\begin{prooftree}
		\hypo{  \typing{\tyenv}{\Loc}{L}{A\funL{\Loc'}B } }
		\hypo{  \typing{\tyenv}{\Loc}{M}{A} }
		%\hypo{  flv{A}\cup flv(\Loc') \subseteq \tyenv  }
		\infer[left label=(T-App)]2{ \typing{\tyenv}{\Loc}{L \ M}{B}   }
	\end{prooftree}}
\\[0.5cm]
&
	{\begin{prooftree}
		\hypo{  \typing{\tyenv,\alpha}{\Loc}{V}{A} }
		\infer[left label=(T-Tabs)]1{ \typing{\tyenv}{\Loc}{\Lambda\alpha.V}{\forall\alpha.A}   }
	\end{prooftree}}
	\
	{\begin{prooftree}
		\hypo{  \typing{\tyenv}{\Loc}{M}{\forall\alpha.A} }
		\infer[left label=(T-Tapp)]1{ \typing{\tyenv}{\Loc}{M[B]}{A\subst{B}{\alpha}}   }
	\end{prooftree}}        
\\[0.5cm]
&
	{\begin{prooftree}
		\hypo{ \typing{\tyenvExtWith{l}}{\Loc}{V}{A} }
		%\hypo{ l \not\in flv(\varenv)\cup flv(\tyenv) \cup flv(\Loc) }
		\infer[left label=(T-Labs)]1{ \typing{\tyenv}{\Loc}{\Lambda l.V}{\forall l.A }}
	\end{prooftree}}
	\
	{\begin{prooftree}
		\hypo{ \typing{\tyenv}{\Loc}{M}{\forall l.A } }
		% \hypo{ flv(\Loc')\subseteq\tyenv  }
		\infer[left label=(T-Lapp)]1{ \typing{\tyenv}{\Loc}{M[\Loc']}{A\subst{\Loc'}{l}}}
	\end{prooftree}}
\\[0.5cm]
 &
	{\begin{prooftree}
		\hypo{ \typing{\tyenv}{Loc}{L}{A} }
		\hypo{ \typing{\tyenv}{Loc}{M}{B} }
		\infer[left label=(T-Pair)]2{ \typing{\tyenv}{Loc}{(L,M)}{ A \times B }}
	\end{prooftree}}
\\[0.5cm]
&
	{\begin{prooftree}
		\hypo{ \typing{\tyenv}{Loc}{M}{A_1 \times A_2} \ \ \ i\in\{1,2\} }
		\infer[left label=(T-Proj-i)]1{ \typing{\tyenv}{Loc}{\pi_i(M)}{ A_i } }
	\end{prooftree}}
\end{tabular}
\end{minipage}
\caption{The type system for \polyrpc}
\Description[The type system for \polyrpc]{The type system for \polyrpc}
\label{fig:polyrpctysystem}
\end{figure}

\subsection{The Type System}
\label{sec:polyrpc:typesystem}

Figure \ref{fig:polyrpctysystem} shows a type system for the
polymorphic RPC calculus~\cite{CHOI:scp2020} that can identify remote
procedure calls at the type level, supporting location
polymorphism. The type language allows function types $A \funL{\Loc}
B$. Then every $\lambda$-abstraction at unknown location gets assigned
$A\funL{l} B$ using some location variable $l$. A universal quantifier
over a location variable, $\forall l. A$, is also introduced to allow
to abstract such occurrences of location variables.

Typing judgments are in the form of $\typing{\tyenv}{\Loc}{M}{A}$,
saying a term $M$ at location $a$ has type $A$ under a type
environment $\tyenv$.  The location annotation, $\Loc$, is either a
location variable or constant.  Typing environments $\tyenv$ have location variables, type
variables, and types of variables, as $\{l_1,
\cdots,l_n,\alpha_1,\cdots,\alpha_k, x_1:A_1, \cdots, x_m:A_m\}$.
%They are used to keep track of a set of free location, type, and value
%variables in the context of a given term.

The typing rules for the polymorphic RPC calculus are defined as
follows.  (T-App) is a refinement of the conventional typing rule for
$\lambda$-applications with respect to the combinations of location
$\Loc$ (where to evaluate the application) and location $\Loc'$ (where
to evaluate the function).  
%% When $\Loc=\Loc'$, one can statically decide
%% that it is a local procedure call.
%% Otherwise, $\Loc$ is different from $\Loc'$.
%% When both locations are constants as $\Loc=a$ and $\Loc'=b$,
%% $L \ M$ is statically found to be a remote procedure call: if
%% $a=\client$ and $b=\server$, it is to invoke a server function from
%% the client, and if $a=\server$ and $b=\client$, it is to invoke a
%% client function from the server.
%% When at least one of them is a location variable, we cannot make a decision statically.
%
For example, (T-App) is applied to $( (\Lambda l. \lamL{l}{g}{g \ 1})
[\server]) \ (\lamL{\client}{x}{x})$ with $\Loc=\client$ and
$\Loc'=\server$, meaning that this application is a remote procedure
call from client to server.
(T-App) is also applied to $g \ 1$ with $\Loc=l$ and $\Loc'=\client$.
We have to check over $l$ at runtime to make a decision whether or not
this is a local procedure call.
The other typing rules are explained in the extended version~\cite{choi2021typedslicingcompilation}.

%% (T-Labs) and (T-Lapp) are similar to the typing rules for type
%% abstraction and type application. (T-Labs) checks if its bound
%% location variable does not appear in the type environment and in the
%% contextual location. (T-Lapp) substitutes $\Loc'$ for all occurrences
%% of a location variable $l$ on $\lambda$-abstractions in $M$.

The type soundness of the type system for the polymorphic RPC
calculus, which was formulated as Theorem \ref{thm:typesoundness} and
was proved by~\cite{CHOI:scp2020}, guarantees that every remote
procedure call thus identified statically will never change to a local
procedure call under evaluation. This enables compilers to generate
call instructions for local calls and network communication for remote
calls safely even though both are in the same syntax of lambda
applications.

\begin{theorem}[Type soundness for \polyrpc~\cite{CHOI:scp2020}] For a  closed term $M$, if \ $\typing{\emptyset}{a}{M}{A}$ and $\evalRPC{M}{a}{V}$, then $\typing{\emptyset}{a}{V}{A}$.
\label{thm:typesoundness}
\end{theorem}

\subsection{The Static Approach to Polymorphic Locations}
\label{sec:polyrpc:staticapproach}

	When a polymorphic application is written in the way that the location of the application, $\Loc$, and the location of the function to run, $\Loc'$, may be location variables, compilers cannot statically determine if the lambda application is for remote calls, local calls, or both.
	The existing slicing compilation method for the typed RPC calculus~\cite{choijfp2019}, that is the simply typed and monomorphic subset of the polymorphic RPC calculus~\cite{CHOI:scp2020}, cannot deal with such a polymorphic lambda application any more.

	The previous study~\cite{CHOI:scp2020} overcame this limitation by translating all polymorphic locations in RPC programs into monomorphic ones by the so called  {\it monomorphisation} translation. This approach is called static because all polymorphic locations can now be resolved at compile-time.

	As stated by the study, in the worst case the monomorphisation translation can potentially lead to code explosion by generating client and server versions for each location abstraction. When there are $n$ location abstractions nested subsequently, $2^n$ monomorphic versions could be generated. This is called a code explosion problem of the static approach to the implementation of the polymorphic RPC calculus.

	To show the code explosion problem in the worst case that the study~\cite{CHOI:scp2020} mentioned, let us consider a small example of {\texttt {S}} and {\texttt {K}} combinators written in PolyRPC to make an identity function.
\begin{lstlisting}
 s : {l1 l2 l3}. [a b c].
       ((a-l1->b-l1->c) -l3-> (a-l2->b) -l3-> a-l3->c)
   = {l1 l2 l3}. \f @ l3  g @ l3  x @ l3. f x (g x) ;
 k : {l}. [a b]. (a -l-> b -l-> a)
   = {l}. \x @ l  y @ l . x ;
 identity : {l}. [a]. (a -l-> a)
   = {l}. \x @ l. s {l l l} (k {l}) (k {l}) x;
 main : Int  = identity {client} 123
\end{lstlisting}
	Let us call this a {\it spine location style} SKI program where every multiple-argument function is applied to all its arguments at the same location. There are several variants including a full freedom SKI program by allowing applying a multiple-argument function to each argument all at different locations. In the full freedom program, the {\texttt {S}} combinator will have a location abstraction with six location variables as {\texttt {\{l11 l12 l2 l31 l32 l33\}} } by replacing the two occurrences of {\texttt {l1}} by {\texttt {l11}} and {\texttt {l12}} and by replacing the three occurrences of {\texttt {l3}} by {\texttt {l31}}, {\texttt {l32}}, and {\texttt {l33}}.

	Here is a simple experimental result with these two programs for code size and location checks. By counting the nodes of a program tree (excluding type nodes), the spine location style SKI program is of size 59, and the full freedom SKI program is of size 68. After applying the monomorphisation, the sizes become 190 and 844. Running each of the two programs applies functions 9 times. Both of the spine location style and full freedom SKI programs do dynamic location checks 3 times. %: for example, in the H-style program,  once at {\texttt {(g x)}} where the current location is {\texttt {l3}} and the location for {\texttt {g}} to run is {\texttt {l2}}, twice at {\texttt {(f x)} where the current location is {\texttt {l3}} and the location for {\texttt {f}} to run is {\texttt {l1}}, and three times at the application of {\texttt {(f x)}} to {\texttt {(g x)}} where the current location is {\texttt {l3}} and the location for the function from {\texttt {(f x)}} to run is {\texttt {l1}}.

	With a preliminary experience with programming PolyRPC, we are not so sure how often the worst case behavior would appear in practice  by nested location abstractions as the existing study~\cite{CHOI:scp2020} is concerned.
        For now, this multi-tier TodoMVC program is the largest program about 300 lines written in PolyRPC. It is of size 1855, increasing up to 2554 after the monomorphisation.
        Some functions may have nested location abstractions naturally. Consider a located thread creation function, {\texttt{fork : \{l1 l2\}. (ProcId-l1->void) -l2-> ProcId}} where {\texttt{l1}} is the location of a child process and \texttt{l2} is the location of the parent process. Running \texttt{fork \{client server\}} in the server would create a client process with a parent process id and it would return an id for communication with the client child process. We would need more programming experience, which is left as a future work.

	In the next section, we will introduce a new polymorphically typed client-server calculus. Basically, this new CS calculus will be based on statically resolved location as done in the typed RPC calculus~\cite{choijfp2019}. In addition, it will also support a dynamic approach offering a way to determine dynamically whether polymorphic-location lambda applications are local or remote procedure calls.
        We are interested in the dynamic approach for several reasons. First, the dynamic approach does not have to do any static translations for polymorphic locations at compile-time. It can allow compilers to deal with polymorphic location programs directly. Second, this approach can handle the worst case behavior of the static approach in case such a bad situation happens. In this respect, the dynamic approach can be viewed as a generalization of the static approach. Third, having the dynamic approach itself is of our theoretical interest as a complementary technology. Fortunately, it is found out that it is easy to add the dynamic approach to the portion of the calculus that uses statically resolved locations.

%%%%%

\section{A Polymorphic Client-Server Calculus}
\label{sec:polycs}

\begin{figure*}[t]
\begin{tabular}{l l l c l }
\multicolumn{5}{l}{\textbf{Types \&  Terms}} \\
   & Type & $A,B,C$ & $ ::=$ &
                               $base \ \ | \ \ A \funL{Loc} B \ \ | \ \ Clo(A) \ \ | \ \ A \times B % $   \\
  %
%$
\ \ | \ \
		\alpha \ \ | \ \ \forall\alpha.A \ \ | \ \ \forall l.A \ \ | \ \ T \ A$  \\[0.1cm]
    %% & Code type & $\CloCodeType$ & $::=$ &
    %% 	$\loctyvars. \ \overline{B}.A$ \\[0.1cm]
%\multicolumn{5}{l}{\textbf{Terms \& Values}} \\
    & Term & $L,M,N$ & $::=$ &
                               $V \ \ | \ \ \llet{x}{M}{N} \ \ | \ \ \pi_i(V) % $ \\
%$
  \ \ | \ \
        V(W) \ \ | \ \ V[A] \ \ | \ \ V[Loc]$\\[0.1cm]
    & Value& $V,W$   & $::=$ &
     $x \ \ | \ \ (V,W) \ \ | \ \ \clo{\overline{W}}{F} \ \ | \ \  \Lambda\alpha.V$ \\
    &      &         & $| $   &
     $\lunit{V} \ \ | \ \ \ldo{x}{M}{N}$ %\\
%    &      &         & $ |$   &
     \ \ | \ \
     $\req{V}{W} \ \ | \ \ \call{V}{W} \ \ | \ \ \gen{Loc}{V}{W}$ \\[0.1cm]
    & Code term & $\CloCode$  & $::=$ &
    	$\loctyvars. \ \overline{z}. \ \OpenCode$ \ \ where $\OpenCode \ ::= \ \lambda x.M \ \ | \ \ \Lambda l.V$ \\
    %% &            & $\OpenCode$  & $::=$ &
    %% 	$\lambda x.M \ \ | \ \ \Lambda l.V$ \\
    & Code name & $F$     & $::=$ &
    	$F_{name}[\loctys]$ \\ % [0.1cm]
%\multicolumn{5}{l}{\textbf{Programs}} \\
    & Program & $prg$ & $\!\!\!::=\!\!\!$ &
     	$(\funstore_\client,\funstore_\server)$ \\
    & Function map & $\funstore$ & $\!\!\!::=\!\!\!$ &
     	$\{ \ F_{name,1}:\CloCodeType_1 = \CloCode_1, \ \cdots, \ F_{name,n}:\CloCodeType_n = \CloCode_n \ \}$ \ \ where $\CloCodeType$ is
    	$\loctyvars. \ \overline{B}.A$
\end{tabular}
\caption{Types and terms in the polymorphic CS calculus}
\Description[Types and terms in the polymorphic CS calculus]{Types and terms in the polymorphic CS calculus}
\label{fig:syntaxofdynamicpolycs}
\end{figure*}

	Slicing compilation is a desirable feature of multi-tier programming languages because it can slim down code sizes as small as necessary at each location and it can avoid potential security leaks resulting from the server code being available at the client.

	The idea of slicing compilation of the polymorphic RPC calculus naturally leads to the introduction of a client-server (CS) calculus where there are two separate programs, one for the client and the other for the server.
	Client-server programs can be modeled as a pair of client and server function maps, written as $(\funstore_\client, \funstore_\server)$ where $\funstore$ maps function names into the codes available at each location.
	Then the slicing compilation is a translation of RPC calculi terms into pairs of the function maps.

	The behavior of the client-server programs will be described using configurations, $Conf$, that are snapshots in the client-server model, written as $\confcs{client \ }{\ server}$.
	Firstly, a locally well-formed behavior is required in the client-server programs:
	the client part is only allowed to look up the client function map to find and run client functions and so is the server part with its own function map. For example, a closure whose function name refers to the server can appear at the client, but an attempt to run the closure at the client would get stuck.
	Secondly, the client and the server should keep a simple communication protocol: when one attempts to send something, the other should be ready to receive it, and  subsequently the roles are changed.

	We design a new typed calculus, named a polymorphic client-server (CS) calculus, {\polycs}, that serves as a target language where the type system guarantees both the locally well-formed behavior and the simple client-server communication protocol. This is contrasted with the existing CS calculi left in an untyped setting~\cite{Cooper:2006:LWP:1777707.1777724,choijfp2019}.

	As in the existing CS calculi, the polymorphic CS calculus syntactically distinguishes local procedure calls, written as $V(W)$, from the remote procedure calls, $\req{V}{W}$ and  $\call{V}{W}$. For example, $\req{V}{W}$ is interpreted as sending a pair of $V$ and $W$ to the server and doing a local procedure call there by $V(W)$, and $\call{V}{W}$ has the same interpretation but for the reverse direction.
	The polymorphic CS calculus will use these three syntactic forms of procedure calls wherever location information is statically known, as in the static approach.

	In addition, {\polycs } supports a new dynamic approach where location information can be examined in runtime.  We introduce a new syntactic form of procedure calls, $\gen{\Loc}{V}{W}$. The semantics of this generic procedure call is to take a location argument $\Loc$, which is the location of the function $V$.
	%Every generic procedure call becomes either a local one or a remote one depending on whether the current location is the same as the location to run the function.
	Suppose the client program has such a generic procedure call. Then it becomes a local procedure call if $\Loc$ is $\client$, and it becomes a remote one if $\Loc$ is $\server$. In the server program, it will have the opposite role.
	At compile-time, however, either $\Loc$ or the current location (or both) may be unknown if they are a location variable. This is where generic procedure calls are necessary to postpone the decision until runtime. The introduction of the generic procedure calls does not break the static approach still ensuring both the locally well-formed behavior and the simple client-server communication protocol.

	There are subtle issues in typing the symmetric RPC communication pattern between the client and the server. The symmetric pattern means that before a remote procedure call finishes to get a result, another remote procedure call in the reverse direction can intervene.
	For example, consider $(\lamL{\server}{g}{(g \ 1)}) \ (\lamL{\client}{x}{x})$. Before calling the server function from the client finishes, the client function is called back from the server through $g$.
	The existing CS calculi have implemented this RPC communication pattern  using {\it trampolined style}~\cite{Ganz:1999:TS:317636.317779} where a single ``scheduler'' loop, called trampoline, manages all transfers of control by remote procedure call. Whenever a computation performs a unit of work followed by remote procedure call, the remaining work is returned to the scheduler. Thus every remote procedure call is wrapped with a trampoline loop that would repeat to process intervening procedure call requests from the other location and that would eventually stop on a result of the remote procedure call.
%        Thus a function type $A\funL{\server} B$ in the RPC calculus would roughly be mapped to $A \funL{\server} (B \ + \ (\exists \beta\gamma. \ (\beta\funL{\client} \gamma) \times \beta))$ in the CS calculus. Return values intend to be either a result value of the function or an intervening procedure call request with a function and an argument that the trampoline method would process. The existential types are needed to express that those requested in the middle has nothing to do with the function.

	A type system for the polymorphic CS calculus is designed to have the calculus be high-level so that the low-level details of the trampoline-based implementation are not explicitly exposed to the syntactic terms and types. For this, monads are used to abstract the trampoline details of remote procedure calls and to focus on their eventual result values. Every term of monad type $T \ A$ may involve remote
        procedure calls that will eventually return a value of type
        $A$. Roughly speaking, all of $\req{V}{W}$,  $\call{V}{W}$,
        and $\gen{\Loc}{V}{W}$ will get assigned this type whenever
        $V$ is a function of type of $A$ to $B$ and $W$ is of type $A$. Thus typing remote procedure calls becomes as simple as typing local procedure calls, which is an advantage of our design decision. Typing the simple client-server communication protocol is going to be simple as will be explained in the following.

	The other issue is about typing concurrency between the client and the server in the polymorphic CS calculus. The RPC calculi are high-level enough to be able to treat them like a sequential calculus over the client and the server. Once communication primitives, such as {\it send} and {\it receive}, were introduced to the low-level implementation of the RPC calculus, some very limited form of concurrency would appear: a sender would have to be ready before a receiver would be so, or vice versa.
        To deal with this, we could resort to some advanced techniques, such as session types, but we decide to remain in a simpler type system.
        %design the CS calculus with the dedicated remote and generic procedure call terms 

%	For example, one would like to introduce universal low-level communication primitives, such as {\it send} and {\it receive}, rather than some dedicated high-level terms, $\req{V}{W}$ and $\call{V}{W}$ if the terms can be interpreted as involving sending and receiving. If the low-level communication primitives were introduced to the polymorphic CS calculus, we would have to consider concurrency in the static and dynamic semantics though the concurrency would exist in a very limited form. To deal with this, we would resort to some advanced techniques, such as the session calculus, but we believe that this would be a sledgehammer to crack a nut. As a result, we decide to have the dedicated terms for remote procedure calls in the calculus.
%	This issue will be revisited later when we discuss how to implement the polymorphic CS calculus.

%
Here is a summary of a few highlights of the polymorphic CS calculus.
First, the type system guarantees that while functions may be passed
to arbitrary locations, every function must run at the declared
location.
Second, it is designed to support the combination of the static
approach relying on monomorphisation and a complementary dynamic
approach for handling polymorphic-location programs directly.
Third, it employs a monadic abstraction for {\it trampolined style}
RPC communication that allows us to treat the polymorphic CS calculus
like a sequential calculus over the client and server.

\subsection{Types and Terms}
\label{sec:polycs:syntax}

\begin{figure*}[t]
\begin{tabular}{l l}
\multicolumn{2}{l}{\textbf{Typing Rules}} \\
&
{\begin{prooftree}
		\hypo{  \tyenv(x)=A }
		\infer[left label=(T-Var)]1{ \typing{\tyenv}{\Loc}{x}{A} }
	\end{prooftree}}
%\\[0.5cm]
%&
\ \ \
{  	\begin{prooftree}
		\hypo{  \typing{\tyenv}{\Loc}{M}{A} }
		\hypo{  \typing{\tyenvExt{x}{A}}{\Loc}{N}{B} }
		\infer[left label=(T-Let)]2{ \typing{\tyenv}{\Loc}{\llet{x}{M}{N}}{B} }
	\end{prooftree}}
%\\[0.5cm]
%&
\ \ \
	{\begin{prooftree}
		\hypo{ \funtyping{}{\tyenv}{\Loc}{F}{\overline{B_i}.A} }
		\hypo{ \typing{\emptyset}{\Loc}{\overline{W_i}}{\overline{B_i}}  }
		\infer[left label=(T-Clo)]2{ \typing{\tyenv}{\Loc}{\clo{\overline{W_i}}{F}}{\cloty{A}} 	}
	\end{prooftree}}
\\[0.5cm]
&
	{\begin{prooftree}
		\hypo{ \typing{\tyenv}{\Loc}{V}{A_1 \times A_2} \ \ \ i\in\{1,2\} }
		\infer[left label=(T-Proj-i)]1{ \typing{\tyenv}{\Loc}{\pi_i(V)}{ A_i } }
	\end{prooftree}}
\ \
	{\begin{prooftree}
		\hypo{ \typing{\tyenv}{\Loc}{V}{A} }
		\hypo{ \typing{\tyenv}{\Loc}{W}{B} }
		\infer[left label=(T-Pair)]2{ \typing{\tyenv}{\Loc}{(V,W)}{ A \times B }}
	\end{prooftree}}
\\[0.5cm]
&
%\\[0.5cm]
%&
	{\begin{prooftree}
		\hypo{ \typing{\tyenvExtWith{\alpha}}{Loc_0}{V}{A} \ \ \mbox{for all} \ \Loc_0}
		\infer[left label=(T-Tabs)]1{ \typing{\tyenv}{Loc}{\Lambda \alpha.V}{\forall\alpha.A }}
	\end{prooftree}}
%\\[0.5cm]
%&
\ \
	{\begin{prooftree}
		\hypo{  \typing{\tyenv}{\Loc}{V}{\forall\alpha.A} }
		\hypo{  \valty{B} }
		\infer[left label=(T-Tapp)]2{ \typing{\tyenv}{\Loc}{V[B]}{A\subst{B}{\alpha}}   }
	\end{prooftree}}
%\\[0.5cm]
%&
\ \ \
  	{\begin{prooftree}
		\hypo{  \typing{\tyenv}{\Loc}{V}{\cloty{\forall l.A}} }
		\infer[left label=(T-Lapp)]1{ \typing{\tyenv}{\Loc}{V[\Loc']}{A\subst{\Loc'}{l}}   }
	\end{prooftree}}
\\[0.5cm]
&
	{\begin{prooftree}
		\hypo{   \typing{\tyenv}{\Loc}{V}{A} }
		\hypo{  \valty{A} }
		\infer[left label=(T-Unit)]2{ \typing{\tyenv}{\Loc}{\lunit{V}}{T A} }
	\end{prooftree}}
%\\[0.5cm]
%&
\ \ \
  	{\begin{prooftree}
		\hypo{  \typing{\tyenv}{\Loc}{M}{T A} }
		\hypo{  \typing{\tyenvExt{x}{A}}{\Loc}{N}{T B} }
		\infer[left label=(T-Bind)]2{ \typing{\tyenv}{\Loc}{\ldo{x}{M}{N}}{T B} }
	\end{prooftree}}
\\[0.5cm]
        
 &
	{\begin{prooftree}
		\hypo{  \typing{\tyenv}{\Loc}{V}{\cloty{A\funL{\Loc}B} } }
		\hypo{  \typing{\tyenv}{\Loc}{W}{A} }
		\infer[left label=(T-App)]2{ \typing{\tyenv}{\Loc}{V(W)}{B}   }
	\end{prooftree}}
\ \ \
	{\begin{prooftree}
		\hypo{  \typing{\tyenv}{\Loc}{V}{\cloty{A\funL{\Loc'}T B} } }
		\hypo{  \typing{\tyenv}{\Loc}{W}{A} }
		\hypo{  \valty{A} }
		\infer[left label=(T-Gen)]3{ \typing{\tyenv}{\Loc}{\gen{\Loc'}{V}{W}}{T B}   }
	\end{prooftree}}
\\[0.5cm]
&
  \mbox{
	\begin{prooftree}
		\hypo{  \typing{\tyenv}{\client}{V}{\cloty{A\funL{\server}T B} } }
		\hypo{  \typing{\tyenv}{\client}{W}{A} }
		\hypo{  \valty{A} }
		\infer[left label=(T-Req)]3{ \typing{\tyenv}{\client}{\req{V}{W}}{T B}   }
	\end{prooftree}
	}
%\\[0.5cm]
%&
\ \ \
	{\begin{prooftree}
		\hypo{  \typing{\tyenv}{\server}{V}{\cloty{A\funL{\client}T B} } }
		\hypo{  \typing{\tyenv}{\server}{W}{A} }
		\hypo{  \valty{A} }
		\infer[left label=(T-Call)]3{ \typing{\tyenv}{\server}{\call{V}{W}}{T B}   }
	\end{prooftree}}
%\end{tabular}
%\begin{tabular}{l l}
\\[0.5cm]
\multicolumn{2}{l}{\textbf{Typing rule for function names}} \\
&
	{\begin{prooftree}
		\hypo{  (F_{name}:\loctyvars. \ \overline{B}.A\funL{\Loc'}B=\loctyvars. \overline{z}.\lambda x.M) \ \in\ \funstore_{\Loc'} }
		\hypo{ \typing{\loctyvars,\overline{z:B},x:A}{\Loc'}{M}{B} }
		\hypo{ \overline{\valty{A_i}} }
		\hypo{ \overline{\valty{B_j}} }
		\infer[left label=(T-F-Abs)]4{ \funtyping{}{\tyenv}{Loc}{F_{name}[\loctys]}{\overline{B\{\subst{\loctys}{\loctyvars}\} }.(A\funL{\Loc'}B)\{\subst{\loctys}{\loctyvars}\} }
		}
	\end{prooftree}}
\\[0.7cm]
&
	{\begin{prooftree}
		\hypo{  (F_{name}:\loctyvars. \ \overline{B}.\forall l.A=\loctyvars. \overline{z}.\Lambda l.V) \ \in\ \funstore_{common} }
		\hypo{ \typing{\loctyvars,\overline{z:B},l}{\Loc_0}{V}{A} \ \ \mbox{for all $\Loc_0$} }
		\hypo{ \overline{\valty{A_i}} }
		\hypo{ \overline{\valty{B_j}} }
		\infer[left label=(T-F-LAbs)]4{ \funtyping{}{\tyenv}{Loc}{F_{name}[\loctys]}{\overline{B\{\subst{\loctys}{\loctyvars}\} }.(\forall l.A)\{\subst{\loctys}{\loctyvars}\} }
		}
	\end{prooftree}}
\end{tabular}
\caption{The type system for the polymorphic CS calculus}
\Description[The type system for the polymorphic CS calculus]{The type system for the polymorphic CS calculus}
\label{fig:cstypesystem}
\end{figure*}

	The polymorphic CS calculus is as shown in Figure 	\ref{fig:syntaxofdynamicpolycs}. A monadic type, $T\ A$, denotes a computation that produces values of type $A$ and may involve remote procedure calls during the computation.
	A term for unit operation, $\lunit{V}$, turns values into monadic ones, and a term for monad composition, $\ldo{x}{M}{N}$, transforms monadic values of type $T\ A$ from $M$ into other monadic values of type $T\ B$ from $N$ after binding the unwrapped value of type $A$ to the variable $x$.
	Three monad terms are introduced for remote and generic procedure calls: $\req{V}{W}$, $\call{V}{W}$, and $\gen{\Loc}{V}{W}$.
	We call these five terms {\it monadic values}.
	The others are called {\it plain terms} and {\it values}.
        %Plain terms are let terms, $\llet{x}{M}{N}$, projections, $\pi_i(V)$, local applications, $V(W)$, type applications, $V[A]$, and location applications, $V[\Loc]$. Plain values are variables $x$, pairs $(V,W)$, type abstractions $\Lambda\alpha.V$, and closures $\clo{\overline{W}}{F}$ where $F$ is the name of code and $\overline{W}$, a sequence of values, is a free variable environment.

	As well as the monadic types, closure types, $Clo(A)$, are introduced for typing closures with functions or location abstractions hiding free variables occurring in them. Closures are allowed to be passed over tiers. In addition, all kinds of types and terms in the polymorphic RPC calculus are adopted.

	Every program in {\polycs} is a pair of client and server function maps, $(\funstore_\client, \funstore_\server)$. Function maps $\funstore$ are defined as mappings of names, $F_{name}$, into pairs of closed types, $\CloCodeType$, and closed codes, $\CloCode$. Such a mapping is described as $F_{name}:\CloCodeType=\CloCode$.

	Every closed code is written as $\loctyvars. \overline{z}.\OpenCode$. The prefix denotes abstractions over location variables, type variables, and free variables occurring in the open code.
	Values for the free variables in the open code are stored in closures.
	Types for the free type variables are not stored in closures but they replace the occurrences of the free type variables in the open code.
        Locations for the free location variables are treated as the same as types.
        The name of code, $F$, in a closure is defined as a name with location and type applications, $F_{name}[\overline{\Loc}\ \overline{A}]$, which represents an instance of a closed code referred by $F_{name}$, $\overline{z}.(\OpenCode\subst{\overline{\Loc}\ \overline{A}}{\loctyvars})$, where the occurrences of location variables and type variables are replaced by the locations and types, respectively.
	Each open code denoted by $\OpenCode$ is defined as either a lambda abstraction or a location abstraction. Because type abstractions will be erased later in the type erasure semantics, we will never construct  any closures for type abstractions treating them as a value, not as open code.

	%Note that locations for the free location variables in the open code are subtle. Later, in the type erasure semantics, we will treat free location variables in the same way as free variables, storing the free locations in closures.

        From the running example in Section~\ref{sec:polyrpc:syntax&semantics}, one can obtain a \polycs program as:
        $
          main=\ldo{ \ h \ } {\clo{\emptyset}{f_1}[\server] } {\req{\ h}{\ \clo{\emptyset}{f_3} \ }}
        $
where $(\funstore_\client,\funstore_\server)$ is 
\begin{center}
\begin{tabular}{l l l l l}
  $f_1 : \emptyset.\emptyset.A_1$ & $\!\!\!=\!\!\!$ & $\emptyset.\emptyset. \ \Lambda l. \ \lunit{ (\clo{\emptyset}{f_2[l]}) }$
  &
  $\in$ & $ \funstore_\client, \funstore_\server$
  \\
  $f_2 : l.\emptyset.A_2$ & $\!\!\!=\!\!\!$ & $l.\emptyset. \ \lambda g. \ \gen{ \ \client}{ \ g}{ \ 1 \ }$
  &
  $\in$ & $ \funstore_\client, \funstore_\server$
  \\
  $f_3 : \emptyset.\emptyset.A_3$ & $\!\!\!=\!\!\!$ & $\emptyset.\emptyset. \ \lambda x. \ \lunit{ x }$
  &
  $\in$ & $ \funstore_\client$
\end{tabular}
\end{center}
such that $A_1=\forall l.T \ \cloty{A_2}$,
$A_2 = \cloty{A_3} \funL{l} T \ Int$, and
$A_3 = Int \funL{\client} T\ Int$.
Note that the empty sequence is denoted by $\emptyset$.

\subsection{A Type System}
\label{sec:polycs:typesystem}

	The purpose of the type system for the polymorphic CS calculus is to guarantee both locally well-formed behavior and a simple client-server communication protocol. As previously, typing judgments for terms are $\typing{\tyenv}{\Loc}{M}{A}$ saying a term $M$ has type $A$ at location $\Loc$ under a type environment $\tyenv$.

	Every client-server program, $(\funstore_\client, \funstore_\server)$ is well-formed
	if there exist $\funstore_\client^\circ$, $\funstore_\server^\circ$, and $\funstore_{common}$  such that all function names are distinct, $\funstore_\client \cup \funstore_\server = \funstore_\client^\circ \uplus \funstore_\server^\circ \uplus \funstore_{common}$, and
	each binding $(F_{name} : \overline{l}\ \overline{\alpha}.\overline{B}.A=\overline{l}\ \overline{\alpha}.\overline{z}.\OpenCode)$ where  $\tyenv= \{\overline{l}, \overline{\alpha}, \overline{z:B}\}$, satisfies either
\begin{itemize}
	\item $A=A_1\funL{Loc}A_2$, $\OpenCode=\lambda x.M$, $\typing{\tyenvExtWith{x:A_1}}{Loc}{M}{A_2}$, and the binding is in $\funstore_{Loc}$; or
	\item $A=\forall l.A_1$, $\OpenCode=\forall l.V$, $\typing{\tyenvExtWith{l}}{Loc}{V}{A_1}$ for all arbitrary locations $Loc$, and the binding is in $\funstore_{common}$.
\end{itemize}
	where for notation, $\funstore_{Loc} = \funstore_a^\circ$ if $Loc=a$, and otherwise, if $Loc$ is a location variable, it is $\funstore_{common}$.

	Intuitively, every client-server program is type-checked under the client and server function maps. The union of the two functions maps has to be decomposed into three disjoint ones: $\funstore_{common}$, $\funstore_\client^\circ$, and $\funstore_\server^\circ$.
        %Each of the three function maps has to contain bindings satisfying some conditions.
        When an open code is a function whose type has a constant location annotation $a$, its binding belongs to $\funstore_a^\circ$. When an open code is associated with a location variable annotation or it is a location abstraction, its binding belongs to the common function map, $\funstore_{common}$. When such a decomposition is possible using the typing rules, the client-server program is said to be well-formed.

	The typing rules in Figure \ref{fig:cstypesystem} are designed to guarantee the locally well-formed behavior.
	Every remote procedure call has one's own location: $\req{V}{W}$ is a server procedure call from the client.
	(T-Req) specifies $\client$ as the location for evaluation and describes the procedure $V$ as a closure with a server function of type $\cloty{A\funL{\server}T B}$. The return type $T B$ denotes that the result values of the remote procedure call are of type $B$ involving a trampoline communication.
	(T-Call) is defined similarly for the reverse direction call, $\call{V}{W}$.

	Given a remote procedure call at the client with a typing derivation concluding $\typing{\emptyset}{\client}{\req{V}{W}}{T B}$, we are able to construct another typing derivation now for a local procedure call at the server concluding with $\typing{\emptyset}{\server}{V(W)}{T B}$ as long as the two values $V$ and $W$ are relocatable, i.e., $\typing{\emptyset}{\client}{V}{\cloty{A\funL{\server}T B}}$ implies $\typing{\emptyset}{\server}{V}{\cloty{A\funL{\server}T B}}$ and $\typing{\emptyset}{\client}{W}{A}$ implies $\typing{\emptyset}{\server}{W}{A}$.

	In fact, plain values can be shown to be all relocatable. To capture relocatable values, we define a predicate over types, $\valty{A}$. If $A$ is one of $\alpha$, $base$, $\cloty{B}$, and $\forall\alpha.B$, then $A$ is relocatable. If both $A$ and $B$ are relocatable, then $A\times B$ is relocatable too. Otherwise, $\valty{T\ A}$ is false for any $A$.
	In other words, integers are relocatable. Every closure can be located at the client or the server regardless of the function location. %When it is invoked in the server, we will do a local call. When it is invoked in the client, we will do a remote call by the request construct.
	%Note that both $\valty{A \funL{Loc} B}$ and $\valty{\forall l.A}$ are undefined because lambda abstractions and location abstractions now appear inside the closures, they are classified as so called {\it OpenCode}, not as values, and so there are no values of such types in {\polycs}.

	In (T-Req) and (T-Call), $V$ has closure type, so it is relocatable. For $W$, the typing rules enforce it to be relocatable by the extra condition.
	In an ill-formed term, $\req{V}{\req{W_1}{W_2}}$, when sent to to the server, one could attempt to invoke $\req{W_1}{W_2}$ at the server violating the well-formed local behavior. This is prevented by the predicate. % saying that no monadic value (of type $T \ A$) should be relocatable.
	This completes a justification for an interplay between typing remote and local procedure calls.

	In (T-Unit), the type of $V$ in $\lunit{V}$ is defined as relocatable because this term is used to return one location to the other. (T-Bind) is straightforward.

	In (T-Tapp), the argument type $B$ of type applications $V[B]$ is relocatable since type variables are defined as relocatable by the predicate. 
%Our type system can be said to be {\it predicative} in that only relocatable types are allowed to instantiate type variables. It would be interesting to design an {\it impredicative} but safe type system.
The system is currently limited in that type abstractions can only be instantiated with relocatable types, but this is not a problem in practice because we translate all types actually occurring as type applications in the source language to relocatable types anyway.  This limitation is due to the fact that we otherwise have no way of determining whether a type variable $\alpha$ is relocatable, for example in (T-Gen), (T-Call) or (T-Req); the solution is to ensure that the types eventually substituted for type variable are always relocatable. It would be interesting to remove this restriction, for example using qualified types~\cite{DBLP:journals/scp/Jones94} to track which type variables actually need to be relocatable.

	(T-Clo) is a typing rule for closures. It uses one of the typing rules (T-F-Abs)  and (T-F-Labs) for two purposes. One is for clarifying which function map the type checker should look at by a similar idea to that used in the function map decomposition. The other is for getting an instantiated type by appropriate location and type applications. The typing rules also enforce that all type arguments and all free type variables are relocatable.

        (T-Tabs) prohibits location-dependent values from $\Lambda\alpha.V$ by having a condition of typing $V$ at arbitrary locations. For example, $\Lambda\alpha.\req{f}{arg}$ is ill-typed because the monadic value $\req{f}{arg}$ is well-typed only at client by (T-Req).
        
	The other typing rules, (T-Let), (T-Proj-i), (T-Lapp), (T-Var), and (T-Pair) are straightforward except (T-Gen). 

	Until now, the typing rules have aimed at ensuring the well-formed behavior by the static approach while (T-Gen) is for the dynamic approach.
	The typing rule for $\gen{\Loc'}{V}{W}$ specifies that $\Loc'$ should be the function location. Thus the function location in the closure type $\cloty{A\funL{\Loc'}T B}$ becomes available in the term-level for examination against the evaluation location, $\Loc$, in runtime.
	The type soundness property shows that the introduction of generic procedure calls and (T-Gen) preserves the statically resolved location information by the static approach.
        Note that (T-App), (T-Req) and (T-Call) can be viewed as specific instances of (T-Gen), demanding no runtime location examination.

        In the running example, a code of function name $f_2$ 
        \[
        f_2 : l.\emptyset.A_2 \ = \ l.\emptyset. \ \lambda g. \ \gen{ \ \client}{ \ g}{ \ 1 \ }
        \ \in \ \funstore_\client, \funstore_\server
        \]
        would be type-checked by the following typing judgment
        \[
        \typing{l,g:\cloty{Int\funL{\client}T \ Int}}{l}{\gen{\client}{g}{1}}{T\ Int}
        \]
        where $A_2 = \cloty{Int\funL{\client}T \ Int} \funL{l} T\ Int$.

	In the next section, we will discuss the well-formed communication of the polymorphic CS calculus and will explain how our {\it stack typing} can guarantee the simple client-server protocol.

\subsection{Runtime Contexts and Typing Rules}
\label{sec:polycs:runtimecontexts}

\begin{figure}[t]
\begin{tabular}{l l c l l l}
%\multicolumn{5}{l}{\textbf{Runtime contexts}} \\
      Eval. context & $E[\ ]$ & $\!\!\!::=\!\!\!$ &
     	$E_{let}[\ ] \ \ | \ \ \ldo{x}{E[\ ]}{M}$ \\[0.1cm]
                         & $E_{let}[\ ]$ & $\!\!\!::=\!\!\!$ &
     	$[\ ] \ \ | \ \ \llet{x}{E_{let}[\ ]}{M}$ \\[0.1cm]
      Stack & $\stack$ & $\!\!\!::=\!\!\!$ &
     	$\emptystack \ \ | \ \ E[\ ];\stack$\\[0.1cm]
      Configuration & $Conf$ & $\!\!\!::=\!\!\!$ &
     	$\confcs{M;\stack_\client \ \ }{\ \ \stack_\server} \ \ | \ \ \confcs{\stack_\client\ \ }{\ \ M;\stack_\server}$ \\[0.1cm]
      Conf. context & $\conf$ & $\!\!\!::=\!\!\!$ &
     	$\confcs{E[\ ];\stack_\client\ \ }{\ \ \stack_\server} \ \ | \ \ \confcs{\stack_\client\ \ }{\ \ E[\ ];\stack_\server}$ \\
\end{tabular}
\caption{The runtime system and contexts}
\Description[The runtime system and contexts]{The runtime system and contexts}
\label{fig:runtimecontextsforpolycscalculus}
\end{figure}

	The communication aspect of {\polycs} involves runtime components and associated contexts as introduced in Figure \ref{fig:runtimecontextsforpolycscalculus}.
	Configurations, $Conf$, are snapshots in the client-server model, written as $\confcs{client \ }{\ server}$. There are two kinds of configurations: $\confcs{M_\client;\stack_\client \ }{\ \stack_\server}$ describes when the client evaluates a term $M_\client$ on the client stack $\stack_\client$ and the server stack $\stack_\server$, and $\confcs{\stack_\client\ }{\ M_\server;\stack_\server}$ is for the reverse roles of the client and the server. Stacks are defined as sequences of evaluation contexts separated by semicolons, $E_1[\ ];\cdots;E_n[\ ]$, and the empty stack is denoted by $\epsilon$. Stacks increase on each remote procedure call, and they decrease on its return. Each evaluation context in a stack denotes a term with a hole waiting for return values from the remote procedure call.

	The simple communication protocol that should be respected by the client and the server is this. When one attempts to send something, the other should be ready to receive it, and after that, the roles should be changed. This should be repeated until the two stacks are empty.

	Figure \ref{fig:configstacktysystem} shows typing rules for stacks and configurations. Stack typing judgements $\stacktyping{ a }{ \stackcs }{ A \Rightarrow B }$ is read as: the client and server respect the communication protocol by the stacks viewed at the location $a$ cooperatively producing a result value of type $B$ at the location whenever a value of type $A$ at the location is sent to the other.
        This generalizes the idea that evaluation contexts $E[\ ]$ can be understood as a function filling a value of type $A$ in the hole and evaluating the completed term to produce a result of type $B$

	In (T-Stk-Client), whenever a term, $E[x]$, completed with a variable $x$ for a value of type $A$ received from the client has type $T C$ at the server, and the stacks except the evaluation context, $\stackcsWith{\stack_\client}{\stack_\server}$, from the server view are well-formed with stack type $T C \Rightarrow B$, the two stacks from the client view will be well-formed with stack type $A \Rightarrow B$. Alternating views in stack typing judgments are changing roles in the communication, i.e., who to send and who to receive.
	%(T-Stk-Client) specifies that when a client is running, the server must wait for receiving from the client by an evaluation context on the server stack. If there is no such evaluation context, then the server stack is not only empty but the client stack should be also empty by (T-Stk-Empty).
	For example, $\confcs{M;E[] \ }{\ \emptystack}$ is an ill-formed configuration. When the client sends the value of $M$ to the server, there is no one to receive. Also, the client will receive nothing from the server through $E[]$ on the client stack.
%	In (T-Stk-Server), the idea of unwinding the stacks alternating between the client and the server is the same as previously. But the unwinding begins with sending a value to the client from the server.
	In (T-Stk-Server), the unwinding begins with sending a value to the client from the server.
	By (T-Stk-Empty), a pair of the two empty stacks is treated as an identity continuation.

        Configurations are well-formed when well-typed terms of type $T A$ fit well-typed pairs of stacks of stack type $T A \Rightarrow B$.
        So, configuration typing rules combine term typing with stack typing. By (T-Client), we assign a type $B$ to each client-running configuration, $\confcs{M;\stack_\client}{\stack_\server}$, if a closed term $M$ has type $T A$ at the client and the stacks $\stackcsWith{\stack_\client}{\stack_\server}$ has type $T A \Rightarrow B$ viewed from the client.
	By (T-Server), we can define a typing rule for server-running configurations, $\confcs{\stack_\client}{M;\stack_\server}$, in the same manner at the server.

        %% Recall that a stack is a sequence of evaluation contexts. Every evaluation context, $E[\ ]$, can be understood as a function filling a value of type $A$ in the hole and evaluating the completed term to produce a result of type $B$. By generalizing this idea, every pair of well-formed client and server stacks can also be understood as a function.

	%% The generalized idea is realized by the three stack typing rules. Stack types, written as $A \Rightarrow B$, are introduced to assign to pairs of client and server stacks.

\begin{figure}[t]
\begin{tabular}{l l }
\multicolumn{2}{l}{\textbf{Stack typing rules}} \\[0.1cm]
&
	{\begin{prooftree}
		\hypo{  }
		\infer[left label=(T-Stk-Empty)]1{ \stacktyping{ a }{ \stackcsWith{\emptystack \ }{\ \emptystack} } { A \Rightarrow A } }
	\end{prooftree}}
\\[0.3cm]
&
	{\begin{prooftree}
		\hypo{  \typing{x:A}{ \server }{E[x]}{T C} }
		\hypo{  \stacktyping{ \server }{ \stackcs }{ TC \Rightarrow B } }
		\infer[left label=(T-Stk-Client)]2{ \stacktyping{ \client }{ \stackcsWith{\stack_\client \ }{ \ E[\ ];\stack_\server} } { A \Rightarrow B } }
	\end{prooftree}}
\\[0.3cm]
&
	{\begin{prooftree}
		\hypo{  \typing{x:A}{ \client }{E[x]}{T C} }
		\hypo{  \stacktyping{ \client }{ \stackcs }{ TC \Rightarrow B } }
		\infer[left label=(T-Stk-Server)]2{ \stacktyping{ \server  }{ \stackcsWith{E[\ ];\stack_\client \ }{ \ \stack_\server} } { A \Rightarrow B } }
	\end{prooftree}}
\\[0.3cm]
\multicolumn{2}{l}{\textbf{Configuration typing rules}} \\[0.1cm]
&
	{\begin{prooftree}
		\hypo{ \typing{ \emptyset }{ \client }{M}{T A} }
		\hypo{ \stacktyping{ \client }{  \stackcs }{TA \Rightarrow B} }
		\infer[left label=(T-Client)]2{ \conftyping{ \confcs{M;\stack_\client \ \ }{\ \ \stack_\server} }{B} }
	\end{prooftree}}
\\[0.3cm]
&
	{\begin{prooftree}
		\hypo{ \typing{ \emptyset }{ \server }{M}{T A} }
		\hypo{ \stacktyping{ \server }{  \stackcs }{TA \Rightarrow B} }
		\infer[left label=(T-Server)]2{ \conftyping{ \confcs{\stack_\client\ \ }{\ \ M;\stack_\server} }{B} }
	\end{prooftree}}
\end{tabular}
\caption{Stack typing and configuration typing}
\Description[Stack typing and configuration typing]{Stack typing and configuration typing}
\label{fig:configstacktysystem}
\end{figure}

        In the running example, a term $(\lamL{\client}{x}{x}) \ 1$ at client intervening a remote call from client to server would correspond to a configuration $\confcs{\ \clo{\emptyset}{f_3}(1);[\ ] \ \ }{\ \ [\ ] \ }$, which is well-formed by 
        \[
	{\begin{prooftree}
		\hypo{ \typing{ \emptyset }{ \client }{\clo{\emptyset}{f_3} (1)}{T\ Int} }
		\hypo{ \stacktyping{ \client}{  \stackcsWith{[\ ]}{[\ ]} }{T\ Int \Rightarrow T\ Int} }
		\infer[left label=(T-Client)]2{ \conftyping{ \confcs{\clo{\emptyset}{f_3} (1);[\ ] \ \ }{\ \ [\ ]} }{T\ Int} }
	\end{prooftree}}
        \]

\subsection{The Semantics and Type Soundness}
\label{sec:polycs:semantics}

\begin{figure*}[t]
\noindent
\begin{minipage}[t]{\textwidth}
\begin{tabular}{l l l l l }
\multicolumn{5}{l}{[Local reduction]}
\\
(E-Local)
&
\multicolumn{3}{l}{
\begin{prooftree}
\hypo{ M \run^a M' }
\infer1{ \conf_a[M] \run \conf_a[M']  }
\end{prooftree}
}
\\[0.3cm]
(E-Let)
&
$\llet{x}{V}{M}$ & $\!\!\!\run^a\!\!\!$ & $M\subst{V}{x}$ &
\\[0.1cm]
(E-Do)
&
$\ldo{x}{\lunit{V}}{M}$ & $\!\!\!\run^a\!\!\!$ & $ M\subst{V}{x}$
\\[0.1cm]
(E-Proj-i)
&
$\pi_i(V_1,V_2)$ & $ \!\!\!\run^a\!\!\!$ & $ V_i$ \ \  where $i=1,2$
\\[0.1cm]
(E-TApp)
&
$(\Lambda \alpha.V)[A]$ & $\!\!\!\run^a\!\!\!$ & $V\subst{A}{\alpha}$
\\[0.1cm]
(E-App)
&
$clo(\overline{W},F)(V)$ & $\!\!\! \run^a \!\!\!$ & $M\subst{\overline{W}}{\overline{z}}\subst{V}{x}$ \\
& & & if $\funcode_a(F)=\overline{z}.\lambda x.M$
\\[0.1cm]
(E-LApp)
&
$clo(\overline{W},F)[b]$ & $\!\!\! \run^a \!\!\!$ & $V\subst{\overline{W}}{\overline{z}}\subst{b}{l}$ \\
& & & if $\funcode_a(F)=\overline{z}.\Lambda l.V$
\\
\end{tabular}
\begin{tabular}{l l l l l}
\multicolumn{5}{l}{[Communication]}
\\
(E-Req)
&
$\confcs{E[\req{V}{W}];\stack_\client \ \ }{\ \ \stack_\server}$
& $\!\!\!\run\!\!\!$
& $\confcs{E[\ ];\stack_\client\ \ }{\ \ V(W);\stack_\server}$
\\[0.1cm]
(E-Call)
&
$\confcs{\stack_\client\ \ }{\ \ E[\call{V}{W}];\stack_\server}$
& $\!\!\!\run\!\!\!$
& $\confcs{V(W);\stack_\client\ \ }{\ \ E[\ ];\stack_\server}$
\\[0.1cm]
(E-Unit-C)
&
$\confcs{\lunit{V};\stack_\client\ \ }{\ \ E[\ ];\stack_\server}$
& $\!\!\!\run\!\!\!$
& $\confcs{\stack_\client\ \ }{\ \ E[\lunit{V}];\stack_\server}$
\\[0.1cm]
(E-Unit-S)
&
$\confcs{E[\ ];\stack_\client\ \ }{\ \ \lunit{V};\stack_\server}$
& $\!\!\!\run\!\!\!$
& $\confcs{E[\lunit{V}];\stack_\client\ \ }{\ \ \stack_\server}$
\\[0.1cm]
(E-Unit-S-E)
&
$\confcs{\emptystack\ \ }{\ \ \lunit{V}}$
& $\!\!\!\run\!\!\!$
& $\confcs{\lunit{V}\ \ }{\ \ \emptystack}$
\\[0.1cm]
(E-Gen-C-C)
&
$\confcs{E[\gen{\client}{V}{W}];\stack_\client\ \ }{\ \ \stack_\server}$
& $\!\!\!\run\!\!\!$
& $\confcs{E[V(W)];\stack_\client\ \ }{\ \ \stack_\server}$
\\[0.1cm]
(E-Gen-S-C)
&
$\confcs{E[\gen{\server}{V}{W}];\stack_\client\ \ }{\ \ \stack_\server}$
& $\!\!\!\run\!\!\!$
& $\confcs{E[\req{V}{W}];\stack_\client\ \ }{\ \ \stack_\server}$
\\[0.1cm]
(E-Gen-C-S)
&
$\confcs{\stack_\client\ \ }{\ \ E[\gen{\client}{V}{W}];\stack_\server}$
& $\!\!\!\run\!\!\!$
& $\confcs{\stack_\client\ \ }{\ \ E[\call{V}{W}];\stack_\server}$
\\[0.1cm]
(E-Gen-S-S)
&
$\confcs{\stack_\client\ \ }{\ \ E[\gen{\server}{V}{W}];\stack_\server}$
& $\!\!\!\run\!\!\!$
& $\confcs{\stack_\client\ \ }{\ \ E[V(W)];\stack_\server}$
\\
\end{tabular}
\end{minipage}
\caption{The semantics for the polymorphic CS calculus}
\Description[The semantics for the polymorphic CS calculus]{The semantics for the polymorphic CS calculus}
\label{fig:polycscalculus}
\end{figure*}

	The semantics for the polymorphic client-server calculus is described by the small-step operational semantics over configurations, $Conf \rightarrow  Conf'$, as shown in Figure \ref{fig:polycscalculus}.
	The basic idea is to evaluate terms to monadic values and then to interpret the monadic values to perform remote procedure calls.
	The semantics is defined by a sequence of configurations whose configuration types are all $T A$ and whose last configuration will normally have a form as $\confcs{\lunit{V};\emptystack}{\emptystack}$ giving a value $V$ of type $A$.

        In the semantics, communication rules manages all transfers of control by remote procedure call while local reduction rules are used to perform each unit of work between two subsequent RPCs.
	For the local reduction rules, configuration contexts, $\conf_a[\ ]$, capture a local reduction at the location $a$ by (E-Local). Then the other local rules are applied.
	Evaluation contexts have two forms: $E[\ ]$ and $E_{let}[\ ]$. Computational evaluation is captured by $E[\ ]$ while plain term evaluation is captured by $E_{let}[\ ]$.
	Configuration contexts are either $\conf_\client[-]$ denoting $\confcs{E[-];\stack_\client}{\stack_\server}$ or $\conf_\server[-]$ denoting $\confcs{\stack_\client}{E[-];\stack_\server}$.
        (E-App) and (E-LApp) use only the function map at the location $a$ for looking up codes. An ill-formed program could get stuck because of the absence of the code to run.

	In the communication rules,
	(E-Req) and (E-Call) send a function and an argument to the other location leaving an evaluation context on the stack at the current location.
        The symmetry of the two rules demands a trampolined style implementation that one direction remote call can be intervened by the other one. 
        Later, our implementation of \polycs will make trampolined style loops explicit.

	The three Unit rules send back the remote procedure call results to the other location.
	The four Gen rules examine the location in the first argument against the current location of the generic procedure call to determine if the call is local or remote.

        The following are evaluation steps for the running example starting with $\confcs{main \ }{\ \epsilon}$.
\begin{flushleft}
\begin{tabular}{l l }
& $\confcs{ \ldo{ \ h \ } {\clo{\emptyset}{f_1}[\server] } {\req{\ h}{\ \clo{\emptyset}{f_3} \ }} \ }{\ \epsilon}$
\\
$\longrightarrow$ & $\confcs{ \ldo{ \ h \ } {\lunit{ (\clo{\emptyset}{f_2[\server]}) } } {\req{\ h}{\ \clo{\emptyset}{f_3} \ }} \ }{\ \epsilon}$
\\
$\longrightarrow$ & $\confcs{\ \req{ \ \clo{\emptyset}{f_2[\server]} \ }{ \ \clo{\emptyset}{f_3} \ } \ }{\ \epsilon \ }$
\\  
$\longrightarrow$ & $\confcs{\ [\ ] \ }{\ \clo{\emptyset}{f_2[\server]} ( \ \clo{\emptyset}{f_3} \ ) \ }$
\\
$\longrightarrow$ & $\confcs{\ [\ ] \ }{\ \gen{\client}{\clo{\emptyset}{f_3}}{1} \ }$
\\  
$\longrightarrow$ & $\confcs{\ [\ ] \ }{\ \call{ \ \clo{\emptyset}{f_3} \ }{1} \ }$
\\
$\longrightarrow$ & $\confcs{\ \clo{\emptyset}{f_3} \ (1);[\ ] \ }{\ [\ ] \ }$
\\  
$\longrightarrow$ & $\confcs{\ \lunit{1};[\ ] \ }{\ [\ ] \ }$
\\
$\longrightarrow$ & $\confcs{\ [\ ] \ }{\ \lunit{1} \ }$
\\
$\longrightarrow$ & $\confcs{\ \lunit{1} \ }{\ \epsilon \ }$
\end{tabular}
\end{flushleft}

	The type soundness property for the polymorphic CS calculus is proven as Theorem \ref{theorem:corollary:typesoundness} by showing the type preservation and the  progress properties.

\begin{theorem}[Type soundness] 	Given a well-formed polymorphic CS program $(\funstore_\client,\funstore_\server)$ with the main term $M$,
	if $\conftyping{\confcs{M;\emptystack}{\emptystack}}{TA}$,
	either $\confcs{M;\emptystack}{\emptystack} \run^* \confcs{\lunit{V};\emptystack}{\emptystack}$
	or it loops indefinitely.
\label{theorem:corollary:typesoundness}
\end{theorem}

%%%%%

\subsection{A Typed Slicing Compilation}
\label{sec:polycs:slicing}

\begin{figure*}[t]
\begin{minipage}[t]{.9\textwidth}    
\begin{tabular}{l l l l l l l l l l l l l l}
\multicolumn{4}{l}{\textbf{Type compilation}} \\
&
$\vcomp{\alpha}$ & $=$ & $\alpha$
&
$\vcomp{base}$ & $=$ & $base$
% \\
&
$\vcomp{A \funL{Loc} B}$ & $=$ & $\cloty{\vcomp{A} \funL{Loc} \ccomp{B}}$
 \\
&
$\vcomp{\forall\alpha.A}$ & $=$ & $\forall\alpha.\ccomp{A}$
&
$\vcomp{A\times B}$ & $=$ & $\vcomp{A}\times \vcomp{B}$
&
$\vcomp{\forall l.A}$ & $=$ & $\cloty{\forall l.\ccomp{A}}$
\\[0.1cm]
&
$\ccomp{A}$ & $=$ & $T \ (\vcomp{A})$
\\[0.1cm]
\end{tabular}

\begin{tabular}{l l l l }
\multicolumn{4}{l}{\textbf{Term \& value compilation}} \\
&
$\vcomp{x}_{\tyenv,Loc,A}$ & $=$ & $x$
\\
&
$\vcomp{\lambda^{Loc'}x.M}_{\tyenv,Loc,A\funL{Loc'}B}$ & $=$ & $\clo{\overline{z}}{F[\overline{l},\overline{\alpha}]}$ where
\ \ \
$\tyenv=\{\overline{l},\overline{\alpha},\overline{z}:\overline{C}\}$, \ $F_{name}$ fresh,
\\
& & &
\ \ \
$(F_{name}
  \ : \ \loctyvars.\overline{\vcomp{C}}.\vcomp{A}\funL{Loc'}\ccomp{B}
  \ = \ \loctyvars.\overline{z}.\lambda x. \ccomp{M}_{\tyenv,x:A,Loc',B}) \in \funstore_{\Loc'}$
%% \\
%% & & &
%% \ \ \
%% $\tyenv=\{\overline{l},\overline{\alpha},\overline{z}:\overline{C}\}$, $F_{name}$ fresh
%% \\
%% & & &
%% \ \ \
%% $\CloCodeType=\loctyvars.\overline{\vcomp{C}}.\vcomp{A}\funL{Loc'}\ccomp{B}$
%% \\
%% & & &
%% \ \ \
%% $\CloCode=\loctyvars.\overline{z}.\lambda x. \ccomp{M}_{\tyenv,x:A,Loc',B}$
%% \\
%% & & &
%% \ \ \
%% $(F_{name}: \CloCodeType = \CloCode) \in \funstore_{\Loc'}$
\\
%& & &
%\ \ \
%$\tyenv=\{\overline{\alpha},\overline{l},\overline{z}:\overline{C}\}$, $F_{name}$ fresh
%\\
%& & &
%\ \ \
%$\CloCodeType=\loctyvars.\overline{\vcomp{C}}.\forall\alpha.\ccomp{A} $
%\\
%& & &
%\ \ \
%$\CloCode=\loctyvars.\overline{z}.\Lambda \alpha. \ccomp{V}_{\tyenv,\alpha,Loc,A}$
%\\
%& & &
%\ \ \
%$(F_{name}:\CloCodeType = \CloCode) \in \funstore_{\Loc}$
%\\
&
$\vcomp{\Lambda l.V}_{\tyenv,Loc,\forall l.A}$ &  $=$ & $\clo{\overline{z}}{F[\overline{l},\overline{\alpha}]}$ where
% \\
% & & &
\ \ \
$\tyenv=\{\overline{l},\overline{\alpha},\overline{z}:\overline{C}\}$, \ $F_{name}$ fresh,
\\
& & &
\ \ \
$(F_{name}
 \ : \ \loctyvars.\overline{\vcomp{C}}.\forall l.\ccomp{A}
 \ = \ \loctyvars.\overline{z}.\Lambda l. \ccomp{V}_{\tyenv,l,Loc,A}) \in \funstore_{\client},\funstore_{\server}$
%% \\
%% & & &
%% \ \ \
%% $\CloCodeType=\loctyvars.\overline{\vcomp{C}}.\forall l.\ccomp{A}$
%% \\
%% & & &
%% \ \ \
%% $\CloCode=\loctyvars.\overline{z}.\Lambda l. \ccomp{V}_{\tyenv,l,Loc,A}$
%% \\
%% & & &
%% \ \ \
%% $(F_{name}: \CloCodeType = \CloCode) \in \funstore_{\client},\funstore_{\server}$
\\
&
$\vcomp{\Lambda\alpha.V}_{\tyenv,Loc,\forall\alpha.A}$ & $=$ &
$\Lambda\alpha.\ccomp{V}_{\tyenv,\alpha,Loc,A}$ % \clo{\overline{z}}{F[\overline{\alpha},\overline{l}]}$ % where
\\
&
$\vcomp{(V,W)}_{\tyenv,Loc,A\times B}$ & $=$ & $(\vcomp{V}_{\tyenv,Loc,A}, \vcomp{W}_{\tyenv,Loc,B}$)
\\
&
$\ccomp{V}_{\tyenv,Loc,A}$ & $=$ & $\lunit{(\vcomp{V}_{\tyenv,Loc,A})}$
\\
&
$\ccomp{L\ M}_{\tyenv,Loc,B}$ & $=$ & $\ldo{f}{\ccomp{L}_{\tyenv,Loc,A\funL{Loc}B}}{
  \ldo{x}{\ccomp{M}_{\tyenv,Loc,A}}{ f(x) } }$
\\
&
$\ccomp{L\ M}_{\tyenv,\client,B}$ & $=$ & $\ldo{f}{\ccomp{L}_{\tyenv,\client,A\funL{\server}B}}{
  \ldo{x}{\ccomp{M}_{\tyenv,\client,A}}{ \req{f}{x} } }$
\\
&
$\ccomp{L\ M}_{\tyenv,\server,B}$ & $=$ & $\ldo{f}{\ccomp{L}_{\tyenv,\server,A\funL{\client}B}}{
  \ldo{x}{\ccomp{M}_{\tyenv,\server,A}}{ \call{f}{x} } }$
\\
&
$\ccomp{L\ M}_{\tyenv,Loc,B}$ & $=$ & $\ldo{f}{\ccomp{L}_{\tyenv,Loc,A\funL{Loc'}B}}{
  \ldo{x}{\ccomp{M}_{\tyenv,Loc,A}}{ \gen{Loc'}{f}{x} } }$
%% \\
%% & & &
%% $\ \ \ \ldo{x}{\ccomp{M}_{\tyenv,Loc,A} \ \ \ \ \ \ }{
%% 		\gen{Loc'}{f}{x}
%% 	}$
\\
&
$\ccomp{M[B]}_{\tyenv,Loc,A\subst{B}{\alpha}}$ & $=$ & $\ldo{f}{\ccomp{M}_{\tyenv,Loc,\forall\alpha.A}}{ f[\vcomp{B}] }$
\\
&
$\ccomp{M[Loc']}_{\tyenv,Loc,A\subst{Loc'}{l}}$ & $=$ & $\ldo{f}{\ccomp{M}_{\tyenv,Loc,\forall l.A}}{ f[Loc'] }$
\\
&
$\ccomp{(L, M)}_{\tyenv,Loc,A\times B}$ & $=$ & $\ldo{x}{\ccomp{L}_{\tyenv,Loc,A}}{
\ldo{y}{\ccomp{M}_{\tyenv,Loc,B}}{
		\lunit{(x,y)}
	}
}$
%% \\
%% & & &
%% $ \ \ \ \ldo{y}{\ccomp{M}_{\tyenv,Loc,B}}{
%% 		\lunit{(x,y)}
%% 	}$
\\
&
$\ccomp{\pi_i(M)}_{\tyenv,Loc,A_i}$ & $=$ & $\ldo{p}{\ccomp{M}_{\tyenv,Loc,A_1\times A_2}}{
\llet{x}{\pi_i(p)}{
		\lunit{(x)}
	}
}$
%% \\
%% & & &
%% $\ \ \ \llet{x}{\pi_i(p)}{
%% 		\lunit{(x)}
%% 	} $
\end{tabular}
\end{minipage}
\caption{A typed compilation of {\polyrpc} into {\polycs}}
\Description[A typed compilation of {\polyrpc} into {\polycs}]{A typed compilation of {\polyrpc} into {\polycs}}
\label{fig:typedcompilation}
\end{figure*}

	Our typed slicing compilation translates {\polyrpc} into {\polycs}. Basically, it is a monadic conversion with a slicing, compiling
	RPC terms of type $A$ into monadic client and server terms of  type $T \ (\vcomp{A})$ denoting a computation of values of type $\vcomp{A}$ that may involve calling remote procedures during the computation.

	Figure \ref{fig:typedcompilation} shows the typed slicing compilation rules.
	It comprises type compilations, $\ccomp{A}$ and $\vcomp{A}$, and term compilations, $\ccomp{M}_{\tyenv,\Loc,A}$ and $\vcomp{V}_{\tyenv,\Loc,A}$.
	The term compilation rules actually take as its input typing derivations for terms, such as typing derivations concluding with typing judgments $\typing{\tyenv}{\Loc}{M}{A}$ or $\typing{\tyenv}{\Loc}{V}{A}$.
	The output is  two function maps, $\funstore_\client$ and $\funstore_\server$, with a main client expression.
	We use a notation,  $(F_{name} : \CloCodeType = \CloCode) \in \funstore_{\Loc}$, for adding the binding  of $F_{name}$ to function stores.
	If $\Loc$ in $\funstore_{\Loc}$ is a location variable, the compilation adds the binding both to the client function map and the server function map.

        The type and term compilation rules are quite straightforward and are in line with the ideas explained until now. Both lambda abstractions and location abstractions are compiled as closures while type abstractions are compiled as themselves that will be erased later. Lambda applications can be compiled with the new generic application by default. But by analyzing the location of the lambda applications ($\Loc$) and a function location ($\Loc'$), it is easy to have optimized compilation with local and remote application terms whenever the relevant location information is statically available, as was done for compiling the typed RPC calculus~\cite{choijfp2019}.
        When $\Loc=\Loc'$, $f(x)$ can replace $\gen{\Loc'}{f}{x}$ in the compilation. When $\Loc=\client\wedge\Loc'=\server$ and $\Loc=\server\wedge\Loc'=\client$, $\req{f}{x}$ and $\call{f}{x}$ will do so, respectively.
        Location applications and type applications are compiled as themselves but only the latter will be erased later. More explanations are available in the extended version~\cite{choi2021typedslicingcompilation}.

        By definition, the slicing compilation rules guarantee a linear bound on the size of target programs, incurring no code explosion problem like the one by the monomorphisation. 

	Now we state the type correctness and the semantic correctness properties of the typed slicing compilation rules as follows. By the type correctness property, every well-typed term in the polymorphic RPC calculus will be compiled into a well-typed term in the CS calculus by the typed slicing  compilation.

\begin{theorem}[Type correctness]
If \ $\typing{\tyenv}{Loc}{M}{A}$ in {\polyrpc} then $\typing{\vcomp{\tyenv}}{Loc}{\ccomp{M}_{\tyenv,\Loc,A}}{ \ccomp{A} }$ in {\polycs}
where $\vcomp{\tyenv}$ is a pointwise extension of the type compilation.
\label{thm:typecorrectslicingcompilation}
\end{theorem}

%%%

	We can also prove the semantic correctness of the slicing typed compilation meaning that whenever a well-typed term evaluates to a value under the semantics of the polymorphic RPC calculus, the compiled term will evaluate to the compiled value.

\begin{theorem}[Semantic correctness]
If \ $\typing{\emptyset}{\client}{M}{A}$  and $\evalRPC{M}{\client}{V}$ then $\confcs{ \ccomp{M}_{\emptyset,\client,A}}{\emptystack} \run^* \confcs{\ccomp{V}_{\emptyset,\client,A}}{\emptystack}$.
\label{thm:semanticcorrectslicingcompilation}
\end{theorem}

%%%%%

\section{Implementation of the Polymorphic CS Calculus}
\label{sec:implementation}

\begin{figure}[t]
\begin{tabular}{l l c l }
     Term & $m,n$ & $\!\!\! ::= \!\!\!$ &
		$v \ \ | \ \ \llet{x}{m}{n} \ \ | \ \ \pi_i(v) \ \ | \ \ v(w)  \ \ | \ \ 
                 p(\overline{v})$ \\
        &         & $|$    & $\case{e}{\overline{c \ \overline{x} \rightarrow m}} $ \\[0.1cm]
     Value& $v,w$   & $\!\!\! ::= \!\!\!$ &
     $x \ \ | \ \ (v,w)
     \ \ | \ \ Con\ \overline{v} % $\\
     % &      &         & $| $   &
      \ | \ \lunit{v} \ \ | \ \ \ldo{x}{m}{n} $ \\
%    &      &         & $ |$   &
     % $
 %    \ \ | \ \ \req{v}{w} \ \ | \ \ \call{v}{w} \ \ | \ \ \gen{v_{loc}}{v}{w}$ \\
      Primitive& $p$   & $\!\!\! ::= \!\!\!$ &
     $send \ \ | \ \ receive$\\[0.1cm]
%    & Code term & $\CloCode$  & $::=$ &
%    	$ \overline{z}.\lambda x.M$ \\[0.1cm]
% \multicolumn{5}{l}{\textbf{Programs}} \\
     Prog. & $prg$ & $\!\!\! ::= \!\!\!$ &
     	$(\funstore_\client,\funstore_\server)$ \\[0.1cm]
     Fun. map & $\funstore$ & $\!\!\! ::= \!\!\!$ &
%     	$\{ \ F_{name,1} = \CloCode_1, \ \cdots, \ F_{name,n} = \CloCode_n \ \}$ \\
%     $\{ \ F_{name,1} = \overline{z}_1\lambda x_1.m_1, \ \cdots, \ F_{name,n} = \overline{z}_n\lambda x_n.m_n \ \}$
     $\{ \ F_{1} = \overline{z}_1\lambda x_1.m_1, \ \cdots, \ F_{n} = \overline{z}_n\lambda x_n.m_n \ \}$
\\[0.1cm]
\end{tabular}
\caption{The syntax for the untyped CS calculus}
\Description[The syntax for the untyped CS calculus]{The syntax for the untyped CS calculus}
\label{fig:syntaxuntypedcs}
\end{figure}

	This section discusses how to implement {\polycs} client and server sliced programs efficiently. Firstly, the programs use types that were necessary for the slicing compilation but are not for execution. In the implementation, we want to erase the types but should retain the locations necessary for runtime examination.
	Secondly, the notion of monads in {\polycs} was useful before as an abstraction and we now need to implement it using low-level primitives.

	For implementation, we introduce an untyped language named {\cs} to be used as a target language for a type erasure translation retaining locations by value representation and exposing the concrete trampolined style communication. After presenting this translation, we will show that execution in \cs mirrors execution in \polycs.

%% \subsection{Untyped Variant of {\polycs}}
%% \label{sec:untypedcscalculus}

	Figure \ref{fig:syntaxuntypedcs} shows the syntax for an untyped CS calculus, which is a first-order functional programming language with networking.
	Terms denoted by $m$ include {\it send} and {\it receive} as communication primitives. Case terms are included to deconstruct data constructor value,
	$Con \ \overline{v}$ where $Con$ is a data constructor and $\overline{v}$ are its arguments. Note that values are denoted by $v$ or $w$.
	For example,
	$Client$ and $Server$ are ordinary data constructors of type $Location$ in {\cs} to represent location constants $\client$ and $\server$ in {\polycs}, respectively.
	Another form of $Closure \ \overline{v} \ F_{name}$ is introduced to {\cs} to implement $\clo{\overline{W}}{F_{name}[\overline{\Loc}\,\overline{\alpha}]}$ in {\polycs} under the assumption that $\overline{v}$ % faithfully
	implements $\overline{\Loc}$ together with $\overline{W}$ and $F_{name}$ is $Con_{F_{name}}$.
	The payloads for remote procedure calls are represented by $Apply \ v \ w$ for $\req{V}{W}$ and $\call{V}{W}$ and by $Ret \ v$ for $\lunit{V}$.

	Function maps now hold codes with no free type and free location variables where some free term variables of the codes  originate from the free location variables.

\begin{figure}

\begin{tabular}{l l c l}
     Configuration & $conf$ & $\!\!\!::=\!\!\!$ &
     	$\confcs{m_{\client} \ \ }{\ \ m_{\server}} $ \\[0.1cm]
     Eval. context & $e[\ ]$ & $\!\!\!::=\!\!\!$ &
     	$e_{let}[\ ] \ \ | \ \ \ldo{x}{e[\ ]}{m}$ \\[0.1cm]
                        & $e_{let}[\ ]$ & $\!\!\!::=\!\!\!$ &
     	$[\ ] \ \ | \ \ \llet{x}{e_{let}[\ ]}{m}$ \\[0.1cm]
%     & Stack & $\stack$ & $\!\!\!::=\!\!\!$ &
%     	$\emptystack \ \ | \ \ E[\ ];\stack$\\[0.1cm]
     Conf. context & $\confuntyped$ & $\!\!\!::=\!\!\!$ &
     	$\confcs{e_1[\ ] \ }{\ e_2[\ldo{x}{receive}{m}]} $\\
                &     & $ | $  &
        $\confcs{e_1[\ldo{x}{receive}{m}] \ }{\ e_2[\ ]}$ \\[0.1cm]
\end{tabular}

\begin{tabular}{l l l c l l}
\multicolumn{6}{l}{[Local reduction]}
\\
&
(e-local)
&
\multicolumn{3}{l}{
\mbox{
\begin{prooftree}
\hypo{ m \run m' }
\infer1{ \confuntyped[m] \run \confuntyped[m']  }
\end{prooftree}
}
}
\\[0.2cm]
&
(e-let)
&
$\llet{x}{v}{m}$ & $\!\!\!\run\!\!\!$ & $m\subst{v}{x}$ &
\\[0.1cm]
&
(e-do)
&
$\ldo{x}{\lunit{v}}{m}$ & $\!\!\!\run\!\!\!$ & $ m\subst{v}{x}$
\\[0.1cm]
&
(e-proj-i)
&
$\pi_i(v_1,v_2)$ & $ \!\!\!\run\!\!\!$ & $ v_i$ \ \  where $i=1,2$
\\[0.1cm]
&
(e-app)
&
$(Closure \ \overline{w} \ f)(v)$ & $\!\!\!\run\!\!\!$ & $m_f\subst{\overline{w}}{\overline{z}}\subst{v}{x}$
\\[0.1cm]
&
&
&
&
\ \ if $\funcode(f)=\overline{z}.\lambda x.m_f$
\\
&
(e-case)
&
$\case{c_i \ \overline{v}}{\overline{c \ \overline{x} \rightarrow m}}$ & $\!\!\!\run\!\!\!$ & $m_i\subst{\overline{v}}{\overline{x}}$
\\[0.2cm]
\multicolumn{6}{l}{[Communication]}
\\
&
(e-comm-c-s)
&
\multicolumn{4}{l}{$\confcs{\ e_1[\ldovoid{send \ v}{m_1} ] \ \ }{\ \ e_2[\ldo{x}{receive}{m_2}] \ } $}
\\
&
&
\multicolumn{4}{l}{$ \ \ \ \run \ \ \ \confcs{\ e_1[m_1] \ \ }{\ \ e_2[m_2\subst{v}{x}] \ }$}
\\[0.1cm]
&
(e-comm-s-c)
&
\multicolumn{4}{l}{$\confcs{\ e_1[\ldo{x}{receive}{m_1} ] \ \ }{\ \ e_2[\ldovoid{send \ v}{m_2}] \ } $}
\\
&
&
\multicolumn{4}{l}{$\ \ \  \run \ \ \ \confcs{\ e_1[m_1\subst{v}{x}] \ \ }{\ \ e_2[m_2] \ } $}
\end{tabular}
\caption{The semantics for the untyped CS calculus}
\Description[The semantics for the untyped CS calculus]{The semantics for the untyped CS calculus}
\label{fig:semanticsuntypedcs}
\end{figure}

	Configurations are in the form of $\confcs{m_\client\ }{\ m_\server}$  client term $m_\client$ and server term $m_\server$ as shown in Figure \ref{fig:semanticsuntypedcs}. Evaluation of an untyped CS program begins with $\confcs{main \ }{\ loop_{body}}$ where the main term is in the client side and $loop_{body}$ is the body of trampoline $loop$. The definition of the trampoline loop is in Figure \ref{fig:compilationtountypedcs}. A loop function at server waits for receiving payloads sent from client, serving them. Communication rules involve the two primitives, for example, as:
        \[ \confcs{ \ldovoid{send \ v}{m_1} \ }{\ \ldo{x}{receive}{m_2} }
        \ \run \ \confcs{m_1 \ }{\ m_2\subst{v}{x} }\]
for sending a value from the client to the server. Actually, each location has a trampoline loop for symmetric communication. The only difference is that $f(arg)$ in the $Apply$ case of the loop function is implemented with one's own function store $\funstore_a$. 

The semantic rules are available in Figure \ref{fig:semanticsuntypedcs}.
	Evaluation contexts $e[\ ]$ to choose a specific rule to execute are actually the same as the previous ones but configuration contexts $\sigma[\ ]$ are notable.  For example, client-side configuration contexts are in the form of $\confcs{e_1[\ ]}{e_2[\ldo{x}{receive}{m}]}$ meaning that server is ready to receive any payloads that would be sent by the client during the evaluation of a term.
	When the term in the client is in the form of $\ldovoid{send \ v}{ m}$, which is an abbreviation of $\ldo{x}{send \ v}{m}$ where $x$ is unused, the client is about to send a payload $v$. Then the communication rule (e-comm-c-s) enables us to move the payload from the client to the server. For the opposite direction, server-side configuration contexts and (e-comm-s-c) will do that.

	Local reduction rules are straightforward. A case reduction rule (e-case) is used to analyze location values in generic applications and to control the trampolined style communication flow.
        An application rule (e-app) extracts a function name from a closure, looks up $\funstore$ for its code, and continues to evaluate after substitutions on free variables and an argument.

%% \subsection{A Compilation and Its Semantic Correctness}
%% \label{sec:correctness}

\begin{figure}[t]
\begin{minipage}[t]{\columnwidth}    
\begin{tabular}{l l l l }
\multicolumn{4}{l}{\textbf{Function store compilation}} \\
&
$\ecomp{\funstore_\client,\funstore_\server}$
&
$=$
&
$\ecomp{\funstore_\client}_\client,\ecomp{\funstore_\server}_\server$
\\
&
$\ecomp{ \{ \overline{ F_{name}:\CloCodeType=\CloCode } \} }_a$ & $=$ & $ \{ \ \overline{ F_{name}=\ecomp{\CloCode}_a} \ \} $
\\
&
$\ecomp{ \overline{l} \, \overline{\alpha}. \, \overline{z}.OpenCode }_a$ & $=$ & $ \overline{z_l}\, \overline{z}.\ecomp{OpenCode}_a $
\\
&
$\ecomp{\lambda x.M}_a$ & $=$ & $\lambda x. \ecomp{M}_a$
\\
&
$\ecomp{\Lambda l.V}_a$ & $=$ & $\lambda x_l. \ecomp{V}_a$
\\[0.1cm]
\end{tabular}
\begin{tabular}{l l}
\multicolumn{2}{l}{\textbf{Location, term, and value compilation}} \\
&
$\ecomp{l} \ = \ x_l$ \ \ \ \ \ \ \ \
$\ecomp{\client} \ = \ Client$ \ \ \ \ \ \ \ \
$\ecomp{\server} \ =\ Server$ 
\\[0.1cm]
&
$\ecomp{x}_a = x$
\ \ \
$\ecomp{(V,W)}_a = ( \ecomp{V}_a, \ecomp{W}_a )$
\ \ \
$\ecomp{\pi_i(V)}_a = \pi_i( \ecomp{V}_a )$
\\
%% &
%% $\ecomp{(V,W)}_a \ = \ ( \ecomp{V}_a, \ecomp{W}_a )$
%% \\
&
$\ecomp{\clo{\overline{W}}{F[\overline{Loc}][\overline{A}]}}_a \ = \ Closure \ (\overline{ \ecomp{\Loc} } \, \overline{ \ecomp{W}_a }) \  F $
\\
&
$\ecomp{\Lambda\alpha.\lunit{V}}_a \ = \ \ecomp{V}_a$ \ \ \ \ \ $\ecomp{V[A]}_a \  = \  \lunit{ \ecomp{V}_a } $
\\
&
$\ecomp{ \lunit{V} }_a \ = \ \lunit{ \ecomp{V}_a }$
\\
&
$\ecomp{ \ldo{x}{M}{N} }_a \ = \ \ldo{x}{ \ecomp{M}_a }{ \ecomp{N}_a }$
\\
&
$\ecomp{ \req{V}{W} }_a \ = \ \ldokeyword \  send \ (Apply \ \ecomp{V}_a  \  \ecomp{W}_a ); \  loop \ ()$
\\
&
$\ecomp{ \call{V}{W} }_a \ = \ \ldokeyword \  send \ (Apply \ \ecomp{V}_a  \  \ecomp{W}_a ); \ loop \ ()$
\\
%% &
%% $\ecomp{ \gen{\Loc}{V}{W} }_\client$ & $\!\!\! = \!\!\!$ & $ \ldokeyword \ x \leftarrow \case{ \ecomp{\Loc} }{}$
%% \\
&
$\ecomp{ \gen{\Loc}{V}{W} }_\client \ = \ \mbox{if}(\ecomp{\Loc},\ecomp{ V(W) }_\client,\ecomp{ \req{V}{W} }_\client)$
%% \\
%% &                                                 &         & $ \ \ \ \ \ \ \ \ \ \ \ \ \ \ \ \{ \ Client \rightarrow \ecomp{ V(W) }_\client,\ Server \rightarrow \ecomp{ \req{V}{W} }_\client \ \}$
%% \\
%&                                                 &         & $ \ \ \ \ \ \ \ \ \ \ \ \ \ \ \ Server \rightarrow \ecomp{ \req{V}{W} }_\client$
%\\
%&                                                 &         & $ \ \ \ \ \ \lunit x$
\\
%% &
%% $\ecomp{ \gen{\Loc}{V}{W} }_\server$ & $\!\!\! = \!\!\!$ & $ \ldokeyword \ x \leftarrow \case{ \ecomp{\Loc} }{}$
%% \\
&
$\ecomp{ \gen{\Loc}{V}{W} }_\server \ = \ \mbox{if}(\ecomp{\Loc}, \ecomp{ \call{V}{W} }_\server, \ecomp{ V(W) }_\server)$
%% \\
%% &                                                 &         & $ \ \ \ \ \ \ \ \ \ \ \ \ \ \ \  \{ \ Client \rightarrow \ecomp{ \call{V}{W} }_\server, \ Server \rightarrow \ecomp{ V(W) }_\server \ \}$
%% \\
%&                                                 &         & $ \ \ \ \ \ \ \ \ \ \ \ \ \ \ \  Server \rightarrow \ecomp{ V(W) }_\server$
%\\
% &                                                 &         & $ \ \ \ \ \ \lunit x$
\\
&
\ \ \ where $\mbox{if}(v,m_1,m_2)=\case{v}{\{ \ Client \rightarrow m_1; \ Server \rightarrow m_2 \ \}}$
\\
&
$\ecomp{ \llet{x}{M}{N} }_a \ = \ \llet{x}{ \ecomp{M}_a }{ \ecomp{N}_a }$
\\
%% &
%% $\ecomp{\pi_i(V)}_a \ = \ \pi_i( \ecomp{V}_a )$
%% \\
&
$\ecomp{V(W)}_a \ = \ \ecomp{V}_a(\ecomp{W}_a)$
\\
&
$\ecomp{V[\Loc]}_a \ = \ \ecomp{V}_a(\ecomp{\Loc}_a)$
\\
%&
%$\ecomp{V(W)}_a \ = \ \case{ \ecomp{V}_a }{ Closure \ \overline{w} \ f \rightarrow m \subst{\overline{w}}{\overline{z}}\subst{ \ecomp{W}_a }{ x }}$
% \\
%&                                                 &         & $ \ \ \ \ \ Closure \ \overline{w} \ f \rightarrow m \subst{\overline{w}}{\overline{z}}\subst{ \ecomp{W}_a }{ x } $
%\\
% &                                                 &         & \ \ \ where $\llbracket\funstore_a\rrbracket_a(f)=\overline{z}.\lambda x.m$
%& \ \ \ where $\llbracket\funstore_a\rrbracket_a(f)=\overline{z}.\lambda x.m$
% \\
% &
% $\ecomp{V[\Loc]}_a \ = \ \case{ \ecomp{V}_a }{ Closure \ \overline{w} \ f \rightarrow m \subst{\overline{w}}{\overline{z}}\subst{ \ecomp{\Loc} }{ x }}$
% \\
%&                                                 &         & $ \ \ \ \ \ Closure \ \overline{w} \ f \rightarrow m \subst{\overline{w}}{\overline{z}}\subst{ \ecomp{\Loc} }{ x } $
%\\
% &                                                 &         & \ \ \ where $\llbracket\funstore_a\rrbracket_a(f)=\overline{z}.\lambda x.m$
% & \ \ \ where $\llbracket\funstore_a\rrbracket_a(f)=\overline{z}.\lambda x.m$
\\
\end{tabular}

\begin{tabular}{l l l l }
\multicolumn{4}{l}{\textbf{Trampoline function}} \\
&
$loop \ u$ & $=$ & $\ldokeyword \ x \ \leftarrow \ receive $
\\
&                                                 &         & $ \ \ \ \ \ \case{x}{} $
\\
&                                                 &         & $ \ \ \ \ \  \ \ \ Apply \ f \ arg \ \rightarrow \ldokeyword \ z \leftarrow  f(arg)$
%% \ send \ (Ret \ z); \ loop \ (), \ \ \ Ret \ y  \ \rightarrow \lunit{y} \ \}
\\
&                                                 &         & $ \ \ \ \ \  \ \ \ \ \  \ \ \ \ \  \ \ \ \ \  \ \ \ \ \  \ \ \ \ \  \ \ \ \ \  \ \  \ \ \ \ send \ (Ret \ z)$
\\
&                                                 &         & $ \ \ \ \ \  \ \ \ \ \  \ \ \ \ \  \ \ \ \ \  \ \ \ \ \  \ \ \ \ \  \ \ \ \ \  \ \  \ \ \ \ loop \ ()$
\\
&                                                 &         & $ \ \ \ \ \  \ \ \ Ret \ y  \ \rightarrow \lunit{y} $
\end{tabular}
\end{minipage}
\caption{Compilation of {\polycs} into {\cs}}
\Description[Compilation of {\polycs} into {\cs}]{Compilation of {\polycs} into {\cs}}
\label{fig:compilationtountypedcs}
\end{figure}

	Figure \ref{fig:compilationtountypedcs} shows compilation rules for locations, terms, values and function stores.
	Firstly, we review how to erase types and to compile locations. Every location variable $l$ is replaced by a term variable $x_l$ while the two location constants, $\client$ and $\server$, are compiled into $Client$ and $Server$, respectively, that we explained previously.

	Compiling a code, $\overline{l}\overline{\alpha}.\overline{z}.OpenCode$, erases the free type variables $\overline{\alpha}$, and changes the free location variables $\overline{l}$ into term variables. The compiled code will have $\overline{z_l} \, \overline{z}$ as free variables.
	Symmetrically, compiling a closure, $\clo{\overline{W}}{F_{name}[\overline{Loc}][\overline{A}]}$, erases the free types $\overline{A}$, and lets the compiled closure hold $\overline{\ecomp{\Loc}} \, \overline{\ecomp{W}}_a$ as  values that come from translating the free locations $\overline{\Loc}$ to variables $\overline{z_l}$ and from the existing  values $\overline{W}$.
        Also, location applications are compiled essentially in the same way as term applications.
        Then the terms compiled from generic applications can examine locations by case term over values representing the locations.

	Term applications $V(W)$ are compiled as an application term. 
	Location applications $V[\Loc]$ are compiled essentially in the same way but with the value representation $\ecomp{\Loc}$ as an argument.

        The compilation rules by definition guarantee a linear bound on the size of target terms too; in compiling $\gen{\Loc}{V}{W}$, terms compiled from $V$ and $W$ can be hoisted out of the conditional. 
        
	The trampoline communication between the client and the server is supported by a key pattern $\ldovoid{send \ v}{loop \ ()}$ as used in compiling remote procedure call terms, $\req{V}{W}$ and $\call{V}{W}$.
        Here $loop$ is a function waiting for receiving either $Apply \ f \ arg$ to call $f(arg)$ locally and to return its result back to the other  location, or $Ret \ y$ to finish the trampoline communication.
	Both of $\req{V}{W}$ and $\call{V}{W}$ are compiled into a term in this pattern but at one's own location enforced by the {\polycs} type system.
	For $\gen{\Loc}{V}{W}$, the compiled term has a case analysis on a value from the compiled location $\ecomp{\Loc}$ to determine whether $V$ is a remote procedure with an argument $W$.

\begin{figure*}[ht]
\begin{tabular}{l l}
  & $\confcs{\ main \ \ }{\ \ loop_{body} \ }$ \ \ where $loop_{body} = \ldokeyword \ x \leftarrow receive; \ \textsfCase \ x \ \textsfOf \ \{ \ Apply \ f \ arg \rightarrow \cdots; \ Ret \ y \rightarrow \lunit{y} \ \}$
\\
$=$
&
$\confcs{\ldo{h}{ \lunit{(Closure \ Server \ f_2)} }{\ \ldokeyword \ \{ \ send \ (Apply \ h \ (Closure \ \emptyset \ f_3)); \ loop \ () \ \}} \ \ }{\ \ loop_{body} \ }$
\\
$\longrightarrow$
& $\confcs{\ \ldokeyword \ \{ \ send \ (Apply \ (Closure \ Server \ f_2) \ (Closure \ \emptyset \ f_3)); \ loop \ () \ \} \ }{\ \ loop_{body} \ }$
\\  
$\longrightarrow^2$
& $\confcs{\ loop_{body} \ \ }{\ \ e[(Closure \ Server \ f_2) \  (Closure \ \emptyset \ f_3)] \ }$
%& $\confcs{\ loop_{body} \ \ }{\ \ e[\textsfCase \ (Closure \ Server \ f_2) \ \textsfOf \ Closure \ w \ f \rightarrow m_{f_2} \subst%{w}{z_l}\subst{(Closure \ \emptyset \ f_3)}{g}] \ }$
\ \ where $e[ \ ] = \ldokeyword \ z \leftarrow [ \ ]; send \ (Ret \ z); \ loop \ ()$
\\  
$\longrightarrow$
& $\confcs{\ loop_{body} \ \ }{\ \ e[m_{f_2} \subst{Server}{z_l}\subst{(Closure \ \emptyset \ f_3)}{g}] \ }$ \ \ where $m_{f_2}$ is the body of $f_2$
\\
$=$
& $\confcs{\ loop_{body} \ \ }{\ \ e[\ldokeyword \ { \ send \ (Apply \ (Closure \ \emptyset \ f_3) \ 1); \ loop \ () \ }]  \ }$
\\  
$\longrightarrow^2$
& $\confcs{\ e[ (Closure \ \emptyset \ f_3) \ (1)] \ \ }{\ \ e[ \ loop_{body} ] \ }$
%& $\confcs{\ e[\textsfCase \ (Closure \ \emptyset \ f_3) \ \textsfOf \ Closure \ \emptyset \ f \rightarrow m_{f_3} \subst{1}{x}] \ \ }{\ \ e[ \ loop_{body} ] \ }$
\\  
$\longrightarrow$
& $\confcs{\ e[m_{f_3} \subst{1}{x}] \ \ }{\ \ e[ \ loop_{body} ] \ }$ \ \ where $m_{f_3}$ is the body of $f_3$
\\  
$=$
& $\confcs{\ e[\lunit{1}] \ \ }{\ \ e[ \ loop_{body} ] \ }$
\\  
$\longrightarrow$
& $\confcs{\ send \ (Ret \ 1); \ loop \ () \ \ }{\ \ e[ \ loop_{body} ] \ }$
\\  
$\longrightarrow^2$
& $\confcs{\ loop_{body} \ \ }{\ \ e[ \ \lunit{1} ] \ }$
\\  
$\longrightarrow$
& $\confcs{\ loop_{body} \ \ }{\ \ send \ (Ret \ 1); \ loop \ () \ }$
\\  
$\longrightarrow^2$
& $\confcs{\ \lunit{1} \ \ }{\ \ loop_{body} \ }$
\end{tabular}
\caption{Evaluation steps for the running example of untyped {\cs} program}
\Description[A running example of an untyped {\cs} program]{A running example of an untyped {\cs} program}
\label{fig:runninguntypedcsprogram}
\end{figure*}
        
        For example, an untyped \cs program can be obtained from compiling the \polycs program in Section~\ref{sec:polycs:syntax},  as follows.
\begin{flushleft}
\begin{tabular}{l}
  $\ldo{ \ h \ } {\lunit{ \ (Closure \ Server \ f_2)}}{ }$ \\
  $\ldokeyword \ \{ \ send \ (Apply \ h \ (Closure \ \emptyset \ f_3)); \ loop \ () \ \}$
\end{tabular}
\ \\ where \ \\
\begin{tabular}{l l}
  $f_1 = \emptyset. \ \lambda x_l. \ \lunit{ \ (Closure \ x_l \ f_2) \ }$
  &
  $\!\!\!\in \funstore_\client, \funstore_\server$
  \\
  $f_2 = z_l.\ \lambda g. \ \textsfIf(Client, \ldokeyword \ \{ \ send \ (Apply \ g \ 1); \ loop \ () \ \}, ...)$
  &
  $\!\!\!\in \funstore_\server$
  \\
  $f_2 = z_l.\ \lambda g. \ \textsfIf(Client, \textsfCase \ g \ \textsfOf \ Closure \ \overline{w} \ f \rightarrow m, ...)$
  &
  $\!\!\!\in \funstore_\client$
  \\
  \ \ \ \ \ \ \ \ \ where 
  $\funstore_\client(f)=\overline{z_f}.\lambda x_f.m_f$, \ \ 
  $m = m_f \subst{\overline{w}}{\overline{z_f}}\subst{1}{x_f}$
  \\
  $f_3 = \emptyset.\ \lambda x. \ \lunit{x}$
  &
  $\!\!\!\in \funstore_\client$
\end{tabular}
\end{flushleft}

        Note that the code of $f_2$ in \polycs is compiled into two different \cs codes because $\gen{\client}{g}{1}$ would be a remote call at server while it would be a local one at client. 

        Figure \ref{fig:runninguntypedcsprogram} shows a running of the untyped CS program example above. In the evaluation steps, note the following configuration
        \[
        \confcs{\ e[(Closure \ \emptyset \ f_3) \ (1)] \ \ }{\ \ e[ \ loop_{body} ] \ }
        \ \ \ \mbox{in \cs}
        \]
        where $e[ \ ] = \ldokeyword \ z \leftarrow [ \ ]; send \ (Ret \ z); \ loop \ ()$.
        This \cs configuration actually mirrors a configuration $\confcs{\ \clo{\emptyset}{f_3}(1);[\ ] \ \ }{\ \ [\ ] \ }$ in \polycs that implements a term $(\lamL{\client}{x}{x}) \ 1$ at client intervening a remote call from client to server.
        This shows an example of how client and server trampoline loops in \cs, $e[\ ] \ | \ e[ loop_{body}]$, implement client and server stacks in \polycs, $[\ ] \ | \ [\ ]$.

        Guided by the execution in {\polycs}, our implementation with {\cs} is shown to respect the well-formed trampoline communication protocol by proving the semantic correctness of the compilation of {\polycs} into {\cs}.
        More details are found in the extended version~\cite{choi2021typedslicingcompilation}.

	%% Note that our formulation treats this function specially, always placing $loop \ ()$ to follow $send \ v$ immediately. So, $loop \ ()$ can be unfolded  to the function body, $\ldo{x}{receive}{\cdots}$, immediately after sending a value.

	Now we can prove the semantic correctness of the compilation of {\polycs} into {\cs}.

\begin{theorem}[Semantic Correctness of Compilation of {\polycs} into {\cs}]
\label{theorem:semanticcorrectnessoftypeandlocationerasuretotheend}
If $\confcs{M}{\emptystack} \ \run^* \ \confcs{\lunit \  V}{\emptystack}$ in {\polycs}
then $\confcs{ \ecomp{M}_\client \ }{ \  loop \ ()} \ \run^* \ \confcs{\ecomp{\lunit \ V}_\client \ }{\ loop \ ()}$  in {\cs}.
\end{theorem}

%%%%%

\section{Related Work and Discussion}
\label{sec:relatedwork}

\paragraph{Polymorphic locations:}
	The polymorphic RPC calculus~\cite{CHOI:scp2020} was the first RPC calculus that supports polymorphic locations useful for writing succinct multi-tier programs.
	There are only  a few publications that are relevant to the notion of polymorphic locations.
        ML5 has what they call {\it world polymorphism} based on modal logic, supporting {\it mobile code} runnable on different tiers represented by different possible worlds~\cite{Murphy:2007:TDP:1793574.1793585,Murphy:2008:MTM:1467784}. The RPC calculi are not about the mobility of code.

	Eliom~\cite{radanneaplas2016,radanneifl2016,Radanne2017,Radanne:2018:TWP:3184558.3185953} provides a macro feature called {\it shared sections}, which makes it possible to write code for the client and for the server at the same time,  the third location called {\it base} such that codeat location base can be used both on the client and on the server, and {\it mixed} declarations from multiple locations in a single module. For the first and second features in Eliom, the polymorphic RPC calculus may serve as a theoretical foundation.
        %seem to be a coarse-grained location polymorphism while the polymorphic RPC calculus allow more fine-grained one.
        Regarding the third feature, it would be interesting how the polymorphic RPC calculus can be extended with ML modules.

        There are many questions left about programming with the RPC calculus. Are polymorphic locations useful? Judging from using similar features in the existing programming languages and our experience, this feature is useful for writing succinct programs.
        Is  type-based control for remote procedure calls good? %The typed RPC calculi have this unique feature.
        Polymorphic locations surely fit the type-based scheme.
        Although without any term-level distinctions, programmers could be confused, even with a term-level sign to signal remote procedure calls, such confusion would arise too. Rather the type-level information could help them to understand the RPC behavior.
        Is more than one location abstraction useful, and if so, for what? PolyRPC is still at an early stage, and so programming experiences with it are too limited to answer this question firmly. We could think of applying PolyRPC to more complex distributed programming, such as for the cloud~\cite{10.1145/2096148.2034690} than the Web only with two locations.
        How can the PolyRPC compiler help programmers to avoid writing location annotations? For now, PolyRPC has a simple extension of  bidirectional type checking~\cite{10.1145/2500365.2500582} where programmers have to write all location applications, which can be burdensome sometimes. We could design some method to supply location arguments deduced from contexts as done in the context-aware programming languages~\cite{coeffects-thesis}.

\paragraph{Typed slicing compilations:}
	The feature of slicing compilation is desirable in multi-tier programming languages because it can reduce code size at each location by stripping code that does not belong to the current location. More importantly, it can avoid unnecessary security leaks resulting from the server code being available at the client on the web browser where every detail of the (compiled JavaScript) code is exposed to reverse engineering.

	Only a few multi-tier programming languages have supported slicing compilation .
		The untyped and monomorphic RPC calculi~\cite{Cooper:2006:LWP:1777707.1777724,choijfp2019} supported a slicing compilation but type information became  unavailable after it while our slicing compilation produces typed polymorphic CS calculus programs.
	Links~\cite{Cooper:2006:LWP:1777707.1777724} has a slicing compilation method for the so called {\it stateless} server scheme but does not use it anymore. The source program is compiled into an intermediate representation tree, and the client portions of the tree are compiled to JavaScript and the server portions are directly interpreted.
	Ur/Web~\cite{10.1145/2784731.2784741} supports a slicing compilation in implementation but there is no formal description of it. % Ur/Web~\cite{Chlipala2015} is the only work that evaluated performance.
	Eliom~\cite{radanneaplas2016,radanneifl2016,Radanne2017,Radanne:2018:TWP:3184558.3185953} has both theory and implementation of a typed slicing compilation generating OCaml programs.
	ScalaLoci~\cite{Weisenburger:2018:DSD:3288538.3276499,weisenburger2020programming}, Hop.js~\cite{Serrano:2016:GH:3022670.2951916}, and Gavial~\cite{Reynders2020} do not separate the client part from the server part for running multi-tier programs.
	The multi-tier calculus~\cite{Neubauer:2005:SPM:1040305.1040324,neubauer2007} is equipped with a typed slicing compilation used to optimize the sliced code correctly. But their slicing compilation scheme is different from ours in that every sliced program from the scheme has the same control structure as the multitier program.
	ML5~\cite{Murphy:2007:TDP:1793574.1793585,Murphy:2008:MTM:1467784} had a formal description and implemented it with no correctness proofs.

\paragraph{Intensional location polymorphism and location representations:}

	The idea of runtime location representations  and checking in our CS calculi is closely connected with the existing runtime type analysis~\cite{Harper:1995:CPU:199448.199475,crary_weirich_morrisett_2002,cheney:tr2003-1901,Xi:10.1145/640128.604150}.
	For example, the use of generic applications on locations is  analogous to the intensional polymorphism using {\texttt {typecase}} on types~\cite{Harper:1995:CPU:199448.199475}, and the use of the location representations is similar to the intentional polymorphism in type-erasure semantics~\cite{crary_weirich_morrisett_2002}.
	A difference is that our study can guarantee the running of functions at the right place in the client-server model.
	With generalized algebraic data types (GADT)~\cite{Xi:10.1145/640128.604150, cheney:tr2003-1901}, how to encode locations was discussed in~\cite{CHOI:scp2020}, and is used in our implementation. But they did not design any type-erasure translation nor prove its correctness.

%%%%%

\section{Conclusion and future work}
\label{sec:conclusion}

In this paper, we provided the first implementation of the polymorphic
RPC calculus by the typed slicing compilation into the polymorphic CS
calculus that is subsequently supported by the type-erasure,
runtime location representation, and explicit communication
primitives.
The combination of static and dynamic resolution of local or remote
procedure calls is new. This is different from the previous approaches
only with dynamic resolution as in Links (the untyped RPC calculus)
and only with static resolution as in the typed or polymorphic RPC
calculus.
We designed an experimental multi-tier programming language system for
the Web, and developed a multi-tier ToDoMVC program as a case study.

As future work, we plan to enhance our bidirectional type checker for
programmers to avoid having to write location parameters explicitly.
An approach would automatically infer all location parameters while
allowing programmers to write some only wherever necessary.
Another direction is to expand the typed slicing compilation to the
remaining untyped stage.
An advanced calculus with GADTs and session types could serve as a
target calculus for the purpose.
As a benefit, conventional optimization methods would be made use of
to optimize our dynamic approach, and the polymorphic RPC calculus
could be implemented on top of the concurrent lambda calculus.

\begin{acks}
We would like to thank Simon Fowler and the anonymous reviewers for
helpful feedback and suggestions for improvement.
This work was supported by ERC Consolidator Grant Skye (grant number
ERC 682315), and by an ISCF Metrology Fellowship grant provided by the
UK government's Department for Business, Energy and Industrial
Strategy (BEIS).
Kwanghoon Choi was supported by the National Research Foundation of
Korea (NRF) grant funded by MoE (No. 2019R1I1A3A01058608).
Sam Lindley was supported by the UKRI Future Leaders Fellowship EHOP
(grant number MR/T043830/1).
\end{acks}

\bibliographystyle{ACM-Reference-Format}
%\bibliography{polyrpc2021}


\newpage

\onecolumn

\appendix

\section{Examples in PolyRPC}

\subsection{A Multi-tier TodoMVC Example in PolyRPC}

This section motivates a reader with an example of a multi-tier
TodoMVC program. This is a web-based program to manage a list of work
items, and is structured by the Model-View-Update (MVU~\cite{fowler:LIPIcs:2020:13171}) design pattern.  It is written in
PolyRPC\footnote{https://github.com/kwanghoon/polyrpc}, which is an
experimental programming language based on the polymorphic RPC
calculus in Section \ref{sec:polyrpc}.  Its running is fully supported
by our dynamic approach to be explained from Section \ref{sec:polycs}.
TodoMVC is known as a `Hello World' program in web programming, and is
useful for comparing different web programming languages and
frameworks.

The multi-tier TodoMVC program consists of a web-based UI for the
clients and a server part that manages a list of work items. The UI
allows a user to ask the server to add a new item, mark an item as
completed, delete an item, and so on. Figure \ref{fig:todomvcrunning}
depicts such a configuration of the client and the server.

For an explanation, we present an abbreviated source code of the fully
functional multi-tier TodoMVC
program\footnote{https://github.com/kwanghoon/todomvc} in Figure
\ref{fig:todomvccode}. With the MVU design pattern, the main declares
a page with the initial model, a view function, and an updating
function in Line 42.

PolyRPC allows programmers to omit writing type abstractions and type
arguments that can be reconstructed automatically by bidirectional
type checking~\cite{10.1145/2500365.2500582}.

The {\texttt {view}} function in Line 18 takes a model as an argument
and returns an HTML value at the client. In PolyRPC, programmers can
specify where to run a function. For example, the {\texttt {view}}
function has type {\texttt{Model-client->Html [Msg]}} where the client
location is annotated to the function type.

A user interacts with this constructed HTML of type {\texttt {Html
    [Msg]}} through event handler actions that generate messages of
type {\texttt {Msg}}: when the user types new characters `M', `o',
`v', `i', `n', `g' in sequence as in Figure \ref{fig:todomvcrunning},
{\texttt {onInput}} in Line 17 generates messages {\texttt {Update
    ``M''}}, {\texttt {Update ``Mo''}}, ... , {\texttt {Update
    ``Moving''}}, respectively.  When the user types an Enter key,
{\texttt {onEnter}} generates a message {\texttt {Submit} to add a new
  item to the list with the typed string.  When the user clicks a
  checkbox associated with an item, {\texttt {onClick}} in Line 8
  generates a message {\texttt {Toggle index}} with the index for the
  item.  When the user presses an X button following an item in Figure
  \ref{fig:todomvcrunning}, {\texttt {onClick}} in Line 11 generates
      {\texttt {Delete index}} for the item.

Every message drives the {\texttt {update}} function in Line 24 to
update the existing model at the client, accordingly. So, the function
is of type {\texttt {Msg-client->Model-client->Model}}. The case
expression in the {\texttt {update}} function analyzes a given message
to act on the model at the client together with the list of items
stored at the server.  Then the {\texttt {view}} function will
reconstruct new HTML using the updated model.

Every model of type {\texttt {Model}} is constructed by {\texttt
  {Content}} with a text string that the user is typing, a list of
visible items of type {\texttt {List [TodoItem]}}, and a reference to
a list of all items at the server of type {\texttt {Ref \{server\}
    [List [TodoItem]]}} where \texttt{Ref \{server\}} is the location
application type of \texttt{Ref} to \texttt{server} using
\texttt{\{-\}} and \texttt{List [TodoItem]} is the type application
type of \texttt{List} to \texttt{TodoItem} using \texttt{[-]} in
PolyRPC.  Then, for example, on each {\texttt {Update str}} message
(in Line 28), the existing text in the model is replaced by a new one
that the user is typing.

To know how to handle the other message types, one has to understand a
locative and polymorphic reference type \texttt {Ref \{Loc\} [A]},
which is an abstract data type parameterized by locations {\texttt
  {Loc}}, as well as types {\texttt {A}}, with three interface
functions:
\begin{itemize}
\item \texttt{ref : \{l\}. [a]. a -l-> Ref \{l\} [a] }
\item \texttt{(!) : \{l\}. [a]. Ref \{l\} [a] -l-> a }
\item \texttt{(:=) : \{l\}. [a]. Ref \{l\} [a] -l-> a -l-> Unit }.
\end{itemize}
where \texttt{\{l\}.A} is a location abstraction type over a location
variable {\texttt {l}} and \texttt{[a].A} is a type abstraction type
over a type variable {\texttt {a}}.
In Line 40, {\texttt {Ref \{server\} [List [TodoItem]]}} is the type
for references to a work item list stored at the server. Only {\texttt
  {ref \{server\}}}, {\texttt {! \{server\}}}, and {\texttt {:=
    \{server\}}} can create, read, and modify them. For notation,
{\texttt {M \{Loc\}}} is the location application of $M$ to $\Loc$.
Just replacing {\texttt{\{server\}}} in the types and terms by
{\texttt {\{client\}}} here would be enough for a fully client-side
version of the todo list. This highlights the advantage of location
polymorphism for writing succinct programs instead of writing the same
but multiple differently located programs.

A key property is that every reference of type \texttt{Ref \{Loc\}
  [A]} is dereferenced only at the right location \texttt{Loc}. This
is guaranteed because the three interface functions are designed to
have the same location annotation as the associated location
annotation.

In Line 30, a new item, {\texttt {TodoItem line False}} with a text
line which a user enters at the client, is added to a list of items
stored at the server by
\begin{center}
{\texttt {ref :=\{server\} (TodoItem line False :: !\{server\} ref)}}
\end{center}
where {\texttt {ref}} is a reference of type {\texttt {Ref \{server\}
    [List [TodoItem]]}}. In the code, {\texttt {!\{server\} ref}}
retrieves the existing list, and then {\texttt {ref := \{server\}}}
modifies the existing list with a new one by the list
constructor {\texttt {(::)}}.

Note that the code explained above is supposed to run at the client
and requests the server two times.  One server request can be avoided
by placing the code inside a server function followed by an immediate
application, as:
\begin{center}
{\texttt {(\textbackslash \_ @ server. \texttt ref :=\{server\} (TodoItem line False :: !\{server\} ref) ) ()}}
\end{center}

Programmers can define user-defined data types with polymorphic
locations. For example, one can define a polymorphic location model
type by abstracting the location {\texttt{server} of the reference
  type in Line 2 as
\begin{itemize}
\item \texttt{data Model = \{l\}. Content String (List [TodoItem]) (Ref {l} [List [TodoItem]])}
\end{itemize}

Accordingly, {\texttt{init}}, {\texttt{view}}, and {\texttt{update}}
can be rewritten to use the location-parameterized models:
\begin{itemize}
\item \texttt{init : \{l\}. Model \{l\}}
\item \texttt{view : \{l\}. Model \{l\} -client-> Html [Msg]}
\item \texttt{update : \{l\}.Msg-client->Model\{l\}-client->Model\{l\}}
\end{itemize}

Then one can write a page function
\begin{itemize}
\item \texttt{page : \{l\}.Page [(Model \{l\}) Msg] = Page (init \{l\}) (view \{l\}) (update \{l\})}
\end{itemize}
where {\texttt{page\{client\}}} is a client only TodoMVC program while
{\texttt{page\{server\}}} is a multi-tier TodoMVC program as the
original example.

Now let us examine again the case expression on messages {\texttt
  {Submit}}, {\texttt {Toggle index}}, and {\texttt {Delete index}} in
the {\texttt {update}} function.  On each {\texttt {Submit}} message
(in Line 29) generated when the user types an Enter key after typing a
text, a new item with the text ({\texttt {line}}) and the incomplete
status ({\texttt {False}}) is added to the list of items at the
server.  On each {\texttt{Toggle index}} message (in Line 32), the
server list is replaced by a new list obtained by toggling the
completion status of the indexed item using {\texttt
  {toggleItem}}. Here, {\texttt {mapOnIndex}} is a selective map
function that applies a given function only to the indexed element. It
is written using {\texttt {mapWithCount}}, which is another variant of
the map function applying its argument function to counts as well as
list elements. The variant map function runs at the server when it is
used to toggle an indexed element (Line 33) while it runs at the
client when it is used to build a list of HTML {\texttt{li}} elements
(Line 13).  On each {\texttt{Delete index}} message (in Line 36), the
indexed item is deleted from the server list using {\texttt {delete}}.

\begin{lstlisting}
mapWithCount : {l}.[a b].(Int -l-> (a-l->Int-l->b) -l-> List [a] -l-> List [b])
   = {l}. \idx @ l  f @ l  xs @ l .
	case xs {
	 Nil => Nil;
	 Cons y ys => Cons (f y idx) (mapWithCount {l} (idx + 1) f ys)
        };
mapOnIndex : {l}. [a b]. (Int -l-> (a -l-> b) -l-> List [a] -l-> List [b])
   = {l}. \targetIdx @ l  f @ l  xs @l .
	mapWithCount {l} 0
	  (\av @ l idx @ l. if targetIdx == idx then f av else av) xs;
\end{lstlisting}

Our dynamic approach works like the static approach with no extra cost
of dynamic location checks for functions, such as {\texttt {header}}
of type {\texttt {String -client-> Html [Msg]}} invoked in the client,
whose caller and callee locations can be statically compared.  A
difference appears when polymorphic location functions, such as
{\texttt {toggleItem}}, are invoked. This invocation may incur a cost
of dynamically comparing two locations.  In Line 33, the current
location is client, and {\texttt {toggleItem [server]}} becomes a
server function by binding {\texttt {server}} to the location variable
$l$ in Line 22.  Therefore, when calling {\texttt
  {\mbox{$\backslash$}ti@l.case ti \{$\cdots$\}}} of type {\texttt
  {TodoItem-l->TodoItem}}, the current location, which is {\texttt
  {client}}, will be dynamically compared with the location bound to
$l$, which is {\texttt {server}}, to determine that this lambda
abstraction is a remote procedure in the context.

\subsection{The SKI Variants in PolyRPC}

	When a polymorphic application is written in the way that the location of the application, $\Loc$, and the location of the function to run, $\Loc'$, may be location variables, compilers cannot statically determine if the lambda application is for remote calls, local calls, or both.
	The existing slicing compilation method for the typed RPC calculus~\cite{choijfp2019}, that is the simply typed and monomorphic subset of the polymorphic RPC calculus~\cite{CHOI:scp2020}, cannot deal with such a polymorphic lambda application any more.

	The previous study~\cite{CHOI:scp2020} overcame this limitation by translating all polymorphic locations in RPC programs into monomorphic ones by the so called  {\it monomorphisation} translation. This approach is called static because all polymorphic locations can now be resolved at compile-time.

	But in the worst case the monomorphisation translation can potentially lead to code explosion by generating client and server versions for each location abstraction. When there are $n$ location abstractions nested subsequently, $2^n$ monomorphic versions could be generated. This is called a code explosion problem of the static approach to the implementation of the polymorphic RPC calculus.

        Various definitions of the S and K combinators are provided here. 

\begin{lstlisting}
s : {l1 l2 l3}. [a b c]. ( (a -l1-> b -l1-> c) -l3-> (a -l2-> b) -l3-> a -l3-> c)
  = {l1 l2 l3}. \f @ l3  g @ l3  x @ l3. f x (g x) ;

k : {l}. [a b]. (a -l-> b -l-> a)
  = {l}. \x @ l  y @ l . x ;
\end{lstlisting}

The full freedom style S and K combinators are ones obtained by annotating all different locations to each arrow of function types.

\begin{lstlisting}
s : {l11 l12 l2 l31 l32 l33}. [a b c].
        ( (a -l11-> b -l12-> c) -l31-> (a -l2-> b) -l32-> a -l33-> c)
  = {l11 l12 l2 l31 l32 l33}. \f @ l31  g @ l32  x @ l33. f x (g x) ;

k : {l1 l2}. [a b]. (a -l1-> b -l2-> a)
  = {l1 l2}. \x @ l1  y @ l2 . x ;
\end{lstlisting}

The client-version is simple. Every location is annotated with a specified location constant.

\begin{lstlisting}
s : [a b c].
       ( (a-client->b-client->c) -client-> (a-client->b) -client-> a -client-> c)
  = \f @ client  g @ client  x @ client. f x (g x) ;

k : [a b]. (a -client-> b -client-> a)
  = \x @ client  y @ client . x ;
\end{lstlisting}

	To show the code explosion problem in the worst case, let us consider a small example of {\texttt {S}} and {\texttt {K}} combinators written in PolyRPC to make an identity function.
	Let us call this a {\it spine location} SKI program where every multiple-argument function is applied to all its arguments at the same location. There are at least two variants: a client only SKI program by replacing all location variables by {\texttt {client}} and a full freedom SKI program by allowing applying a multiple-argument function to each argument all at different locations. In the full freedom program, the {\texttt {S}} combinator will have a location abstraction with six location variables as {\texttt {\{l11 l12 l2 l31 l32 l33\}} } by replacing the two occurrences of {\texttt {l1}} by {\texttt {l11}} and {\texttt {l12}} and by replacing the three occurrences of {\texttt {l3}} by {\texttt {l31}}, {\texttt {l32}}, and {\texttt {l33}}.

	Here is a simple experimental result with these three programs for code size and location checks. By counting the nodes of a program tree (excluding type nodes), the client SKI program is of size 48, the spine location style SKI program is of size 59, and the full freedom SKI program is of size 68. After applying the monomorphisation, the sizes become 48, 190, and 844. Running each of the three programs applies functions 9 times to result in the integer, 123. Both of the spine location style and full freedom SKI programs do dynamic location checks 3 times: for example, in the spine location style program,  once at {\texttt {(g x)}} where the current location is {\texttt {l3}} and the location for {\texttt {g}} to run is {\texttt {l2}}, twice at {\texttt {(f x)} where the current location is {\texttt {l3}} and the location for {\texttt {f}} to run is {\texttt {l1}}, and three times at the application of {\texttt {(f x)}} to {\texttt {(g x)}} where the current location is {\texttt {l3}} and the location for the function from {\texttt {(f x)}} to run is {\texttt {l1}}.

        The single location-version is also possible. Every location is annotated with the same location variable for each function. To use this version with server functions $f$, caller of this version has to do the eta conversion, $\lamL{\client}{x}{f \ x}$ to adjust it to the interface.

\begin{lstlisting}
s : {l}. [a b c].
       ( (a -l-> b -l-> c) -l-> (a -l-> b) -l-> a -l-> c)
  = \f @ l  g @ l  x @ l. f x (g x) ;

k : {l}. [a b]. (a -l-> b -l-> a)
  = \x @ l  y @ l . x ;
\end{lstlisting}

	In this work, we take a dynamic approach that may require runtime location checking only when polymorphic locations are used. The dynamic approach retains all statically resolved locations as in the typed RPC calculus but offers a way to determine dynamically whether polymorphic-location lambda applications are local or remote procedure calls.
	Therefore, this dynamic approach does not have to do any static translations for polymorphic locations at compile-time, which resolves the potential code explosion problem of the static approach at the expense of runtime cost for polymorphic locations.

	The SKI programs used in the experiment are small, and  so whether the worst case behavior would appear often in practice is still left as a question. For example, the multi-tier TodoMVC program, which is about 300 lines written in PolyRPC and is the largest program for now,  is of size 1855, increasing up to 2554 after the monomorphisation.

	Nonetheless we firstly argue that the dynamic approach is preferred as it can be viewed as a generalization of the static approach that can also handle the worst case behavior. A compiler can apply the monomorphisation up to $k$ nested location abstractions for an arbitrary constant $k \geq 0$ and can use the dynamic approach to handle the more deeply nested ones than $k$.
	Secondly, it is also true that it is easy to make up such an example program showing the worst case behavior like the SKI programs. Also, even if the worst case behavior would not appear frequently, each duplicated code can be large. With more or less practical example like the multi-tier TodoMVC program, the used polymorphic location functions are of size 257, and the size increases up to 852 after the monomorphisation.

\section{The Polymorphic RPC Calculus}

This section reminds the reader of the polymorphic RPC calculus~\cite{CHOI:scp2020}. It is a polymorphically typed call-by-value
$\lambda$-calculus with location annotations on $\lambda$-abstractions
specifying where to run. The calculus offers the notion of polymorphic
location to write polymorphically located functions succinctly, which
is convenient for programmers.

\subsection{The Syntax and the Semantics}

	Figure \ref{fig:polyrpc} shows the syntax and semantics of the polymorphic RPC calculus, {\polyrpc} that allows programmers to use the same syntax of $\lambda$-application for both local and remote calls, and allows them to compose differently located functions arbitrarily.
	An important feature is the notion of location variable $l$ for which a  location constant $a$ can be substituted. A syntactic object $\Loc$ is either a location constant or a location variable.
	Assuming the client-server model in the calculus, location constants are either $\client$ denoting client or $\server$ denoting server.

	In the syntax, $M$ denotes terms, and $V$ denotes values. Every $\lambda$-abstraction $\lamL{\Loc}{x}{M}$ has a location annotation of $\Loc$. By substituting a location $b$ for a location variable annotation,  $(\lamL{l}{x}{M})\subst{b}{l}$ becomes a monomorphic $\lambda$-abstraction $\lamL{b}{x}{(M\subst{b}{l})}$. This location variable is abstracted by the location abstraction construct $\Lambda l.V$, and it is instantiated by the location application construct $M[\Loc]$. The rest of the syntax are the same as those in the polymorphically typed call-by-value $\lambda$-calculus extended with the feature of pairs. Variables are denoted by $x$. Term applications are denoted by $L \ M$. Type abstractions are $\Lambda\alpha.V$ for a type variable $\alpha$, and type applications are $M[A]$ for a type $A$. Pairs are $(L,M)$, and projections of the first and second element of the pairs are $\pi_i(M)$ for $i=1,2$.

	The semantics of {\polyrpc} is defined in the style of a big-step operational semantics whose evaluation judgments, $\evalRPC{M}{a}{V}$, denote that a term $M$ evaluates to a value $V$ at location $a$.
	In the semantics, location annotated $\lambda$-abstractions, type abstractions, and location abstractions are all values. So, (Abs), (Tabs), and (Labs) are straightforwardly defined as an identity evaluation relation over them.
	(App) defines local calls when $a=b$ and remote calls when $a\not=b$ in the same syntax of lambda applications. The evaluation of an application $L \ M$ at location $a$ performs $\beta$-reduction at location $b$, where a $\lambda$-abstraction $\lamL{b}{x}{N}$ from $L$ has as an annotation, with a value $W$ from $M$, and it continues to evaluate the $\beta$-reduced term $N\subst{W}{x}$, which is a substitution of $W$ for $x$ in $N$, at the same location.
	(Tapp) defines the evaluation of type applications $M[A]$ as: $M$ evaluates to a type abstraction $\Lambda\alpha.V$, and  all occurrences of the type variable $\alpha$ are replaced by the type $A$ as $V\subst{A}{\alpha}$.
	(Lapp) similarly defines the evaluation of location applications $M[\Loc]$. The only difference is the use of location substitution in $V\subst{\Loc}{l}$ where $V$ is the body of a location abstraction $\Lambda l. V$ from evaluating $M$. (Pair) and (Proj-i) are the standard evaluation rules for creating pairs and projecting one of the pair elements.

\subsection{The Type System}
%\label{sec:polyrpc:typesystem}

	Figure \ref{fig:polyrpctysystem} shows a  type system for the polymorphic RPC calculus~\cite{CHOI:scp2020} that can  identify remote procedure calls at the type level, supporting location polymorphism. The type language allows function types $A \funL{\Loc} B$. Then every $\lambda$-abstraction at unknown location  gets assigned $A\funL{l} B$ using some location variable $l$. A universal quantifier over a location variable, $\forall l. A$, is also introduced to allow to abstract such occurrences of location variables.

	Typing judgments are in the form of $\typing{\tyenv}{\Loc}{M}{A}$, saying a term $M$ at location $a$ has type $A$ under a type environment $\tyenv$.  The location annotation, $\Loc$, is either a location variable or constant.
	Typing environments $\tyenv$ have location variables, type variables, and types of variables, as $\{l_1, \cdots,l_n,\alpha_1,\cdots,\alpha_k, x_1:A_1, \cdots, x_m:A_m\}$.
	They are used to keep track of a set of free location, type, and value variables in the context of a given term.

	The typing rules for the polymorphic RPC calculus are defined as follows.
	(T-Var) is defined as usual.
	(T-Abs) assigns $\lambda$-abstraction a function type with the same location as its annotation. Note that a location on the typing judgment in the conclusion changes to the annotated location in the premise for the body of  $\lambda$-abstraction.
	(T-Tabs) and (T-Tapp) are the standard typing rules for type abstraction and type application. $A\subst{B}{\alpha}$ is a substitution of $B$ for each occurrence of $\alpha$ in $A$.
	(T-Labs) and (T-Lapp) are similar to the typing rules for type abstraction and type application. (T-Labs) checks if its bound location variable does not appear in the type environment and in the contextual location. (T-Lapp) substitutes $\Loc'$ for all occurrences of a location variable $l$ on $\lambda$-abstractions in $M$  by $A\subst{\Loc'}{l}$.
	(T-Pair) and (T-Proj-i) are straightforward extensions of the standard typing rules for pairs and projections with the notion of locations.

	(T-App) is a refinement of the conventional typing rule for $\lambda$-applications with respect to the combinations of location $\Loc$ (where to evaluate the application) and location $\Loc'$ (where to evaluate the function).
	For simplicity, assume both locations are constants as $\Loc=a$ and $\Loc'=b$.
	When $a$ is different from $b$, $L \ M$ is statically found to be a remote procedure call: if $a=\client$ and $b=\server$, it is to invoke a server function from the client, and if $a=\server$ and $b=\client$, it is to invoke a client function from the server. Otherwise, one can statically decide that it is a local procedure call.

	The type soundness of the type system for the polymorphic RPC calculus, which was formulated as Theorem \ref{thm:typesoundness} and was proved by Choi et al.~\cite{CHOI:scp2020},  guarantees that every remote procedure call thus identified statically will never change to a local procedure call under evaluation. This enables compilers to generate call instructions for local calls and network communication for remote calls both in the same syntax of lambda applications safely.

\section{Definitions in the polymorphic CS calculus}

\subsection{Definitions of free variables, type variables, location variables and well-formed typing rules}
	Let us start with $fv(M)$, a definition for a set of free variables over terms as:
	\begin{eqnarray*}
		fv(x) &=& \{x\} \\
		fv((V,W)) &=& fv(V) \cup fv(W) \\
		fv(\clo{\overline{W}}{F}) &=& \bigcup_i fv(W_i) \\
		fv(\Lambda\alpha.V) &=& fv(V) \\
		fv(\lunit V) &=& fv(V) \\
		fv(\ldo{x}{M}{N}) &=& fv(M) \cup (fv(N) \backslash \{x\}) \\
		fv(\req{V}{W}) &=& fv(V) \cup fv(W)\\
		fv(\call{V}{W}) &=& fv(V) \cup fv(W)\\
		fv(\gen{\Loc}{V}{W}) &=& fv(V) \cup fv(W)\\
		\\
		fv(\llet{x}{M}{N}) &=& fv(M) \cup (fv(N) \backslash \{x\}) \\
		fv(\pi_i(V)) &=& fv(V) \\
		fv(V(W)) &=& fv(V) \cup fv(W) \\
		fv(V[A]) &=& fv(V) \\
		fv(V[\Loc]) &=& fv(V)
	\end{eqnarray*}

	Then let us define $ftv(M)$ and $ftv(A)$ for a set of free type variables over terms and a set of free type variables over types, respectively, as:
	\begin{eqnarray*}
		ftv(x) &=& \{\} \\
		ftv((V,W)) &=& ftv(V) \cup ftv(W) \\
		ftv(\clo{\overline{W}}{F_{name}[\overline{\Loc}\,\overline{A}]}) &=& \bigcup_i ftv(W_i) \cup \bigcup_j ftv(A_j)\\
		ftv(\Lambda\alpha.V) &=& ftv(V) \backslash \{\alpha\} \\
		ftv(\lunit V) &=& ftv(V) \\
		ftv(\ldo{x}{M}{N}) &=& ftv(M) \cup ftv(N) \\
		ftv(\req{V}{W}) &=& ftv(V) \cup ftv(W)\\
		ftv(\call{V}{W}) &=& ftv(V) \cup ftv(W)\\
		ftv(\gen{\Loc}{V}{W}) &=& ftv(V) \cup ftv(W)\\
		\\
		ftv(\llet{x}{M}{N}) &=& ftv(M) \cup ftv(N) \\
		ftv(\pi_i(V)) &=& ftv(V) \\
		ftv(V(W)) &=& ftv(V) \cup ftv(W) \\
		ftv(V[A]) &=& ftv(V) \cup ftv(A) \\
		ftv(V[\Loc]) &=& ftv(V)
	\end{eqnarray*}

	\begin{eqnarray*}
		ftv(base) &=& \{\} \\
		ftv(A \funL{\Loc} B) &=& ftv(A) \cup ftv(B) \\
		ftv(\cloty{A}) &=& ftv(A) \\
		ftv(A \times B) &=& ftv(A) \cup ftv(B) \\
		ftv(\alpha) &=& \{\alpha\} \\
		ftv(\forall\alpha.A) &=& ftv(A) \backslash \{\alpha\} \\
		ftv(\forall l.A) &=& ftv(A) \\
		ftv(T\ A) &=& ftv(A)
	\end{eqnarray*}

	Now we define sets of free location variables over locations, types, and terms in the form as $flv(-)$.
	\begin{eqnarray*}
		flv(a) &=& \{ \} \\
		flv(l)  &=& \{ l\} \\
	\end{eqnarray*}

	\begin{eqnarray*}
		flv(base) &=& \emptyset \\
		flv(A\funL{\Loc}B) &=& flv(A)\cup flv(\Loc)\cup flv(B) \\
		flv(\cloty{A}) &=& flv(A) \\
		flv(A\times B) &=& flv(A) \cup flv(B) \\
		flv(\alpha) &=& \emptyset \\
		flv(\forall \alpha.A) &=& flv(A) \\
		flv(\forall l.A) &=& flv(A) \backslash \{l\} \\
		flv(T \ A) &=& flv(A)
	\end{eqnarray*}

	\begin{eqnarray*}
		flv(x) &=& \{\} \\
		flv((V,W)) &=& flv(V) \cup flv(W) \\
		flv(\clo{\overline{W}}{F_{name}[\overline{\Loc}\,\overline{A}]}) &=& \bigcup_i flv(W_i) \cup \bigcup_j flv(\Loc_j)\\
		flv(\Lambda\alpha.V) &=& flv(V) \\
		flv(\lunit V) &=& flv(V) \\
		flv(\ldo{x}{M}{N}) &=& flv(M) \cup flv(N) \\
		flv(\req{V}{W}) &=& flv(V) \cup flv(W)\\
		flv(\call{V}{W}) &=& flv(V) \cup flv(W)\\
		flv(\gen{\Loc}{V}{W}) &=& flv(\Loc)\cup flv(V) \cup flv(W)\\
		\\
		flv(\llet{x}{M}{N}) &=& flv(M) \cup flv(N) \\
		flv(\pi_i(V)) &=& flv(V) \\
		flv(V(W)) &=& flv(V) \cup flv(W) \\
		flv(V[A]) &=& flv(V) \cup flv(A) \\
		flv(V[\Loc]) &=& flv(V) \cup flv(\Loc)
		\ \\
	\end{eqnarray*}

	Note that $fv(-)$, $ftv(-)$, and $flv(-)$ are defined for $\OpenCode$ as:
	\begin{eqnarray*}
	fv(\lambda x.M) &=& fv(M) \backslash \{x\} \\
	fv(\Lambda l.V) &=& fv(V) \\
	\ \\
	ftv(\lambda x.M) &=& ftv(M) \\
	ftv(\Lambda l.V) &=& ftv(V) \\
	\ \\
	flv(\lambda x.M) &=& flv(M) \\
	flv(\Lambda l.V) &=& flv(V) \backslash \{ l \}
	\ \\
	\end{eqnarray*}

	The domain of typing environment, $dom(\tyenv)$, is defined as a union of type, location, and term variables as $\{ \alpha_1,\cdots,\alpha_k,l_1, \cdots,l_n, x_1, \cdots, x_m \}$, and the range, $rng(\tyenv)$, is $\{A_1,\cdots,A_m\}$.

	We can extend the definitions above to typing environments.
	For free variables, $fv(\tyenv)$ is:
	\[ fv(\{ l_1, \cdots,l_n,\alpha_1,\cdots,\alpha_k, x_1:A_1, \cdots, x_m:A_m \}) = \{ x_1, \cdots,x_m \}
	\]

	For free type variables, $ftv(\tyenv)$ is
	\[ ftv(\{ l_1, \cdots,l_n,\alpha_1,\cdots,\alpha_k, x_1:A_1, \cdots, x_m:A_m \}) = \{ \alpha_1, \cdots,\alpha_k \} \cup \bigcup_{1\leq i \leq m} ftv(A_i)
	\]

	For free location variables, $flv(\tyenv)$, which is a union of location variables there and free location variables occurring in types associated with variables as
	\[ flv(\{ l_1, \cdots,l_n,\alpha_1,\cdots,\alpha_k, x_1:A_1, \cdots, x_m:A_m \}) = \{ l_1, \cdots, l_n \} \cup \bigcup_{1\leq i \leq m} flv(A_i)
	\]

			In the type system for the polymorphic CS calculus, we will consider only well-formed typing judgments where there are no unbound variables, no unbound type variables, and no unbound free location variables. That is, given $\typing{\tyenv}{\Loc}{M}{A}$, we will safely assume three things.
			First, $fv(M) \subseteq dom(\tyenv)$.

			Second, $ \bigcup_{A_i \in rng(\tyenv)} ftv(A_i) \cup ftv(M) \cup ftv(A) \subseteq dom(\tyenv)$ where $ftv(A)$ or $ftv(M)$ are the sets of free type variables occurring in the type and the term respectively. They can be defined straightforwardly.

			Third, $\bigcup_{A_i \in rng(\tyenv)} flv(A_i) \cup flv(\Loc) \cup flv(M) \cup flv(A) \subseteq dom(\tyenv)$.

%	For example, in (T-App), location variables occurring in $A$ and $\Loc'$ are from any location abstractions enclosing the application term, in (T-Tabs), a bound type variable $\alpha$ does not occur as a free type variable in $\tyenv$, and
%	in (T-Labs), a bound  location  variable $l$ never occurs as a free location variable in $\tyenv$ and $\Loc$.

\subsection{Definitions of the Substitutions in the polymorphic CS calculus}

		We present the definitions of substitutions over locations, types, and terms:
	\begin{center}
	\begin{tabular}{|l || l | l | l | } \hline
	             & Values & Types & Locations \\ \hline\hline
	Terms & $M\subst{W}{x}$ & $V\subst{B}{\alpha}$ & $V\subst{\Loc'}{l}$ \\ \hline
	Types &                                 & $A\subst{B}{\alpha}$ & $A\subst{\Loc'}{l}$ \\ \hline
	Locations &                          &                                           & $\Loc\subst{\Loc'}{l}$ \\ \hline
	\end{tabular}
	\end{center}

	The definition of $M\subst{W}{x}$ replacing all occurrences of $x$ in $M$ by $W$ is as follows.
	\begin{eqnarray*}
	(y)\subst{W}{x}  & = & \left\{\mbox{\begin{tabular}{l l}
													$W$ &  if  $x=y$ \\
	                                                         				$y$ & otherwise
												\end{tabular}
											} \right.	\\
       (V_1,V_2)\subst{W}{x} &=& (V_1\subst{W}{x} , V_2\subst{W}{x} ) \\
       \clo{\overline{V}}{F}\subst{W}{x}  &=& \clo{\overline{V\subst{W}{x} }}{F} \\
       (\Lambda\alpha.V)\subst{W}{x} &=& \Lambda\alpha.(V\subst{W}{x} ) \\
       (\lunit V)\subst{W}{x}  &=& \lunit (V\subst{W}{x} ) \\
       (\ldo{y}{M}{N})\subst{W}{x}  &=&  \left\{\mbox{\begin{tabular}{l l}
													$\ldo{y}{M\subst{W}{x} }{N }$ &  if  $x=y$ \\
	                                                         				$\ldo{y}{M\subst{W}{x} }{N\subst{W}{x} }$ & otherwise
												\end{tabular}
											} \right.	\\
       (\req{V_f}{V_{arg}})\subst{W}{x}  &=& \req{V_f\subst{W}{x} }{V_{arg}\subst{W}{x} } \\
       (\call{V_f}{V_{arg}})\subst{W}{x}  &=& \call{V_f\subst{W}{x} }{V_{arg}\subst{W}{x} } \\
       (\gen{\Loc}{V_f}{V_{arg}})\subst{W}{x}  &=& \gen{\Loc}{V_f\subst{W}{x} }{V_{arg}\subst{W}{x} } \\
       \ \\
       (\llet{y}{M}{N})\subst{W}{x}  &=&  \left\{\mbox{\begin{tabular}{l l}
													$\llet{y}{M\subst{W}{x} }{N }$ &  if  $x=y$ \\
	                                                         				$\llet{y}{M\subst{W}{x} }{N\subst{W}{x} }$ & otherwise
												\end{tabular}
											} \right.	\\
       (\pi(V))\subst{W}{x}   &=& \pi(V\subst{W}{x}  ) \\
       (V_f(V_{arg}))\subst{W}{x}   &=& (V_f\subst{W}{x}  (V_{arg}\subst{W}{x}  )) \\
       (V[A])\subst{W}{x}  &=& (V\subst{W}{x} )[A] \\
       (V[\Loc])\subst{W}{x}   &=& (V\subst{W}{x})[\Loc]
	\end{eqnarray*}

	The definition of $V\subst{B}{\alpha}$ replacing all occurrences of $\alpha$ in $V$ by $B$ is:
	\begin{eqnarray*}
	(x)\subst{B}{\alpha}  & = & x	\\
       (V,W)\subst{B}{\alpha} &=& (V\subst{B}{\alpha}  , W\subst{B}{\alpha})  \\
       \clo{\overline{V}}{F_{name}[\overline{\Loc}\,\overline{A}]}\subst{B}{\alpha}   &=& \clo{\overline{V\subst{B}{\alpha}  }}{F_{name}[\overline{\Loc}\,\overline{A\subst{B}{\alpha}}]} \\
       (\Lambda\beta.V)\subst{B}{\alpha} &=& \left\{\mbox{\begin{tabular}{l l}
											       $\Lambda\beta.V$ &  if  $\alpha=\beta$ \\
	                                                         				$\Lambda\beta.(V\subst{B}{\alpha})$ & otherwise
												\end{tabular}
											} \right.	\\
       (\lunit V)\subst{B}{\alpha}  &=& \lunit (V\subst{B}{\alpha} ) \\
       (\ldo{y}{M}{N})\subst{B}{\alpha}  &=& 	\ldo{y}{M\subst{B}{\alpha}}{N\subst{B}{\alpha}}  \\
       (\req{V}{W})\subst{B}{\alpha}  &=& \req{V\subst{B}{\alpha} }{W\subst{B}{\alpha} } \\
       (\call{V}{W})\subst{B}{\alpha}  &=& \call{V\subst{B}{\alpha} }{W\subst{B}{\alpha} } \\
       (\gen{\Loc}{V}{W})\subst{B}{\alpha}  &=& \gen{\Loc}{V\subst{B}{\alpha} }{W\subst{B}{\alpha} } \\
       \ \\
       \end{eqnarray*}

       	The definition of $V\subst{\Loc}{l}$ replacing all occurrences of $l$ in $V$ by $\Loc$ is:
	\begin{eqnarray*}
	(x)\subst{\Loc}{l}  & = & x	\\
       (V,W)\subst{\Loc}{l} &=& (V\subst{\Loc}{l}, W\subst{\Loc}{l})  \\
       \clo{\overline{V}}{F_{name}[\overline{\Loc}\,\overline{A}]}\subst{\Loc}{l}   &=& \clo{\overline{\subst{\Loc}{l}  }}{F_{name}[\overline{\Loc_i\subst{\Loc}{l}}\,\overline{A}]} \\
       (\Lambda\beta.V)\subst{\Loc}{l} &=& \Lambda\beta.(V\subst{\Loc}{l})	\\
       (\lunit V)\subst{\Loc}{l}  &=& \lunit (V\subst{\Loc}{l} ) \\
       (\ldo{y}{M}{N})\subst{\Loc}{l}  &=& 	\ldo{y}{M\subst{\Loc}{l}}{N\subst{\Loc}{l}}  \\
       (\req{V}{W})\subst{\Loc}{l}  &=& \req{V\subst{\Loc}{l} }{W\subst{\Loc}{l} } \\
       (\call{V}{W})\subst{\Loc}{l}  &=& \call{V\subst{\Loc}{l} }{W\subst{\Loc}{l} } \\
       (\gen{\Loc}{V}{W})\subst{\Loc}{l}  &=& \gen{\Loc_0\subst{\Loc}{l}}{V\subst{\Loc}{l} }{W\subst{\Loc}{l} }
       \ \\
       \end{eqnarray*}

       The definition of $A\subst{B}{\alpha}$ replacing all occurrences of $\alpha$ in $A$ by $B$ is:
       	\begin{eqnarray*}
		base\subst{B}{\alpha} &=& base \\
		(A_1\funL{\Loc}A_2) \subst{B}{\alpha}&=&  A_1 \subst{B}{\alpha}\funL{\Loc}A_2 \subst{B}{\alpha}\\
		(\cloty{A}) \subst{B}{\alpha} &=& \cloty{A \subst{B}{\alpha}} \\
		(A_1\times A_2) \subst{B}{\alpha} &=& A_1 \subst{B}{\alpha} \times A_2 \subst{B}{\alpha} \\
		\beta  \subst{B}{\alpha} &=&  \left\{\mbox{\begin{tabular}{l l}
											       $B$ &  if  $\alpha=\beta$ \\
	                                                         				$\beta$ & otherwise
												\end{tabular}
											} \right.\\
		(\forall \beta.A)\subst{B}{\alpha}  &=& \left\{\mbox{\begin{tabular}{l l}
											       $\forall \beta.A$ &  if  $\alpha=\beta$ \\
	                                                         				$\forall \beta.(A\subst{B}{\alpha})$ & otherwise
												\end{tabular}
											} \right. \\
		(\forall l.A)\subst{B}{\alpha}  &=& \forall l. (A\subst{B}{\alpha}) \\
		(T \ A)\subst{B}{\alpha}  &=& T \ (A\subst{B}{\alpha} )
		\ \\
	\end{eqnarray*}

	The definition of $A\subst{\Loc}{l}$ replacing all occurrences of $l$ in $A$ by $\Loc$ is this.
	\begin{eqnarray*}
	base \subst{\Loc}{l}  & = & base \\
	(A \funL{\Loc} B) \subst{\Loc'}{l}  & = & A \subst{\Loc'}{l} \funL{ \Loc \subst{\Loc'}{l} } B \subst{\Loc'}{l} \\
	(\cloty{A}) \subst{\Loc'}{l} &=& \cloty{A \subst{\Loc'}{l}} \\
	(A_1\times A_2) \subst{\Loc'}{l} &=& A_1 \subst{\Loc'}{l} \times A_2 \subst{\Loc'}{l} \\
	\alpha \subst{\Loc}{l} & = & \alpha \\
	(\forall\alpha.A) \subst{\Loc'}{l}  & = & \forall\alpha. (A \subst{\Loc'}{l} ) \\
	(\forall l.A)\subst{\Loc}{l'}  & = & \left\{\mbox{`\begin{tabular}{l l}
													$\forall l.A$ &  if  $l=l'$ \\
	                                                         				$\forall l.(A\subst{\Loc'}{l'})$ & otherwise
												\end{tabular}

											} \right. \\
	(T \ A)\subst{\Loc'}{l'}  &=& T \ (A\subst{\Loc'}{l'} ) \\
	\end{eqnarray*}

	Lastly, the definition of $\Loc\subst{\Loc'}{l}$ is:
      	\begin{eqnarray*}
      \Loc\subst{\Loc'}{l} & = & \left\{\mbox{\begin{tabular}{l l}
													$\Loc$ & if $\Loc=a$ \\
	                                                         				$\Loc'$ & if $\Loc=l'$ and $l=l'$ \\
	                                                         				$\Loc$ & if $\Loc=l'$ and $l\not=l'$
												\end{tabular}

											}
								\right.
	\end{eqnarray*}

\section{Type soundness and semantic correctness}
        
\subsection{Type soundness}

\setcounter{lemma}{0}
\renewcommand{\thelemma}{4.\arabic{lemma}}

\setcounter{theorem}{0}
\renewcommand{\thetheorem}{4.\arabic{theorem}}

	In this section, the type soundness property for the polymorphic RPC calculus is proven by showing the type preservation property in Lemma \ref{theorem:typepreservationincs} and the  progress property in Lemma \ref{theorem:progressincs}.

\begin{theorem}[Type soundness] 	Given a polymorphic CS program $(\funstore_\client,\funstore_\server)$ with the main term $M$,
	if $\conftyping{\confcs{M;\emptystack}{\emptystack}}{TA}$,
	either $\confcs{M;\emptystack}{\emptystack} \run^* \confcs{\lunit{V};\emptystack}{\emptystack}$
	or it loops indefinitely.
\label{appendix:theorem:corollary:typesoundness}
\end{theorem}

	Given a polymorphic CS program $(\funstore_\client,\funstore_\server)$ with the main term $M$,
	the type soundness property is that if $\conftyping{\confcs{M;\emptystack}{\emptystack}}{TA}$ and the evaluation does not loop infinitely, we will have $\confcs{M;\emptystack}{\emptystack} \run^* \confcs{\lunit{V};\emptystack}{\emptystack}$.
	This property is proven by Theorem \ref{theorem:corollary:typesoundness} whose proof is immediate from the type preservation by Lemma \ref{theorem:typepreservationincs} and the progress by Lemma \ref{theorem:progressincs}.

	In Lemma \ref{theorem:typepreservationincs}, the proof uses Lemma \ref{lemma:typepreservationforterms} proving type preservation in the local reduction part, and shows the property in the communication part.
	%Lemma \ref{lemma:substitutionsincs} is used by the theorem and the lemma, showing type preservation properties over value substitutions over terms, type substitutions over terms, location substitutions over terms, and value substitutions over open codes.
	Proving the theorem uses a lemma showing type preservation properties over value substitutions over terms, type substitutions over terms, location substitutions over terms, and value substitutions over open codes.

       	Proving type preservation with whole terms requires proving the same property with their subterms identified by evaluation judgments in Lemma \ref{theorem:typepreservationincs}.
%	The theorem uses the subterm typeability lemma, Lemma. \ref{lemma:subtermtypeabilityincs}, to show that every subterm focused through an evaluation judgment in a well-typed term is well-typed, and
	The theorem uses a subterm typeability lemma to show that every subterm focused through an evaluation judgment in a well-typed term is well-typed, and
	% it also uses the subterm replacement lemma, Lemma \ref{lemma:subtermreplacementincs}, to show that every evaluation judgment whose hole is filled with a type-matched subterm produces a well-typed term.
	it also uses a subterm replacement lemma to show that every evaluation judgment whose hole is filled with a type-matched subterm produces a well-typed term.

\begin{lemma}[Type preservation for terms] If \ $\typing{\emptyset}{a}{M}{A}$ and $M \run N$ then $\typing{\emptyset}{a}{N}{A}$.
\label{lemma:typepreservationforterms}
\end{lemma}
\begin{proof} We prove this lemma by case analysis on $M$.

	\ \\
{\bf i)} $M=\llet{x}{V}{L}$: Let us have $N=L\subst{V}{x}$.

	By (T-Let), (1):$\typing{\emptyset}{a}{V}{B}$ and (2):$\typing{\{x:B\}}{a}{L}{A}$.

	By Lemma \ref{lemma:substitutionsincs} over value substitutions with (1) and (2), $\typing{\emptyset}{a}{L\subst{V}{x}}{A}$.

	\ \\
{\bf ii)} $M=\ldo{x}{\lunit{V}}{L}$:  Let us have $N=L\subst{V}{x}$.

	By (T-Bind), (1):$\typing{\emptyset}{a}{\lunit{V}}{T\ C}$ and (2):$\typing{\{x:C\}}{a}{L}{T\ B}$ where $A=T\ B$.

	By (T-Unit) with (1), (3):$\typing{\emptyset}{a}{V}{C}$.

	By Lemma \ref{lemma:substitutionsincs} over value substitutions with (2) and (3), $\typing{\emptyset}{a}{L\subst{V}{x}}{T\ B}$.

	\ \\
{\bf iii)} $M=\pi_i(V_1,V_2)$:  Let us have  $N=V_i$ ($i=1 \ or \ 2$).

	By (T-Proj-i), (1):$\typing{\emptyset}{a}{(V_1,V_2)}{A_1\times A_2}$ where $A=A_i$.

	By (T-Pair) with (1), (2)-i:$\typing{\emptyset}{a}{V_i}{A_i}$.

	\ \\
{\bf iv)} $M=(\Lambda\alpha.V)[B]$: Let us have $N=V\subst{B}{\alpha}$.

	By (T-Tapp), (1):$\typing{\emptyset}{a}{\Lambda\alpha.V}{A_0}$ and (2):$A=A_0\subst{B}{\alpha}$.

	By (T-Tabs) with (1), $\typing{\alpha}{\Loc_0}{V}{A_0}$ for arbitray $\Loc_0$, and so (3):$\typing{\alpha}{a}{V}{A_0}$ for $a$.

	By Lemma \ref{lemma:substitutionsincs} over value substitutions with (3), (4):$\typing{\emptyset}{a}{V\subst{B}{\alpha}}{A_0\subst{B}{\alpha}}$.

	\ \\
{\bf v)} $M=\clo{\overline{W}}{F}(V)$: (1):$\funstore(F)=\overline{z}.\lambda x.M_0$ and $N=M_0\subst{\overline{W}}{\overline{z}}\subst{V}{x}$.

	By (T-App), (2):$\typing{\emptyset}{a}{\clo{\overline{W}}{F}}{\cloty{A_1\funL{a}A_2}}$ and (3):$\typing{\emptyset}{a}{V}{A_1}$ where $A=A_2$.

	By (T-Clo) with (2), (4):$\funtyping{}{\emptyset}{a}{F}{\overline{B}.(A_1\funL{a}A_2)}$ and (5):$\typing{\emptyset}{a}{\overline{W}}{\overline{B}}$.

	By Lemma \ref{lemma:substitutionsincs} with (4) and (5), (6):$\typing{x:A_1}{a}{M_0\subst{\overline{W}}{\overline{z}}}{A_1\funL{a}A_2}$.

	By Lemma \ref{lemma:substitutionsincs} over value substitutions with (6) and (3), $\typing{\emptyset}{a}{M_0\subst{\overline{W}}{\overline{z}}\subst{V}{x}}{A_2}$.

	\ \\
{\bf vi)} $M=\clo{\overline{W}}{F}[b]$: This is provable as for (v). (T-LAbs) and Lemma \ref{lemma:substitutionsincs} over location substitutions will be used instead of (T-Abs) and the lemma over open code substitution.

\end{proof}

	Since the communication rules move values from one location to another, proving the type preservation property requires well-typed values at one location to be well-typed at another location too. This property is proven by the value relocation lemma, Lemma \ref{lemma:valuerelocationincs}, which Lemma \ref{theorem:typepreservationincs} uses for the communication part.

\begin{lemma}[Value relocation in \polycs] If $\typing{\emptyset}{a}{V}{A}$ and $\valty{A}$ then $\typing{\emptyset}{b}{V}{A}$.
\label{lemma:valuerelocationincs}
\end{lemma}
\begin{proof} Basically, we prove this lemma by induction on the structure of values.
  Because of two conditions, the empty typing environment $\emptyset$ and $A$ is a relocatable type, $V$ can be one of $(V_1,V_2)$, $\clo{\overline{W}}{F}$, and $\Lambda\alpha.V_0$. The other cases cannot satisfy the two conditions.
  
   The base case is with $V=\Lambda\alpha.V_0$. Note that $\valty{\forall\alpha.A_0}$ does not always imply $\valty{A_0}$.
   By (T-Tabs) with the condition, $A=\forall\alpha.A_0$ and (1):$\typing{\alpha}{\Loc_0}{V_0}{A_0}$ for arbitrary $\Loc_0$. We have only to use (T-Tabs) with (1) to derive $\typing{\emptyset}{b}{\Lambda\alpha.V_0}{\forall\alpha.A_0}$ for location $b$.

  By induction, we prove the two remaining cases for $(V_1,V_2)$ and $\clo{\overline{W}}{F}$.

  The case $V=(V_1,V_2)$ is proved with (T-Pair) and the definition of relocatable types.

  The case $V=\clo{\overline{W}}{F}$ is also provedy by induction with (T-Clo). Note that the values of free variables have relocatable types by (T-F-Abs) or (T-F-LAbs).

\end{proof}

\begin{lemma}[Substitution in \polycs]
\label{lemma:substitutionsincs}
\ \\
\begin{itemize}
\item[(1)] If $\typing{\tyenv,x:B}{Loc}{M}{A}$ and $\typing{\tyenv}{Loc}{V}{B}$ then $\typing{\tyenv}{Loc}{M\subst{V}{x}}{A}$.
\item[(2)] If $\typing{\tyenv,\alpha}{Loc}{M}{A}$ and $\fvtyping{\tyenv}{B}$ then $\typing{\tyenv}{Loc}{M\subst{B}{\alpha}}{A\subst{B}{\alpha}}$.
\item[(3)] If $\typing{\tyenv,l}{Loc}{M}{A}$ and $\fvtyping{\tyenv}{Loc'}$ then $\typing{\tyenv}{Loc}{M\subst{Loc'}{l}}{A\subst{Loc'}{l}}$.
%% \item[(4)] If $\funtyping{\funstore}{\tyenv}{Loc}{F}{\overline{B}.A}$ and
%% 		$\funstore(F)=\overline{z}.OpenCode$ and
%% 		$\typing{\tyenv}{Loc}{\overline{W}}{\overline{B}}$ then
%% 		$\codetyping{\tyenv}{Loc}{OpenCode\subst{\overline{W}}{\overline{z}}}{A}$.
\end{itemize}
\end{lemma}
\begin{proof} We prove this lemma by induction on the structure of terms and $OpenCode$s.
\end{proof}

	The following type preservation theorem states that every evaluation step does not change the type of configurations. This key lemma is proven by case analysis on configurations with a subterm typeability lemma and a subterm replacement lemma.

\begin{lemma}[Subterm typeability] When a term $E[M]$ or a configuration $\conf[M]$ is well-typed, there exists some subderivation showing that the subterm $M$ is well-typed as follows.
\begin{itemize}
\item[(1)]
If $\typing{\emptyset}{Loc}{E[M]}{A}$ then there exists $B$ s.t. $\typing{\emptyset}{Loc}{M}{B}$. If there are no $Loc$ and $B$ s.t. $\typing{\emptyset}{Loc}{M}{B}$ then there are no $A$ s.t. $\typing{\emptyset}{Loc}{E[M]}{A}$.
\item[(2)]
If $\conftyping{\conf[M]}{A}$ then there exist $a$ and $B$ s.t. $\typing{\emptyset}{a}{M}{B}$. If there are no $a$ and $B$ s.t. $\typing{\emptyset}{a}{M}{B}$ then there are no $A$ s.t. $\conftyping{\conf[M]}{A}$.
\end{itemize}
\label{lemma:subtermtypeabilityincs}
\end{lemma}
\begin{proof} By configuration typing rules, (2) reduces to (1). We prove (1) by induction on the depth of $E$. When $E=[\ ]$, it is trivial since $E[M]=M$. Otherwise $E$ is either $\ldo{x}{E_0[\ ]}{N}$ or $\llet{x}{E_0[\ ]}{N}$. Then $E_0[M]$ is well-typed under the empty typing environment at $\Loc$ by (T-Bind) or (T-Let). Since typing $M$ is in the left subderivation, the location annotation and the typing environment are not changed. By induction, the lemma is proved.
\end{proof}

\begin{lemma}[Subterm replacement] If
\begin{itemize}
\item[(1)] $\mathcal{D}$ is a derivation concluding $\typing{\emptyset}{Loc}{E[M]}{A}$,
\item[(2)] $\mathcal{D}_1$ is a subderivation of $\mathcal{D}$ concluding $\typing{\emptyset}{Loc}{M}{B}$,
\item[(3)] $\mathcal{D}_1$ occurs in $\mathcal{D}$ in the position corresponding to the hole $[-]$ in $E$, and
\item[(4)] $\typing{\emptyset}{Loc}{N}{B}$
\end{itemize}
then $\typing{\emptyset}{Loc}{E[N]}{A}$.
\label{lemma:subtermreplacementincs}
\end{lemma}
\begin{proof} We prove this lemma by induction on the depth of $E$.

\ \\
{\bf i)} $E=[\ ]$: Since $E[M]=M$, $A=B$. $E[N]=N$. So, $\typing{\emptyset}{\Loc}{E[N]}{A}$.

\ \\
{\bf ii)} $E=\ldo{x}{E_0[\ ]}{L}$:

	By (T-Bind), $\typing{\emptyset}{Loc}{\ldo{x}{E_0[M]}{L}}{T\ C_1}$ where $T C_1=A$.

	(1):$\typing{\emptyset}{Loc}{E_0[M]}{T C_2}$ and (2):$\typing{\{x:C_2\}}{Loc}{L}{T C_1}$.

	By I.H. with (1) replacing the first condition, (3):$\typing{\emptyset}{Loc}{E_0[N]}{T C_2}$.

	By (T-Bind) with (3) and (2), $\typing{\emptyset}{Loc}{\ldo{x}{E_0[N]}{L}}{T C_1}$.

\ \\
{\bf iii)} $E=\llet{x}{E_0[\ ]}{L}$: This case is proved by the same way as for ii) except the use of (T-Let) instead of (T-Bind).
\end{proof}

\begin{lemma}[Type preservation for configurations] If \ $\conftyping{Conf}{A}$ and $Conf \run Conf'$ then $\conftyping{Conf'}{A}$.
\label{theorem:typepreservationincs}
\end{lemma}
\begin{proof} We prove this lemma by case analysis on $Conf$.

\ \\
%%%
{\bf i)} $Conf=\conf[M]$: $Conf' = \conf[M']$ and (1):$M \run M'$.

	i-1) $\conf[M]=\confcs{E[M];\stack_\client}{\stack_\server}$:

	By (T-Client), (2):$\typing{\emptyset}{\client}{E[M]}{T B}$ and (3):$\stacktyping{\client}{ \stackcsWith{\stack_\client}{\stack_\server} }{T B \Rightarrow A}$.

	By the subterm typeability (Lemma \ref{lemma:subtermtypeabilityincs}) with (2), (4):$\typing{\emptyset}{\client}{M}{C}$.

	By the type preservation for terms (Lemma \ref{lemma:typepreservationforterms}) with (4) and (1), (5):$\typing{\emptyset}{\client}{M'}{C}$.

	By the subterm replacement (Lemma \ref{lemma:subtermreplacementincs}) with (2) and (5), (6):$\typing{\emptyset}{\client}{E[M']}{T B}$.

	By (T-Client) with (6) and (3), $\conftyping{\confcs{E[M'];\stack_\client}{\stack_\server}}{A}$.

	i-2) $\conf[M]=\confcs{\stack_\client}{E[M];\stack_\server}$:  This is proved by the same way as for i-1) except the use of (T-Server) instead of (T-Client).

	\ \\
%%%
{\bf ii)} $Conf=\confcs{ E[\req{V}{W}];\stack_\client }{ \stack_\server }$: $Conf'=\confcs{ E[\ ];\stack_\client }{ V(W);\stack_\server }$.

	By (T-Client), (1):$\typing{\emptyset}{\client}{E[\req{V}{W}]}{T B}$ and (2):$\stacktyping{\client}{ \stackcsWith{\stack_\client}{\stack_\server} }{T B \Rightarrow A}$.

	By the subterm typeability (Lemma \ref{lemma:subtermtypeabilityincs}) with (1), (3):$\typing{\emptyset}{\client}{\req{V}{W}}{T B'}$.

	By (T-Req) with (3), (4):$\typing{\emptyset}{\client}{V}{\cloty{A'\funL{\server}T B'}}$, (5):$\typing{\emptyset}{\client}{W}{A'}$ and (6):$\valty{A'}$.

	By the value relocation (Lemma \ref{lemma:valuerelocationincs}) with (4) and $\valty{\cloty{A'\funL{\server}T B'}}$, (7):$\typing{\emptyset}{\server}{V}{\cloty{A'\funL{\server}T B'}}$. In the same way with (5) and (6), (8):$\typing{\emptyset}{\server}{W}{A'}$.

	By (T-App) with (7) and (8), (9):$\typing{\emptyset}{\server}{V(W)}{T B'}$.

	From (1) and (3), it is easy to derive (10):$\typing{\{x:T B'\}}{\client}{E[x]}{T B}$.

	By (T-Stk-Server) with (10) and (2),  (11):$\stacktyping{\server}{ \stackcsWith{E[\ ];\stack_\client}{\stack_\server} }{ T B' \Rightarrow A}$.

	By (T-Server) with (9) and (11), $\conftyping{ \confcs{E[\ ];\stack_\client}{V(W);\stack_\server} }{A}$

\ \\
%%%
{\bf iii)} $Conf=\confcs{ \stack_\client }{ E[\call{V}{W}];\stack_\server }$: $Conf'=\confcs{ V(W);\stack_\client }{ E[\ ];\stack_\server }$.

	This case is proved by the same way as the one for the previous case except the use of (T-Server), (T-Call), (T-Stk-Client), and (T-Client) instead of (T-Client), (T-Req), (T-Stk-Server), and (T-Server).

\ \\
%%%
{\bf iv)} $Conf=\confcs{ \lunit{V};\stack_\client }{ E[\ ];\stack_\server }$. $Conf'=\confcs{ \stack_\client }{ E[\lunit{V}];\stack_\server }$.

	By (T-Client), (1):$\typing{\emptyset}{\client}{\lunit{V}}{T B}$ and (2):$\stacktyping{\client}{ \stackcsWith{\stack_\client}{E[\ ];\stack_\server} }{T B \Rightarrow A}$.

	By (T-Stk-Client) with (2), (3):$\typing{\{x:T B\}}{\server}{E[x]}{T C}$ and (4):$\stacktyping{\server}{ \stackcsWith{\stack_\client}{\stack_\server} }{T C \Rightarrow A}$.

	By (T-Unit) with (1), (5):$\typing{\emptyset}{\client}{V}{B}$ and (6):$\valty{B}$.

	By the value relocation (Lemma \ref{lemma:valuerelocationincs}) with (5) and (6), (7):$\typing{\emptyset}{\server}{V}{B}$.

	By (T-Unit) with (7), (8):$\typing{\emptyset}{\server}{\lunit{V}}{T B}$.

	By the value substitution (Lemma \ref{lemma:substitutionsincs}), (3) and (8), (9):$\typing{\emptyset}{\server}{E[\lunit{V}]}{T C}$.

	By (T-Server) with (9) and (4), $\conftyping{ \confcs{ \stack_\client }{ E[\lunit{V}];\stack_\server } }{ A }$.

\ \\
%%%
{\bf v)} $Conf=\confcs{ E[\ ];\stack_\client }{ \lunit{V};\stack_\server }$: $Conf'=\confcs{ E[\lunit{V}];\stack_\client }{ \stack_\server }$.

	This case is proved by the same way as the one for the previous case except the use of (T-Server), (T-Stk-Client), and (T-Client) instead of (T-Client), (T-Stk-Server), and (T-Server).

\ \\
%%%
{\bf vi)}  $Conf=\confcs{ \emptystack }{ \lunit{V} }$: $Conf'=\confcs{ \lunit{V} }{ \emptystack }$.

	By (T-Server), (1):$\typing{\emptyset}{\server}{\lunit{V}}{T B}$ and (2):$\stacktyping{\server}{ \stackcsWith{\emptystack}{\emptystack} }{T B \Rightarrow A}$.

	By (T-Stk-Empty) with (2), (3):$TB=A$.

	By (T-Unit) with (1), (4):$\typing{\emptyset}{\server}{V}{B}$. and (5):$\valty{B}$.

	By the value relocation (Lemma \ref{lemma:valuerelocationincs}) with (4) and (5), (6):$\typing{\emptyset}{\client}{V}{B}$.

	By (T-Unit) with (6), (7):$\typing{\emptyset}{\client}{\lunit{V}}{T B}$.

	By (T-Client) with (7) and (2), $\conftyping{ \confcs{\lunit{V}}{\emptystack} }{ A }$.

\ \\
%%%
{\bf vii)}  $Conf=\confcs{ E[\gen{\client}{V}{W}];\stack_\client }{ \stack_\server }$: $Conf'=\confcs{ E[V(W)];\stack_\client }{ \stack_\server }$.

	By (T-Client), (1):$\typing{\emptyset}{\client}{ E[\gen{\client}{V}{W}] } {T B}$ and (2):$\stacktyping{\client}{ \stackcsWith{\stack_\client}{\stack_\server} }{T B \Rightarrow A}$.

	By the subterm typeability (Lemma \ref{lemma:subtermtypeabilityincs}) with (1), (3):$\typing{\emptyset}{\client}{\gen{\client}{V}{W}}{T B'}$.

	By (T-Gen) with (3), (4):$\typing{\emptyset}{\client}{V}{\cloty{A'\funL{\client}T B'}}$  (5):$\typing{\emptyset}{\client}{W}{A'}$, and (6):$\valty{A'}$.

	By (T-App) with (4) and (5), (7):$\typing{\emptyset}{\client}{V(W)}{T B'}$.

	By the subterm replacement (Lemma \ref{lemma:subtermreplacementincs}) with (1), (3), and (7), (8):$\typing{\emptyset}{\client}{ E[V(W)] } {T B}$.

	By (T-Client) with (8) and (2), $\conftyping{ \confcs{ E[V(W)];\stack_\client }{ \stack_\server } }{A}$.

\ \\
%%%
{\bf viii)}  $Conf=\confcs{ E[\gen{\server}{V}{W}];\stack_\client }{ \stack_\server }$: $Conf'=\confcs{ E[\req{V}{W}];\stack_\client }{ \stack_\server }$.

	By (T-Client), (1):$\typing{\emptyset}{\client}{ E[\gen{\server}{V}{W}] } {T B}$ and (2):$\stacktyping{\client}{ \stackcsWith{\stack_\client}{\stack_\server} }{T B \Rightarrow A}$.

	By the subterm typeability (Lemma \ref{lemma:subtermtypeabilityincs}) with (1), (3):$\typing{\emptyset}{\client}{\gen{\server}{V}{W}}{T B'}$.

	By (T-Gen) with (3), (4):$\typing{\emptyset}{\client}{V}{\cloty{A'\funL{\server}T B'}}$  (5):$\typing{\emptyset}{\client}{W}{A'}$, and (6):$\valty{A'}$.

	By (T-Req) with (4), (5), and (6), (7):$\typing{\emptyset}{\client}{\req{V}{W}}{T B'}$.

	By the subterm replacement (Lemma \ref{lemma:subtermreplacementincs}) with (1), (3), and (7), (8):$\typing{\emptyset}{\client}{ E[\req{V}{W}] } {T B}$.

	By (T-Client) with (8) and (2), $\conftyping{ \confcs{ E[\req{V}{W}];\stack_\client }{ \stack_\server } }{A}$.

\ \\
%%%
{\bf viiii)}  $Conf=\confcs{ \stack_\client }{ E[\gen{\client}{V}{W}];\stack_\server }$:
$Conf'=\confcs{ \stack_\client }{ E[\call{V}{W}];\stack_\server }$.

	This case is proved by the same way as the one for viii) except the use of (T-Server) and (T-Call) instead of (T-Client) and (T-Req).

\ \\
%%%
{\bf x)}  $Conf=\confcs{ \stack_\client }{ E[\gen{\server}{V}{W}];\stack_\server }$:
$Conf'=\confcs{ \stack_\client }{ E[V(W)];\stack_\server }$.

	This case is proved by the same way as the one for vii) except the use of (T-Server)  instead of (T-Client).
\end{proof}

	The following progress lemma says that every well-typed configuration makes progress until the configuration has a unit value with the empty client and server stacks. This theorem is proven by induction on the structure of $Conf$ and its terms $M$.
	When $M$ is a unit value, $\lunit{V}$, the proof involves a case analysis on the client and server stacks.
	When $M$ is a do-binding, the proof uses induction over the one or more nested do-bindings.
	When $M$ is a let-binding, the proof uses induction over the one or more nested let-bindings.

\begin{lemma}[Progress] If \ $\conftyping{Conf}{T A}$ then either $\exists Conf'. \ Conf \run Conf'$ or $Conf=\confcs{\lunit{V}}{\emptystack}$.
\label{theorem:progressincs}
\end{lemma}
\begin{proof} This theorem is proved by induction on the structure of $Conf$ and its term $M$ where $Conf$ is either $\confcs{ M;\stack_\client }{ \stack_\server }$ or $\confcs{ \stack_\client }{ M;\stack_\server }$.

\ \\
%%%
{\bf i)} $M=x$: This term should be closed because of the empty typing environment in the configuration typing under the condition.

\ \\
%%%
{\bf ii)} $M=(V_1,V_2)$: The type of this term, $B\times C$ cannot be in the form of $T A$ that the configuration typing expects in the condition.

\ \\
%%%
{\bf iii)} $M=\clo{\overline{W}}{F}$: The type of this term, $\cloty{B}$, cannot have the form of $T A$ that the configuration typing expects in the condition.

\ \\
%%%
{\bf iv)} $M=\lunit{V_0}$: We do case analysis on the client stack and the server stack.

	iv-1) $\stack_\client=\emptystack$ and $\stack_\server=\emptystack$: This case is proved since $Conf=\confcs{\lunit{V}}{\emptystack}$.

	\ \\

	iv-2) $\stack_\client\not=\emptystack$ and $\stack_\server=\emptystack$: $Conf$ cannot be in the form of $\confcs{M;\stack_\client}{\stack_\server}$ because (T-Client) is inapplicable to the configuration typing in the condition. If it were applicable, we would have to derive $\stacktyping{\client}{ \stackcsWith{\stack_\client}{\emptystack} }{T A_0\Rightarrow T A}$. But this is impossible by (T-Stk-Client) that expects some nonempty server stack.
	Therefore, $Conf=\confcs{\stack_\client}{\lunit{V_0};\stack_\server}$, so we have (1):$\conftyping{ \confcs{\stack_\client}{\lunit{V_0};\stack_\server} }{T A}$.

	By (T-Server) with (1), (2):$\typing{\emptyset}{\server}{\lunit{V_0}}{T A_0}$ and (3):$\stacktyping{\server}{ \stackcsWith{ \stack_\client }{ \stack_\server } }{TA_0 \Rightarrow TA}$.

	By (T-Stk-Server) with (3), (4):$\stack_\client=E[\ ];\stack_\client'$, (5):$\typing{\{x:TA_0\}}{\client}{E[x]}{TC}$, and (6):$\stacktyping{\client}{ \stackcsWith{\stack_\client'}{\stack_\server} }{TC \Rightarrow TA}$.

	$Conf=\confcs{\stack_\client}{\lunit{V_0};\stack_\server}=\confcs{E[\ ];\stack_\client'}{\lunit{V_0};\stack_\server} \run \confcs{E[\lunit{V_0}];\stack_\client'}{\stack_\server} = Conf'$.

		\ \\

	iv-3) $\stack_\client=\emptystack$ and $\stack_\server\not=\emptystack$: This case is proved by the same way as the one for the previous case except the use of (T-Client) and (T-Stk-Client) instead of (T-Server) and (T-Stk-Server).

	$Conf=\confcs{\lunit{V_0};\stack_\client}{\stack_\server}=\confcs{\lunit{V_0};\stack_\client}{E[\ ];\stack_\server'} \run \confcs{\stack_\client}{E[\lunit{V_0}];\stack_\server'} = Conf'$ where $\stack_\server=E[\ ];\stack_server'$.

			\ \\

	iv-4) $\stack_\client\not=\emptystack$ and $\stack_\server\not=\emptystack$: We will have $Conf \run Conf'$ as follows: either

	$Conf=\confcs{\lunit{V_0};\stack_\client}{E[\ ];\stack_\server'} \run \confcs{\stack_\client}{E[\lunit{V_0}];\stack_\server'}=Conf'$ where $\stack_\server=E[\ ];\stack_\server'$, or

	$Conf=\confcs{E[\ ];\stack_\client'}{\lunit{V_0};\stack_\server} \run \confcs{E[\lunit{V_0}];\stack_\client'}{\stack_\server}=Conf'$ where $\stack_\client=E[\ ];\stack_\client'$.

\ \\
%%%
{\bf v)} $M=\pi_i(V_0)$: $Conf=\conf[\pi_i(V_0)]$.

	By (T-Client) or (T-Server), (1):$\typing{\emptyset}{a}{\pi_i(V_0)}{T A_0}$.

	By (T-Proj-i) with (1), (2):$\typing{\emptyset}{a}{V_0}{A_1 \times A_2}$ where $i=1 \ \mbox{or} \ 2$ and $A_i=T A_0$.

	By (T-Pair) with (2), $V_0=(V_1, V_2)$.

	$Conf=\conf[\pi_i(V_1,V_2)] \run \conf[V_i]=Conf'$.

\ \\
%%%
{\bf vi)} $M=V_0[B_0]$: $Conf=\conf[V_0[A_0]]$.

	By (T-Client) or (T-Server), (1):$\typing{\emptyset}{a}{V_0[B_0]}{T \ A_0}$.

	By (T-Tapp) with (1), (2):$\typing{\emptyset}{a}{V_0}{\forall\alpha.A_1}$.

	By (T-Tabs) with (2), (3):$V_0=\Lambda\alpha.V_1$. So, we will have an evaluation step as:

	$Conf=\conf[V_1[B_0]] = \conf[(\Lambda\alpha.V_1)[B_0]] \run \conf[V_1\subst{B_0}{\alpha}]= Conf'$.

\ \\
%%%
{\bf vii)} $M=V_0(V_1)$: $Conf=\conf[V_0(V_1)]$.

	By (T-Client) or (T-Server), (1):$\typing{\emptyset}{a}{V_0(V_1)}{TA_0}$.

	By (T-App) with (1), (2):$\typing{\emptyset}{a}{V_0}{\cloty{A_1\funL{a}TA_0}}$ and (3):$\typing{\emptyset}{a}{V_1}{A_1}$.

	By (T-Clo) with (2), (4):$V_0=\clo{\overline{W}}{F}$, (5):$\funtyping{}{\emptyset}{Loc}{F}{\overline{B}.(A_1\funL{a}TA_0)}$ and (6):$\typing{\emptyset}{Loc}{\overline{W}}{\overline{B}}$.

	By (T-F-Abs) with (5), $F$ is $F_{name}[\loctys]$ and $F_{name}:\loctyvars.\overline{B_2}.A_2 = \loctyvars.\overline{z}.OpenCode \in \funstore$ such that $((\overline{B_2}.A_2)[\loctys/\loctyvars])=\overline{B}.(A_1\funL{a}TA_0)$.

	By (T-Abs), $\overline{z}.OpenCode$ is $\overline{z}.\lambda x. M_0$ where each $z_i$ in $\overline{z}$ has type $B_i$ in $\overline{B}$. Therefore, $\funstore(F)=\overline{z}.\lambda x. M_0$.

	$Conf=\conf[V_0(V_1)] = \conf[\clo{\overline{W}}{F}(V_1)] \run \conf[M_0\subst{\overline{W}}{\overline{z}}\subst{V_1}{x}]= Conf'$.

\ \\
%%%
{\bf viii)} $M=V_0[Loc_0]$: $Conf=\conf[V_0[Loc_0]]$.

	This is proved by the same way as the one for vii) except the use of (T-Lapp) instead of (T-App) for $V_0$ to have type $\cloty{\forall l.A_1}$. This will eventually lead to $\funstore(F)=\overline{z}.\Lambda l. V_1$. Because of this, we will have an evaluation step as:

	$Conf=\conf[V_0[Loc_0]] = \conf[\clo{\overline{W}}{F}[Loc_0]] \run \conf[V_1\subst{\overline{W}}{\overline{z}}\subst{Loc_0}{l}]= Conf'$.

\ \\
%%%
{\bf viiii)} $M=\req{V_0}{V_1}$:

	By (T-Client), we have $\typing{\emptyset}{a}{\req{V_0}{V_1}}{T A}$ where $a$ is either $\client$ or $\server$.
	But by (T-Req), $a$ must be $\client$. So, $Conf=\confcs{\req{V_0}{V_1};\stack_\client}{\stack_\server}$.

	$Conf=\confcs{\req{V_0}{V_1};\stack_\client}{\stack_\server} \run \confcs{\stack_\client}{V_0(V_1);\stack_\server} = Conf'$.

\ \\
%%%
{\bf x)} $M=\call{V_0}{V_1}$:

	By (T-Server), we have $\typing{\emptyset}{a}{\call{V_0}{V_1}}{T A}$ where $a$ is either $\client$ or $\server$.
	But by (T-call), $a$ must be $\server$. So, $Conf=\confcs{\stack_\client}{\call{V_0}{V_1};\stack_\server}$.

	$Conf=\confcs{\stack_\client}{\call{V_0}{V_1};\stack_\server} \run \confcs{V_0(V_1);\stack_\client}{\stack_\server} = Conf'$.

\ \\
%%%
{\bf xi)} $M=\gen{Loc'}{V_0}{V_1}$: Since $M$ is closed by the configuration typing in the condition, $Loc'$ is either $\client$ or $\server$. Then it is straightforward to have one of four evaluation steps as follows:

	$Conf=\confcs{\gen{\client}{V_0}{V_1};\stack_\client\ \ }{\ \ \stack_\server}
	\run
	\confcs{V_0(V_1);\stack_\client\ \ }{\ \ \stack_\server}=Conf'$

	$Conf=\confcs{\gen{\server}{V_0}{V_1};\stack_\client\ \ }{\ \ \stack_\server}
	\run
	\confcs{\req{V_0}{V_1};\stack_\client\ \ }{\ \ \stack_\server}=Conf'$

	$Conf=\confcs{\stack_\client\ \ }{\ \ \gen{\client}{V_0}{V_1};\stack_\server}
	\run
	\confcs{\stack_\client\ \ }{\ \ \call{V_0}{V_1};\stack_\server}=Conf'$

	$Conf=\confcs{\stack_\client\ \ }{\ \ \gen{\server}{V_0}{V_1};\stack_\server}
	\run
	\confcs{\stack_\client\ \ }{\ \ V_0(V_1);\stack_\server}=Conf'$

\ \\
%%%
{\bf xii)} $M=\llet{x}{L}{N}$:

	We analyze the term $M$ by sub-induction on the structure of $L$.Assuming that $E_0[\ ]=\llet{x}{[\ ]}{N}$, $M=E_0[L]$.
	\ \\

	xii-1) $L$ is $V$. We can use (E-Let), proving this case.
	\ \\

	xii-2) $L$ is $\llet{y}{L'}{N'}$. Assuming $E_1[\ ]= \llet{y}{[\ ]}{N'}$, we have $E_2=E_0[E_1[\ ]]$. Then $M=E_2[L']$. By the sub-induction, we prove the case xii-2).
	\ \\

	xii-3) $L$ is one of $\pi(V)$, $V(W)$, $V[A]$, and $V[\Loc]$. We prove this case by the same way as v), vi), vii), and viii).
	\ \\

{\bf xiii)} $M=\ldo{x}{L}{N}$: $Conf$ is $\confcs{M;\stack_\client}{\stack_\server}$ or $\confcs{\stack_\client}{M;\stack_\server}$.

	We analyze the term $M$ by sub-induction on the structure of $L$.Assuming that $E_0[\ ]=\ldo{x}{[\ ]}{N}$, $M=E_0[L]$.
	\ \\

	xiii-1) $L$ cannot be a variable because of  the empty typing environment in $\conftyping{Conf}{T A}$. In fact, the same argument is given in the case i).
	\ \\

	xiii-2) $L$ cannot be one of $(V,W)$ and $\clo{\overline{W}}{F}$ either because of the following reason. By Lemma \ref{lemma:subtermtypeabilityincs} with the condition of this lemma, the subterm $L$ is typed. By (T-Bind), the type of $L$ must be in the form of monadic type. In fact, the same arguments are given in the case ii) and iii).
	\ \\

	xiii-3) $L$ is one of $\lunit{V}$, $\req{V}{W}$, $\call{V}{W}$, and $\gen{\Loc}{V}{W}$. We prove this case by the same way as iv), viiii), x), and xi).
	\ \\

	xiii-4) $L$ is $\ldo{y}{L'}{N'}$. Assuming $E_1=\ldo{y}{[\ ]}{N'}$, we have $E_2=E_0[E_1[\ ]]$ such that $M=E_2[L']$. By the sub-induction, we prove the case xiii-4).
	\ \\

	xiii-5) $L$ is one of $\pi(V)$, $V(W)$, $V[A]$, and $V[\Loc]$. We prove this case by the same way as v), vi), vii), and viii).
	\ \\

	xiii-6) $L$ is $\llet{y}{L'}{N'}$. Assuming $E_{let,1}=\llet{y}{[\ ]}{N'}$, we have $E_2=E_0[E_{let,1}[\ ]]$ such that $M=E_2[L']$. We prove this case by the same way as xii).
\end{proof}

\subsection{A type-based slicing compilation}

	Our typed slicing compilation translates the polymorphic RPC calculus into the polymorphic CS calculus.
	The basic idea is to compile terms of type $A$ in the RPC calculus into monadic terms of  type $T \ (\vcomp{A})$ in the CS calculus. That is, the compiled term denotes a computation that may call remote procedures during the evaluation and that returns a value of type $\vcomp{A}$.

	Note that $\vcomp{-}$ is the translation of types, $A$, in the source calculus into plain value types, $\vcomp{A}$, in the target calculus, and $\ccomp{-}$ is another translation of types in the source calculus into monadic types, $T \ (\vcomp{A})$ in the target calculus.

	Figure \ref{fig:typedcompilation} shows our typed slicing compilation of {\polyrpc} into {\polycs}.
	It comprises type compilations, $\ccomp{A}$ and $\vcomp{A}$, and term compilations (or typing derivation compilations), $\ccomp{M}_{\tyenv,\Loc,A}$ and $\vcomp{V}_{\tyenv,\Loc,A}$.

	For convenience, the compilation of typing environments, $\vcomp{\tyenv}$, is defined as: $\vcomp{\tyenv,x:A} = \vcomp{\tyenv},x:\vcomp{A}$; $\vcomp{\tyenv,\alpha}  =  \vcomp{\tyenv},\alpha$; $\vcomp{\tyenv,l} = \vcomp{\tyenv},l$; and $\vcomp{\emptyset} =  \emptyset$.

	The type compilation rules, $\ccomp{-}$ and $\vcomp{-}$, are defined in terms of type structure. As already seen, $\ccomp{A}=T(\vcomp{A})$.
	% For type variables, $\vcomp{-}$ returns the type variables themselves.
	% It does the same for base types.
	% For pair types, it simply does a recursion over the first and the second types.
	For function types, $\vcomp{A\funL{\Loc}B}$ is defined as $\cloty{\vcomp{A}\funL{\Loc}\ccomp{B}}$. The located function type is translated into a closure type with the same location. The argument type is translated into a plain value type, and the return type is translated into a monadic type.
	For polymorphic location types, $\vcomp{\forall l.A}$ translates the polymorphic location into a closure type as $\cloty{\forall l.\ccomp{A}}$.
	For polymorphic types, $\vcomp{\forall\alpha.A}$ is simply translated into a polymorphic type $\forall\alpha.\ccomp{A}$ where the body type becomes a computational one.
	Later, this form is convenient for the type erasure translation that will erase type abstraction and application terms.

	The term compilation rules, $\ccomp{M}_{\tyenv,\Loc,A}$ and $\vcomp{V}_{\tyenv,\Loc,A}$, take as its input typing derivations for terms, such as typing derivations concluding with typing judgments $\typing{\tyenv}{\Loc}{M}{A}$ or $\typing{\tyenv}{\Loc}{V}{A}$.
	The output of the term compilation is  two function maps, $\funstore_\client$ and $\funstore_\server$ and a main expression at the client.
	We use a notation,  $(F_{name} : \CloCodeType = \CloCode) \in \funstore_{\Loc}$, for adding the binding  of $F_{name}$ to function stores.
	If $\Loc$ in $\funstore_{\Loc}$ is a location variable,
the compilation adds the binding both to the client function map and the server function map.

	For applications, the compilation $\ccomp{L\ M}_{\tyenv,\Loc,B}$  comprises the compilations of $L$ and $M$ to produce a closure bound to $f$ of type $\vcomp{A\funL{\Loc'}B}$, which is $\cloty{\vcomp{A}\funL{\Loc'}\ccomp{B}}$, and an argument value bound to $x$ of type $\vcomp{A}$ through do bindings. Then it generates a generic application, $\gen{\Loc'}{f}{x}$.
	Note that $\Loc$ is the application location, and $\Loc'$ is the function location. By analyzing the two locations, we can optimize the application terms whenever the relevant location information is statically available, as was done for compiling the typed RPC calculus~\cite{choijfp2019}.
	For type applications, $M[B]$, we first compile $M$ to produce a closure bound to $f$ of type $\vcomp{\forall\alpha.A}$, which is $\forall\alpha.\ccomp{A}$. We then apply $f$ to the compiled type $\vcomp{B}$.  The compilation rule for location applications is similar to the one for type applications.

	The compilation $\vcomp{\lambda^{Loc'}x.M}_{\tyenv,Loc,A\funL{Loc'}B}$ generates a closure, $\clo{\overline{z}}{F_{name}[\overline{l},\overline{\alpha}]}$ for a fresh name $F_{name}$. The free  location variables, type variables, value variables are collected as $\overline{l}$,  $\overline{\alpha}$, and $\overline{z}$, respectively. The function name refers to the code compiled from $M$ prefixed with the free location, type, value variables, and an abstraction with a bound variable $x$. Thus the code type is $\loctyvars. \overline{\vcomp{C}}.\vcomp{A}\funL{\Loc'}\ccomp{B}$.
	The compilation rule for location abstractions is similar to the compilation rule for lambda abstractions.

	The compilation rule for type abstractions produces a type abstraction whose body is a computational term. It does not produce a closure term.

\subsection{Type correctness of the type-based slicing compilation}

\begin{lemma}[Relocatable value types]
For all types $A$ in {\polyrpc}, $\valty{ \vcomp{A} }$.
\label{lemma:plainvaluetypes}
\end{lemma}
\begin{proof} This lemma is proved by induction on the structure of type $A$. As base cases, the lemma is true for $\alpha$ and $base$.

	As inductive cases, $\vcomp{-}$ for function types, polymorphic types, and polymorphic location types, produces closure types, which themselves are plain value types.

	For pair types $A\times B$, we know $\valty{\vcomp{A}}$ and $\valty{\vcomp{B}}$ by induction. Also the pair type $\vcomp{A}\times \vcomp{B}$ itself is also a plain value type.
\end{proof}

\begin{lemma}[Type and location substitution under type compilation] Type and location substitutions are preserved under the type compilations.

\begin{itemize}
\item
$\ccomp{A\subst{B}{\alpha}}=\ccomp{A}\subst{\vcomp{B}}{\alpha}$.
\item
$\ccomp{A\subst{Loc}{l}}=\ccomp{A}\subst{Loc}{l}$.
\end{itemize}
\label{lemma:substundertypecomp}
\end{lemma}
\begin{proof} This lemma is proved by induction on the structure of the type $A$.
\end{proof}

\begin{theorem}[Type correctness]
If \ $\typing{\tyenv}{Loc}{M}{A}$ in {\polyrpc} then $\typing{\vcomp{\tyenv}}{Loc}{\ccomp{M}_{\tyenv,\Loc,A}}{ \ccomp{A} }$ in {\polycs}.
%\label{thm:typecorrectslicingcompilation}
\end{theorem}
\begin{proof} We prove this theorem by induction on the height of the typing derivation.

\ \\
%%%
{\bf i)} $M=x$: By (T-Var) in {\polyrpc}, (1):$\typing{\tyenv}{Loc}{x}{A}$ and (2):$\tyenv(x)=A$.

	By (2), $\vcomp{\tyenv}(x)=\vcomp{A}$. So, (3):$\typing{\vcomp{\tyenv}}{Loc}{x}{\vcomp{A}}$ by (T-Var) in {\polycs}.

	By (T-Unit) with (3), $\typing{\vcomp{\tyenv}}{Loc}{\lunit{x}}{\ccomp{A}}$ where $\ccomp{x}_{\tyenv,Loc,A}=\lunit{x}$.
\ \\
%%%
{\bf ii)} $M=\forall\alpha.V_0$: By (T-Tabs) in {\polyrpc}, (1):$\typing{\tyenv}{\Loc}{\Lambda\alpha.V}{\forall\alpha.A_0}$ and (2):$\typing{\tyenv,\alpha}{\Loc_0}{V}{A_0}$ for arbitrary $\Loc_0$.

	By I.H. with (2), (3):$\typing{\vcomp{\tyenv,\alpha}}{\Loc_0}{\ccomp{V}_{\tyenv,\alpha,\Loc_0,A_0}}{\ccomp{A_0}}$ for arbitrary $\Loc_0$. 

	By (T-Tabs) in {\polycs} with (3) instantiated with $\Loc$ for $\Loc_0$, (4):$\typing{\vcomp{\tyenv}}{\Loc}{\Lambda\alpha.\ccomp{V}_{\tyenv,\alpha,\Loc,A_0}}{\forall\alpha.\ccomp{A_0}}$ where $\vcomp{\forall\alpha.V}_{\tyenv,\Loc,\forall\alpha.A_0}=\Lambda\alpha.\ccomp{V}_{\tyenv,\alpha,\Loc,A_0}$.

	By (T-Unit) with (4), (5):$\typing{\vcomp{\tyenv}}{\Loc}{\lunit \vcomp{\forall\alpha.V}_{\tyenv,\Loc,\forall\alpha.A_0}}{T \ (\forall\alpha.\ccomp{A_0})}$ where $T \ (\forall\alpha.\ccomp{A_0})=\ccomp{\forall\alpha.A_0}$ and $\lunit \vcomp{\forall\alpha.V}_{\tyenv,\Loc,\forall\alpha.A_0}=\ccomp{\forall\alpha.V}_{\tyenv,\Loc,\forall\alpha.A_0}$.

\ \\
%%%
{\bf iii)} $M=\lamL{Loc'}{x}{M_0}$: By (T-Abs) in {\polyrpc},
	(1):$\typing{\tyenv}{Loc}{\lamL{Loc'}{x}{M_0}}{B\funL{Loc'}C}$,
	(2):$\typing{\tyenv,x:B}{Loc'}{M_0}{C}$,
	(3):$A=B\funL{Loc'}C$.

	By I.H. with (2),
	(4):$\typing{\vcomp{\tyenv,x:B}}{Loc'}{\ccomp{M_0}_{(\tyenv,x:B),Loc',C}}{\ccomp{C}}$
	where $\vcomp{\tyenv,x:B}=\vcomp{\tyenv},x:\vcomp{B}$.

	By (T-Abs) in {\polycs},
	(5):$\codetyping{\vcomp{\tyenv}}{Loc}{\lambda x.\ccomp{M_0}_{(\tyenv,x:B),Loc',C}}{\vcomp{B}\funL{Loc'}\ccomp{C}}$.

	Suppose $\tyenv=\{\overline{l},\overline{\alpha},\overline{z}:\overline{B_i}\}$ without the loss of generality. Also $F_{name}$ is assumed to be fresh.

	By (5), (6):$\codetyping{\overline{l},\overline{\alpha},\overline{z}:\overline{\vcomp{B_i}}}{Loc}{\lambda x.\ccomp{M_0}_{(\tyenv,x:B),Loc',C}}{\vcomp{B}\funL{Loc'}\ccomp{C}}$.

	Then $(F_{name}:Ty = Code) \ \in \ \funstore_{\Loc'}$ where

\ \ \ \ \
\begin{tabular}{l l l}
$Ty$ & $=$ & $\loctyvars.\overline{\vcomp{B_i}}.(\vcomp{B}\funL{Loc'}\ccomp{C})$ and \\
$Code$ & $=$ & $\overline{z}.(\lambda x. \ccomp{M_0}_{(\tyenv,x:B),Loc',C})$.
\end{tabular}

	By (T-F-Abs), (7):$\funtyping{}{\vcomp{\tyenv}}{Loc}{F_{name}}{Ty}$ and
	(8):$\funtyping{}{\vcomp{\tyenv}}{Loc}{F_{name}[\loctyvars]}{ \overline{\vcomp{B_i}}.(\vcomp{B}\funL{Loc'}\ccomp{C}) }$.

	By (T-Var), (9):$\typing{\vcomp{\tyenv}}{Loc}{\overline{z}}{\overline{\vcomp{B_i}}}$.

	By (T-Clo) with (8) and (9), (10):$\typing{\vcomp{\tyenv}}{Loc'}{\clo{\overline{z}}{F_{name}[\overline{\alpha},\overline{l}]}}{ \cloty{\vcomp{B}\funL{Loc'}\ccomp{C}} }$
	where

\begin{tabular}{l l l}
$\vcomp{\lamL{Loc'}{x}{M_0}}_{\tyenv,Loc,B\funL{Loc'}C}$ & $=$ & $\clo{\overline{z}}{F_{name}[\overline{\alpha},\overline{l}]}$ \\
$\vcomp{B\funL{Loc'}C}$ & $=$ & $\cloty{\vcomp{B}\funL{Loc'}\ccomp{C})}$
\end{tabular}

	By (T-Unit) with (10), $\typing{\vcomp{\tyenv}}{Loc}{\lunit{(\clo{\overline{z}}{F_{name}[\loctyvars]}) } }{T (\cloty{\vcomp{B}\funL{Loc'}\ccomp{C})})}$.

\ \\
%%%
{\bf iv)} $M=\forall l.V_0$: This case is proved by the same way as the one for iii) except the use of (T-LAbs) instead of (T-Abs).

\ \\
%%%
{\bf v)} $M=(V_1,V_2)$: By (T-Pair) in {\polyrpc}, (1)-i:$\typing{\tyenv}{Loc}{V_i}{A_i}$ for $i=1 \ \mbox{or} \ 2$.

	By I.H. with (1)-i, (2)-i:$\typing{\vcomp{\tyenv}}{Loc}{\ccomp{V_i}_{\tyenv,Loc,A_i}}{\ccomp{A_i}}$.

	By (T-Unit) with (2)-i, (3)-i:$\typing{\vcomp{\tyenv}}{Loc}{\vcomp{V_i}_{\tyenv,Loc,A_i}}{\vcomp{A_i}}$.

	By (T-Pair) in {\polycs} with (3)-1 and (3)-2, $\typing{\vcomp{\tyenv}}{Loc}{(\vcomp{V_1}_{\tyenv,Loc,A_1},\vcomp{V_2}_{\tyenv,Loc,A_2})}{\vcomp{A_1}\times\vcomp{A_2}}$. So,
	(4):$\typing{\vcomp{\tyenv}}{Loc}{\vcomp{(V_1,V_2)}_{\tyenv,Loc,A_1}}{\vcomp{A_1\times A_2}}$.

	By (T-Unit) with (4),
	$\typing{\vcomp{\tyenv}}{Loc}{\lunit{ (\vcomp{(V_1,V_2)}_{\tyenv,Loc,A_1}) }}{T (\vcomp{A_1\times A_2}) }$.

\ \\
%%%
{\bf vi)} $M=L \ N$: By (T-App) in {\polyrpc},
	(1):$\typing{\tyenv}{Loc}{L}{B\funL{Loc'}A}$ and
	(2):$\typing{\tyenv}{Loc}{N}{B}$.

	By I.H. with (1), $\typing{\vcomp{\tyenv}}{Loc}{\ccomp{L}_{\tyenv,Loc,B\funL{Loc'}A}}{\ccomp{B\funL{Loc'}A}}$.

	By I.H. with (2), $\typing{\vcomp{\tyenv}}{Loc}{\ccomp{N}_{\tyenv,Loc,B}}{\ccomp{B}}$.

	$\ccomp{B\funL{Loc'}A}=T (\vcomp{B\funL{Loc'}A})=T (\cloty{\vcomp{B}\funL{Loc'}\ccomp{A}})$.

	By (T-Gen),
	(3):$\typing{\vcomp{\tyenv}, f: \cloty{\vcomp{B}\funL{Loc'}\ccomp{A}}, x:\vcomp{B}}{Loc}{\gen{Loc'}{f}{x}}{\ccomp{A}}$. Note that $\valty{ \vcomp{B} }$ by Lemma \ref{lemma:plainvaluetypes}.

	By (T-Bind) with (1), (2), and (3), it is straightforward to construct:

	$\typing{\vcomp{\tyenv}}{Loc}{
		\ldo{f}{\ccomp{L}_{\tyenv,Loc,B\funL{Loc'}A}}
			{\ldo{x}{\ccomp{N}_{\tyenv,Loc,B}}{\gen{Loc'}{f}{x}}} }{\ccomp{A}}$.

\ \\
%%%
{\bf vii)} $M=L[B]$: By (T-TApp) in {\polyrpc},
	$\typing{\tyenv}{Loc}{L[B]}{C\subst{B}{\alpha}}$ and
	(1):$\typing{\tyenv}{Loc}{L}{\forall\alpha.C}$.

	By I.H. with (1), (2):$\typing{\vcomp{\tyenv}}{Loc}{\ccomp{L}_{\tyenv,Loc,\forall\alpha.C}}{\ccomp{\forall\alpha.C}}$.

	By definition, (3):$\ccomp{\forall\alpha.C}=T (\vcomp{\forall\alpha.C}) = T (\forall\alpha. \ccomp{C})$.

	By Lemma \ref{lemma:plainvaluetypes}, we have (4):$\valty{ \vcomp{B} }$.

	By (T-TApp) with (4) in {\polycs}, (5):$\typing{\vcomp{\tyenv},f:\forall\alpha. \ccomp{C}}{Loc}{ f[\vcomp{B}] }{\ccomp{C}\subst{\vcomp{B}}{\alpha}}$.

	By (T-Bind) with (2) and (5),
	(6):$\typing{\vcomp{\tyenv}}{Loc}{\ldo{f}{\ccomp{L}_{\tyenv,Loc,\forall\alpha.C}}{f[\vcomp{B}]}}{\ccomp{C}\subst{\vcomp{B}}{\alpha}}$.

	By the type substitution under type compilation (Lemma \ref{lemma:substundertypecomp}),
	$\typing{\vcomp{\tyenv}}{Loc}{\ldo{f}{\ccomp{L}_{\tyenv,Loc,\forall\alpha.C}}{f[\vcomp{B}]}}{\ccomp{C\subst{B}{\alpha}}}$.

\ \\
%%%
{\bf viii)} $M=L[Loc']$: By (T-LApp) in {\polyrpc},
	$\typing{\tyenv}{Loc}{L[Loc']}{C\subst{Loc'}{l}}$ and
	(1):$\typing{\tyenv}{Loc}{L}{\forall l. C}$.

	By I.H. with (1), (2):$\typing{\vcomp{\tyenv}}{Loc}{\ccomp{L}_{\tyenv,Loc,\forall l.C}}{\ccomp{\forall l.C}}$.

	By definition, (3):$\ccomp{\forall l.C}=T(\vcomp{\forall l. C})=T(\cloty{\forall l. \ccomp{C}})$.

	By (T-LApp), (4):$\typing{\vcomp{\tyenv},f:\cloty{\forall l.\ccomp{C}}}{Loc}{f[Loc']}{\ccomp{C}\subst{Loc'}{l}}$.

	By (T-Bind) with (2) and (4),
	(5):$\typing{\vcomp{\tyenv}}{Loc}{\ldo{f}{\ccomp{L}_{\tyenv,Loc,\forall l.C}}{f[Loc']}}{\ccomp{C}\subst{Loc'}{l}}$.

	By the location substitution under the type compilation (Lemma \ref{lemma:substundertypecomp}),
	$\typing{\vcomp{\tyenv}}{Loc}{\ldo{f}{\ccomp{L}_{\tyenv,Loc,\forall l.C}}{f[Loc']}}{\ccomp{C\subst{Loc'}{l}}}$.

\ \\
%%%
{\bf ix)} $M=(L_1,L_2)$: By (T-Pair) in {\polyrpc}, (1)-i:$\typing{\tyenv}{Loc}{L_i}{A_i}$ for $i=1 \ \mbox{or} \ 2$.

	By I.H. with (1)-i, (2)-i:$\typing{\vcomp{\tyenv}}{Loc}{\ccomp{L_i}_{\tyenv,Loc,A_i}}{\ccomp{A_i}}$.

	By (T-Var), (T-Pair), (T-Unit), and (T-Bind) in {\polycs} with (2)-1 and (2)-2, it is straightforward to construct:

	$\typing{\vcomp{\tyenv}}{Loc}{\ldo{x_1}{\ccomp{L_1}_{\tyenv,Loc,A_1}}{\ldo{x_2}{\ccomp{L_2}_{\tyenv,Loc,A_2}}}{\lunit{(x_1,x_2)}}} {T (\vcomp{A_1}\times\vcomp{A_2}) }$

\ \\
%%%
{\bf x)} $M=\pi_i(N)$: By (T-Proj-i) in {\polyrpc}, (1):$\typing{\tyenv}{Loc}{\pi_i(N)}{A_i}$ where $A=A_i$, and (2):$\typing{\tyenv}{Loc}{N}{A_1\times A_2}$.

	By I.H. with (2), (3):$\typing{\vcomp{\tyenv}}{Loc}{\ccomp{N}_{\tyenv,Loc,A_1\times A_2}}{\ccomp{A_1\times A_2}}$.

	By (T-Var), (T-Let), (T-Proj-i), and (T-Unit) in {\polycs}, it is easy to construct
	(4):$\typing{\vcomp{\tyenv},p:\vcomp{A_1\times A_2}}{Loc}{\llet{x}{\pi_i(p)}{\lunit{x}}}{\ccomp{A_i}}$.

	By (T-Bind) with (3) and (4),
	(5):$\typing{\vcomp{\tyenv}}{Loc}{\ldo{p}{\ccomp{N}_{\tyenv,Loc,A_1\times A_2}}{(\llet{x}{\pi_i(p)}{\lunit{x}})} }{\ccomp{A_i}}$.

\end{proof}

\subsection{Semantic correctness of the type-based slicing compilation}

\begin{lemma}[Value compilation under relocation]
$\vcomp{V}_{\emptyset,a,A} = \vcomp{V}_{\emptyset,b,A}$.
\label{lemma:relocationundertermcomp}
\end{lemma}
\begin{proof} This lemma is proved by induction on the height of the typing derivation over $V$.
\end{proof}

\begin{lemma}[Substitution under compilation]
$\ccomp{M}_{\{x:B\},Loc,A}\subst{\vcomp{V}_{\emptyset,Loc,B}}{x} $ $=$ \\
      $ \ccomp{M\subst{V}{x}}_{\emptyset,Loc,A}$.
\label{lemma:substundertermcomp}
\end{lemma}
\begin{proof} This lemma is proved by induction on the height of the typing derivation over $M$.
\end{proof}

\begin{theorem}[Semantic correctness]
If \ $\typing{\emptyset}{\client}{M}{A}$  and $\evalRPC{M}{\client}{V}$ then $\confcs{ \ccomp{M}_{\emptyset,\client,A}}{\emptystack} \run^* \confcs{\ccomp{V}_{\emptyset,\client,A}}{\emptystack}$.
%\label{thm:semanticcorrectslicingcompilation}
\end{theorem}
\begin{proof} We first generalize this theorem as:
\begin{center}
	If \ (1):$\typing{\emptyset}{a}{M}{A}$ and (2):$\evalRPC{M}{a}{V}$
	then $\conf_a[\ccomp{M}_{\emptyset,a,A}] \run^* \conf_a[\ccomp{V}_{\emptyset,a,A}]$.
\end{center}

	We prove the generalized theorem by induction on the height of the derivation tree $\evalRPC{M}{a}{V}$.
	Let $\conf_a$ is either $\confcs{E_0[\ ];\stack_\client}{\stack_\server}$ or
	$\confcs{\stack_\client}{E_0[\ ];\stack_\server}$.

\ \\
%%%
{\bf Base case)} One of (Abs), (Tabs), and (Labs) is used in the derivation tree of height 1 in (2). Then $M=V$. The generalized theorem is proved trivially.

\ \\
%%%
{\bf Inductive case)} Let us consider the cases using one of (Pair), (App), (Tapp), (Lapp), and (Proj-i) in the bottom of the derivation tree (2).

\ \\
{\bf (Pair)} $M=(L,N)$. By (1), (3):$\typing{\emptyset}{a}{(L,N)}{A_1\times A_2}$ where $A=A_1\times A_2$.

	By (T-Pair) with (3), (4):$\typing{\emptyset}{a}{L}{A_1}$ and (5):$\typing{\emptyset}{a}{N}{A_2}$.

	By (Pair) with (2), (6):$\evalRPC{L}{a}{V_1}$ and (7):$\evalRPC{N}{a}{V_2}$.

	By I.H. with (4) and (6), (8):$\conf_a^L[\ccomp{L}_{\emptyset,a,A_1}] \run^* \conf_a^L[\ccomp{V_1}_{\emptyset,a,A_1}]$.

	By I.H. with (5) and (7), (9):$\conf_a^N[\ccomp{N}_{\emptyset,a,A_2}] \run^* \conf_a^N[\ccomp{V_2}_{\emptyset,a,A_2}]$. \\

	By the def. of the compilation,

	$\ccomp{(L,N)}_{\emptyset,a,A_1\times A_2}=
		\ldo{x}{ \ccomp{L}_{\emptyset,a,A_1} }{
			(\ldo{y}{ \ccomp{N}_{\emptyset,a,A_2} }{ \lunit{(x,y)} }) }$. \\

	Let $\conf_a^L$ be either $\confcs{E^L[\ ];\stack_\client}{\stack_\server}$ or
	$\confcs{\stack_\client}{E^L[\ ];\stack_\server}$ where

	$E^L[\ ] = E_0[\ldo{x}{ [\ ] }{
			(\ldo{y}{ \ccomp{N}_{\emptyset,a,A_2} }{ \lunit{(x,y)} }) }]$. \\

	Let $\conf_a^N$ be either $\confcs{E^N[\ ];\stack_\client}{\stack_\server}$ or
	$\confcs{\stack_\client}{E^N[\ ];\stack_\server}$ where

	$E^N[\ ] = E_0[ \ldo{y}{ [\ ] }{ \lunit{(\vcomp{V_1}_{\emptyset,a,A_1},y)} } ]$.  \\

	Then, we can construct the following evaluation steps:

\begin{tabular}{l l l}
& $\conf_a^L[\ccomp{L}_{\emptyset,a,A_1}] $
& by (8) \\
$\run^*$
&
$\conf_a^L[\ccomp{V_1}_{\emptyset,a,A_1}]$
& by (E-Do) \\
$ \run $
& $\conf_a^N[\ccomp{N}_{\emptyset,a,A_2}]$
&  by (9)\\
$\run^*$
&
$\conf_a^N[\ccomp{V_2}_{\emptyset,a,A_2}]$
& by (E-Do) \\
$\run$
&
$\conf_a[ \lunit{(\vcomp{V_1}_{\emptyset,a,A_1},\vcomp{V_2}_{\emptyset,a,A_2})} ]$
\end{tabular}

\ \\
{\bf (App)} $M=L \ N $. (1):$\typing{\emptyset}{a}{L \ N}{A}$, (2):$\typing{\emptyset}{a}{L}{B\funL{b}A}$, and (3):$\typing{\emptyset}{a}{N}{B}$.

	By (App) in {\polyrpc} with (2), (4):$\evalRPC{L}{a}{ \lamL{b}{x}{M_0} }$, (5):$\evalRPC{N}{a}{W}$, and (6):$\evalRPC{M_0\subst{W}{x}}{b}{V}$.

	By I.H. with (2) and (4),
	(7):$\conf_a^L[\ccomp{L}_{\emptyset,a,B\funL{b}A}] \run^* \conf_a^L[\ccomp{\lamL{b}{x}{M_0}}_{\emptyset,a,B\funL{b}A}]$.

	By I.H. with (3) and (5),
	(8):$\conf_a^N[\ccomp{N}_{\emptyset,a,B}] \run^* \conf_a^N[\ccomp{W}_{\emptyset,a,B}]$.

	By the type soundness for {\polyrpc} with (2) and (4),
	(9):$\typing{\emptyset}{a}{\lamL{b}{x}{M_0}}{B\funL{b}A}$.

	By the type soundness for {\polyrpc} with (3) and (5),
	(10):$\typing{\emptyset}{a}{W}{A}$.

	By (T-Abs) in {\polyrpc} with (9), (11):$\typing{\{x:B\}}{b}{M_0}{A}$.

	By the value substitution (Lemma \ref{lemma:substundertermcomp}), (12):$\typing{\emptyset}{b}{M_0\subst{W}{x}}{A}$.

	By I.H. with (12) and (6), (13):$\conf_b[\ccomp{M_0\subst{W}{x}}_{\emptyset,b,A}] \run^* \conf_b[\ccomp{V}_{\emptyset,b,A}]$. \\

	By the def. of the compilation,

	$\ccomp{L \ M}_{\emptyset,a,A} =
		\ldo{f}{\ccomp{L}_{\emptyset,a,B\funL{b}A}}{
			(\ldo{x}{\ccomp{N}_{\emptyset,a,B}}{
				\gen{b}{f}{x})
			}
		}$

	Let $\conf_a^L$ be either $\confcs{E^L[\ ];\stack_\client}{\stack_\server}$ or
	$\confcs{\stack_\client}{E^L[\ ];\stack_\server}$ where

	$E^L[\ ] = E_0[\ldo{f}{ [\ ] }{
		(\ldo{x}{ \ccomp{N}_{\emptyset,a,B} }{ \gen{b}{f}{x} }) }]$. \\

	Let $\conf_a^N$ be either $\confcs{E^N[\ ];\stack_\client}{\stack_\server}$ or
	$\confcs{\stack_\client}{E^N[\ ];\stack_\server}$ where

	$E^N[\ ] = E_0[ \ldo{x}{ [\ ] }{ \gen{b}{\clo{\emptyset}{F}}{x} } ]$.  \\

	Then, we can construct the following evaluation steps:

\begin{tabular}{l l l}
&
$\conf_a^L[\ccomp{L}_{\emptyset,a,B\funL{b}A}]$
& by (7)
\\
$ \run^* $
&
$\conf_a^L[\ccomp{\lamL{b}{x}{M_0}}_{\emptyset,a,B\funL{b}A}]$
& by (E-Do)
\\
$\run$
&
$\conf_a^N[\ccomp{N}_{\emptyset,a,B}]$
& by (8)
\\
$ \run^* $
&
$\conf_a^N[\ccomp{W}_{\emptyset,a,B}]$
& by (E-Do)
\\
$\run$
&
$\conf_a[\gen{b}{\clo{\emptyset}{F}}{ \vcomp{W}_{\emptyset,a,B}}]$
&
\\
\end{tabular}
\ \\

	{\bf i) $a=b$}: $\conf_a = \conf_b$.

\begin{tabular}{l l l}
&
$\conf_a[\gen{a}{\clo{\emptyset}{F}}{ \vcomp{W}_{\emptyset,a,B}}]$
& by (E-Gen-C-C) or (E-Gen-S-S)
\\
$\run$
&
$\conf_a[(\clo{\emptyset}{F})(\vcomp{W}_{\emptyset,a,B})]$
& by (E-App)
\\
&
& and $\funstore(F)=\{\overline{z}\}\lambda x.\ccomp{M_0}_{\{x:B\},b,A}$
\\
$\run$
&
$\conf_a[\ccomp{M_0}_{\{x:B\},b,A}\subst{\vcomp{W}_{\emptyset,a,B}}{x}]$
& by Lemma \ref{lemma:substundertermcomp}
\\
$=$
&
$\conf_a[\ccomp{M_0\subst{W}{x}}_{\emptyset,b,A}]$
& by (13)
\\
$\run^*$
&
$ \conf_a[\ccomp{V}_{\emptyset,a,A}] $
&
\end{tabular}
\ \\

	{\bf ii) $a=\client$ and $b=\server$}:

\begin{tabular}{l l l}
&
$\conf_\client[\gen{\server}{\clo{\emptyset}{F}}{ \vcomp{W}_{\emptyset,\client,B}}]$
& by (E-Gen-S-C)
\\
&
$\conf_\client[\req{\clo{\emptyset}{F}}{ \vcomp{W}_{\emptyset,\client,B}}]$
&
\\
$=$
&
$\confcs{E_0[\req{\clo{\emptyset}{F}}{ \vcomp{W}_{\emptyset,\client,B}}];\stack_\client}{\stack_\server}$
& by (E-Req)
\\
$\run$
&
$\confcs{E_0[];\stack_\client}{(\clo{\emptyset}{F})(\vcomp{W}_{\emptyset,\client,B}); \stack_\server}$
& by (E-App)
\\
&
& and $\funstore(F)=\overline{z}.\lambda x.\ccomp{M_0}_{\{x:B\},\server,A}$
\\
$\run$
&
$\confcs{E_0[];\stack_\client}{\ccomp{M_0}_{\{x:B\},\server,A}\subst{\vcomp{W}_{\emptyset,\server,B}}{x}; \stack_\server}$
& by Lemma \ref{lemma:relocationundertermcomp}
\\
$=$
&
$\confcs{E_0[];\stack_\client}{\ccomp{M_0}_{\{x:B\},\server,A}\subst{\vcomp{W}_{\emptyset,\client,B}}{x}; \stack_\server}$
& by Lemma \ref{lemma:substundertermcomp}
\\
$=$
&
$\confcs{E_0[];\stack_\client}{\ccomp{M_0\subst{W}{x}}_{\emptyset,\server,A}; \stack_\server}$
& by (13)
\\
$\run^*$
&
$\confcs{E_0[];\stack_\client}{\ccomp{V}_{\emptyset,\server,A}; \stack_\server}$
& by def. of comp.
\\
$=$
&
$\confcs{E_0[];\stack_\client}{\lunit{(\vcomp{V}_{\emptyset,\server,A})}; \stack_\server}$
& by (E-Unit-S)
\\
$\run$
&
$\confcs{E_0[\lunit{(\vcomp{V}_{\emptyset,\server,A})}];\stack_\client}{\stack_\server}$
& by Lemma \ref{lemma:relocationundertermcomp}
\\
$=$
&
$\confcs{E_0[\lunit{(\vcomp{V}_{\emptyset,\client,A})}];\stack_\client}{\stack_\server}$
& by def. of $\conf_a$ and $\ccomp{V}$
\\
$=$
&
$\conf_a[\ccomp{V}_{\emptyset,\client,A}]$
&
\end{tabular}
\ \\

	{\bf iii) $a=\server$ and $b=\client$}: This case is proved by the same way as the one for ii) except the use of $\conf_\server$, (E-Gen-C-S), (E-Call), and (E-Unit-C) instead of $\conf_\server$, (E-Gen-S-C), (E-Req), and (E-Unit-S).

\ \\
{\bf (Tapp)} $M=N[B]$. (1):$\typing{\emptyset}{a}{N[B]}{C\subst{B}{\alpha}}$ and (2):$\typing{\emptyset}{a}{N}{\forall\alpha.C}$.

	By (Tapp) in {\polyrpc}, (3):$\evalRPC{N}{a}{ \Lambda\alpha.V_0 }$.

	By I.H. with (2) and (3),
	(4):$\conf_a^N[\ccomp{N}_{\emptyset,a,\forall\alpha.C}] \run^* \conf_a^N[\ccomp{\Lambda\alpha.V_0}_{\emptyset,a,\forall\alpha.C}]$

	By the def. of the compilation,

	$\ccomp{N[B]}_{\emptyset,a,C\subst{B}{\alpha}}=
		\ldo{f}{ \ccomp{N}_{\emptyset,a,\forall\alpha.C} }
			{ f[B] }$.

	Let $\conf_a^N$ be either  either $\confcs{E^N[\ ];\stack_\client}{\stack_\server}$ or
	$\confcs{\stack_\client}{E^N[\ ];\stack_\server}$ where

	$E^L[\ ] = E_0[\ldo{f}{ [\ ] }{ f[B] }]$. \\

	Then we can construct the following evaluation steps:

\begin{tabular}{l l l}
&
$\conf_a^N[\ccomp{N}_{\emptyset,a,\forall\alpha.C}]$
& by (4)
\\
$\run^*$
&
$\conf_a^N[\ccomp{\Lambda\alpha.V_0}_{\emptyset,a,\forall\alpha.C}]$
& by def. of comp. and $\ccomp{-}$
\\
$=$
&
$\conf_a^N[\ldo{f}{ \lunit{(\vcomp{\Lambda\alpha.V_0}_{\emptyset,a,\forall\alpha.C})} }
			{ f[B] }]$
& by (E-Do)
\\
$\run$
&
$\conf_a^N[(\vcomp{\Lambda\alpha.V_0}_{\emptyset,a,\forall\alpha.C})[B]]$
& by def. of $\vcomp{-}$
\\
$=$
&
$\conf_a^N[(\Lambda\alpha.\ccomp{V_0}_{\emptyset,\alpha,a,C})[B]]$
& by (E-TApp)
\\
$\run$
&
$\conf_a^N[\ccomp{V_0}_{\emptyset,\alpha,a,C}\subst{\vcomp{B}}{\alpha}]]$
& by Lemma \ref{lemma:substundertermcomp}
\\
$=$
&
$\conf_a^N[\ccomp{V_0\subst{B}{\alpha}}_{\emptyset,a,C\subst{B}{\alpha}}]]$
\\
\end{tabular}

\ \\
{\bf (Lapp)} $M=N[Loc]$. (1):$\typing{\emptyset}{a}{N[Loc]}{C\subst{Loc}{l}}$ and (2):$\typing{\emptyset}{a}{N}{\forall l.C}$.

	By (Lapp) in {\polyrpc}, (3):$\evalRPC{N}{a}{ \Lambda l.V_0 }$.

	By I.H. with (2) and (3),
	(4):$\conf_a^N[\ccomp{N}_{\emptyset,a,\forall l.C}] \run^* \conf_a^N[\ccomp{\Lambda l.V_0}_{\emptyset,a,\forall l.C}]$

	By the def. of the compilation,
	$\ccomp{N[Loc]}_{\emptyset,a,C\subst{Loc}{l}}=
		\ldo{f}{ \ccomp{N}_{\emptyset,a,\forall l.C} }
			{ f[Loc] }$.

	Let $\conf_a^N$ be either  $\confcs{E^N[\ ];\stack_\client}{\stack_\server}$ or
	$\confcs{\stack_\client}{E^N[\ ];\stack_\server}$ where
	$E^L[\ ] = E_0[\ldo{f}{ [\ ] }{ f[Loc] }]$. \\

	Then we can construct the following evaluation steps:

\begin{tabular}{l l l}
&
$\conf_a^N[\ccomp{N}_{\emptyset,a,\forall l.C}]$
& by (4)
\\
$\run^*$
&
$\conf_a^N[\ccomp{\Lambda l.V_0}_{\emptyset,a,\forall l.C}]$
&  by def. of comp. and $\ccomp{-}$
\\
$=$
&
$\conf_a^N[\ldo{f}{ \lunit{(\vcomp{\Lambda l.V_0}_{\emptyset,a,\forall l.C})} }
			{ f[Loc] }]$
&  by (E-Do)
\\
$\run$
&
$\conf_a^N[ (\vcomp{\Lambda l.V_0}_{\emptyset,a,\forall l.C})[Loc] ]$
& by def. of $\vcomp{-}$
\\
&
& and $\funstore(F)=\Lambda l.\ccomp{V_0}_{l,a,C}$
\\
$=$
&
$\conf_a^N[(\clo{\emptyset}{F})[Loc]]$
& by (E-LApp)
\\
$\run$
&
$\conf_a^N[\ccomp{V_0}_{l,a,C}\subst{Loc}{l}]]$
& by Lemma \ref{lemma:substundertermcomp}
\\
$=$
&
$\conf_a^N[\ccomp{V_0\subst{Loc}{l}}_{\emptyset,a,C\subst{Loc}{l}}]]$
\\
\end{tabular}

\ \\
{\bf (Proj-i)} $M=\pi_i(N)$. By (T-Proj-i) with (1), (3):$\typing{\emptyset}{a}{N}{A_1\times A_2}$

	By (Proj-i) in {\polyrpc} with (2), (4):$\evalRPC{N}{a}{(V_1,V_2)}$.

	By I.H. with (3) and (4), (5):$\conf_a^N[\ccomp{N}_{\emptyset,a,A_1\times A_2}] \run^* \conf_a^N[\ccomp{(V_1,V_2)}_{\emptyset,a,A_1\times A_2}]$.

	By the def. of the compilation,
	$\ccomp{\pi_i(N)}_{\emptyset,a,A_i}=
		\ldo{p}{\ccomp{N}_{\emptyset,a,A_1\times A_2}}{
			\llet{x}{\pi_i(p)}{ \lunit{x} }
		}$

	Let $\conf_a^N$ be either $\confcs{E^N[\ ];\stack_\client}{\stack_\server}$ or $\confcs{\stack_\client}{E^N[\ ];\stack_\server}$ where
	$E^N[\ ] = E_0[ \ldo{p}{[\ ]}{
			\llet{x}{\pi_i(p)}{ \lunit{x} }
		} ]$.

	Let $\conf_a'$ be either $\confcs{E'[\ ];\stack_\client}{\stack_\server}$ or $\confcs{\stack_\client}{E'[\ ];\stack_\server}$ where
	$E'[\ ] = E_0[ \llet{x}{[\ ]}{ \lunit{x} } ]$.

	Then we can construct the following evaluation steps:

\begin{tabular}{l l l}
&
$\conf_a^N[\ccomp{N}_{\emptyset,a,A_1\times A_2}]$
& by (5)
\\
$\run^*$
&
$\conf_a^N[\ccomp{(V_1,V_2)}_{\emptyset,a,A_1\times A_2}]$
& by def. of $\ccomp{-}$
\\
$=$
&
$\conf_a^N[\lunit{(\vcomp{(V_1,V_2)}_{\emptyset,a,A_1\times A_2})}]$
& by (E-Do) and def. of $\conf_a'$
\\
$\run$
&
$\conf_a'[\pi_i(\vcomp{(V_1,V_2)}_{\emptyset,a,A_1\times A_2})]$
& by def. of $\vcomp{-}$
\\
$=$
&
$\conf_a'[\pi_i(\vcomp{V_1}_{\emptyset,a,A_1},\vcomp{V_2}_{\emptyset,a,A_2})]$
& by (E-Proj-i)
\\
$=$
&
$\conf_a[\llet{x}{\vcomp{V_i}_{\emptyset,a,A_i}}{ \lunit{x} }]$
& by (E-Let)
\\
$=$
&
$\conf_a[ \lunit{(\vcomp{V_i}_{\emptyset,a,A_i})} ]$
& by def. of $\ccomp{-}$
\\
$=$
&
$\conf_a[ \ccomp{V_i}_{\emptyset,a,A_i} ]$
&
\\
\end{tabular}

\end{proof}

\section{An Optimization of Locative Application Terms in the Polymorphic CS calculus}

	Definition \ref{opt:transform} shows a transformation $\optrpc{-}$ of {\polycs} by specializing generic application terms to local or remote applications whenever possible.
	We present it as a program transformation for {\polycs} rather than as a slicing compilation integrated with such an optimization.

\begin{definition}[Optimization] \ \\
\begin{tabular}{l c l}
$\optrpc{\llet{x}{M}{N}}_{\tyenv,Loc,B}$ & $=$ &
	$\llet{x}{\optrpc{M}_{\tyenv,Loc,A}}{\optrpc{N}_{\tyenv,x:A,Loc,B}}$
\\
$\optrpc{\pi_i(V)}_{\tyenv,Loc,A_i}$ & $=$ &
	$\pi_i( \optrpc{V}_{\tyenv,Loc,A_1\times A_2} )$
\\
$\optrpc{V(W)}_{\tyenv,Loc,B}$ & $=$ &
	$\optrpc{V}_{\tyenv,Loc,\cloty{A\funL{Loc}B}} (\optrpc{W}_{\tyenv,Loc,A})$
\\
$\optrpc{V[B]}_{\tyenv,Loc,A\subst{B}{\alpha}}$ & $=$ &
	$\optrpc{V}_{\tyenv,Loc,\cloty{\forall\alpha.A}}[B]$
\\
$\optrpc{V[Loc']}_{\tyenv,Loc,A\subst{Loc'}{l}}$ & $=$ &
	$\optrpc{V}_{\tyenv,Loc,\cloty{\forall l.A}}[Loc']$
\\
$\optrpc{x}_{\tyenv,Loc,A}$ & $=$ & $x$
\\
$\optrpc{(V,W)}_{\tyenv,Loc,A_1\times A_2}$ & $=$ &
	$(\optrpc{V}_{\tyenv,Loc,A_1},\optrpc{W}_{\tyenv,Loc,A_2})$
\\
$\optrpc{\clo{\overline{W_i}}{F}}_{\tyenv,Loc',\cloty{B}}$ & $=$ &
	$\clo{\overline{\optrpc{W_i}_{\tyenv,Loc,B_i}}}{F}$
\\
$\optrpc{\Lambda\alpha.V}_{\tyenv,\Loc',\forall\alpha.A}$ & $=$ &
        $\Lambda\alpha.\optrpc{V}_{\tyenv,\Loc',A}$
\\
$\optrpc{\lunit{V}}_{\tyenv,Loc,T A}$ & $=$ &
	$\lunit{(\optrpc{V}_{\tyenv,Loc,A})}$
\\
$\optrpc{\ldo{x}{M}{N}}_{\tyenv,Loc,T B}$ & $=$ &
	$\ldo{x}{\optrpc{M}_{\tyenv,Loc,TA}}{\optrpc{N}_{\tyenv,x:A,Loc,TB}}$
\\
$\optrpc{\req{V,W}}_{\tyenv,\client,TB}$ & $=$ &
	$\req{\optrpc{V}_{\tyenv,\client,\cloty{A\funL{\server}T B}}}
			{\optrpc{W}_{\tyenv,\client,A}}$
\\
$\optrpc{\call{V,W}}_{\tyenv,\server,TB}$ & $=$ &
	$\call{\optrpc{V}_{\tyenv,\server,\cloty{A\funL{\client}T B}}}
			{\optrpc{W}_{\tyenv,\server,A}}$
\\
$\optrpc{\gen{Loc}{V}{W}}_{\tyenv,Loc,A}$ &  $=$ &
	$\optrpc{V}_{\tyenv,Loc,\cloty{A\funL{Loc}T B}}(\optrpc{W}_{\tyenv,Loc,A})$
\\
$\optrpc{\gen{\server}{V}{W}}_{\tyenv,\client,A}$ &  $=$ &
	$\req{\optrpc{V}_{\tyenv,\client,\cloty{A\funL{\server}T B}}}
			{\optrpc{W}_{\tyenv,\client,A}}$
\\
$\optrpc{\gen{\client}{V}{W}}_{\tyenv,\server,A}$ &  $=$ &
	$\call{\optrpc{V}_{\tyenv,\server,\cloty{A\funL{\client}T B}}}
			{\optrpc{W}_{\tyenv,\server,A}}$
\\
$\optrpc{\gen{Loc'}{V}{W}}_{\tyenv,Loc,A}$ &  $=$ &
	$\gen{Loc'}{\optrpc{V}_{\tyenv,Loc,\cloty{A\funL{Loc'}T B}}}
						{\optrpc{W}_{\tyenv,Loc,A}}$
\\
& & if $Loc\not=Loc'$ and also $Loc$ or $Loc'$ is a variable
\end{tabular}
\label{opt:transform}
\end{definition}

	In the following fact, we argue that the optimization program transformation is correct in terms of the semantics for {\polycs}.

\begin{fact}[The correctness of the optimization]
If $\confcs{M}{\emptystack} \run^* \confcs{V}{\emptystack}$ then
$\confcs{\optrpc{M}}{\emptystack} \run^* \confcs{\optrpc{V}}{\emptystack}$
\label{thm:correctoptimization}
\end{fact}
\begin{proof} We first generalize this theorem as: if $Conf \run Conf'$ then $\optrpc{Conf} \run^* \optrpc{Conf'}$.

	$\optrpc{Conf}$ can be defined by $\optrpc{M}$ and $\optrpc{\stack}$.
	Then for each evaluation step $Conf \run Conf'$ not using (E-Gen-X-X) rules, we will exactly have one step as $\optrpc{Conf} \run \optrpc{Conf'}$.
	When $Conf \run Conf'$  uses one of the (E-Gen-X-X) rules, we will have either zero step or one step described as $\optrpc{Conf} \run^n \optrpc{Conf'}$ where $n=0,1$.
	Whenever one of the first three specialization transformation rules is applied to $Conf$, $\optrpc{Conf'}$ becomes identical to $\optrpc{Conf}$
	and so it will be zero step. Otherwise, it will be one step.
\end{proof}

\section{The Untyped CS Calculus}

	This section discusses how to implement client and server sliced programs in the polymorphic CS calculus. There are two things to motivate this section. Firstly, the {\polycs} client and server programs use types that were necessary for the slicing compilation but are not for the runtime execution. In the implementation, we want to erase the types but should retain the locations necessary for supporting the dynamic location polymorphism. %This will clearly show why we call our idea a dynamic approach to the polymorphic RPC calculus.

	Secondly, the notion of monads in {\polycs} was useful for the design of typing remote procedure calls to make it as simple as typing local procedure calls by abstracting some details of the trampoline interaction between the client and the server that should, however, be explicitly implemented using {\it send} and {\it receive} communication operations.

	For an implementation of the polymorphic CS calculus, we introduce an untyped language named {\cs} to be used as a target language for a type erasure translation of {\polycs} retaining locations by value representation and exposing the concrete trampoline communication. After presenting the type erasure translation, we will show that execution in the untyped CS calculus mirrors execution in the polymorphic CS calculus.

\subsection{Untyped CS calculus}
\label{sec:untypedcscalculus}

	Figure \ref{fig:syntaxuntypedcs} and Figure \ref{fig:semanticsuntypedcs} show the syntax and semantics for the untyped CS calculus, which can be viewed as a conventional first-order functional programming language equipped with network libraries.
	For syntax, terms denoted by $m$ include communication primitives, {\it send} and {\it receive}. Case terms are also included to deconstruct data constructor values.
	Values denoted by $v$ or $w$ contain a new form of $Con \ \overline{v}$ where $Con$ is a data constructor and $\overline{v}$ are its arguments.
	Here are examples:
	\begin{itemize}
	\item For locations, $Client$ and $Server$ in {\cs} are data constructor values to represent location constants $\client$ and $\server$ in {\polycs}, respectively.
	\item A form of $Closure \ \overline{v} \ F_{name}$ is introduced to {\cs} to implement $\clo{\overline{W}}{F_{name}[\overline{\Loc}\,\overline{\alpha}]}$ in {\polycs} under the assumption that $\overline{v}$ % faithfully
	implements $\overline{\Loc}$ together with $\overline{W}$ and $F_{name}$ is $Con_{F_{name}}$.
	\item In the remote procedure calls and returns, the payloads to send and receive are represented by $Apply \ v \ w$ for $\req{V}{W}$ and $\call{V}{W}$ and by $Ret \ v$ for $\lunit{V}$.
	\end{itemize}

	Function stores now deal with codes $\overline{z}\lambda x.m$ with no free type and free location variables where the part of the free term variables $\overline{z}$  replace the free location variables.

	For semantics, configurations are in the form of $\confcs{m_\client\ }{\ m_\server}$ with a single term at each location. Execution involves local reduction rules and communications rules.
	Evaluation contexts $e[\ ]$ to choose a specific rule to execute are actually the same as the previous ones but configuration contexts $\sigma[\ ]$ are different.  For example, client-side configuration contexts are in the form of $\confcs{e_1[\ ]}{e_2[\ldo{x}{receive}{m}]}$ meaning that the client is running a term inside the evaluation context and the server is waiting to receive a payload $x$ from the client.
	When the term in the client is in the form of $\ldovoid{send \ v}{ m}$, an abbreviation of $\ldo{x}{send \ v}{m}$ where $x$ is unused, ready to send a payload $v$, the communication rule (e-comm-c-s) sends the payload from the client to the server. For the opposite direction, server-side configuration contexts and (e-comm-s-c) will do that.

	In the local reduction rules, there is no type application rule like (E-TApp). The location application rule (E-LApp) is now supported by (e-app). A case reduction rule (e-case) is introduced.

\subsection{A Compilation and Its Semantic Correctness}
\label{sec:correctness}

	Figure \ref{fig:compilationtountypedcs} shows compilation rules for locations, terms, values and function stores.
	Firstly, we review how to erase types and to compile locations. Every location variable $l$ is replaced by a term variable $x_l$ while the two location constants, $\client$ and $\server$, are compiled into $Client$ and $Server$, respectively.

	Compiling a code, $\overline{l}\overline{\alpha}\overline{z}.OpenCode$, erases the free type variables $\overline{\alpha}$, and changes the free location variables $\overline{l}$ into term variables. The compiled code will have $\overline{z_l}\cdot \overline{z}$ as free variables.
	Symmetrically, compiling a closure, $\clo{\overline{W}}{F_{name}[\overline{Loc}][\overline{A}]}$, erases the free types $\overline{A}$, and lets the compiled closure hold $\overline{\ecomp{\Loc}} \cdot \overline{\ecomp{W}}_a$ as free values that come from turning the free locations $\overline{\Loc}$  and from the existing free values $\overline{W}$.

	Compiling type abstraction and application is subtle. For example,
	\[
	\ecomp{ (\Lambda\alpha.\lunit V)[A]}_a = \lunit (\ecomp{\Lambda\alpha.\lunit V}_a) = \lunit  \ecomp{V}_a.
	\]
	To have this simple type erasure compilation, we limit the polymorphism by type abstractions to the form of values as $\lunit{V}$, which is a computational value but has no {\it effects}. This form is analogous to {\it syntactic values} that permit the simple ML polymorphism~\cite{wright95}.
	One reason is that a type erasure compilation without having such a limitation is possible but must be complex. Because $\Lambda\alpha.V$ is a plain value but $V$ is a computational value, $\ecomp{V}_a$ cannot be in the same context as where $\ecomp{\Lambda\alpha.V}_a$ is.
	The other reason comes from the fact that there is no gain with having an arbitrary (computational) value in the body of a type abstraction. Our slicing compilation always generates  type abstractions in the limited form.
	A final remark on this topic is that our theory of the polymorphic CS calculus and a slicing compilation works with or without the limitation.

	Term applications $V(W)$ are compiled as a case term that extracts free values $\overline{w}$ and a function name from $\ecomp{V}_a$ to apply the function to a location argument value from $\ecomp{W}$ after substituting the free values for the free variables $\overline{z}$ in the function body $m$.
	Location applications $V[\Loc]$ are compiled essentially in the same way but with the value representation $\ecomp{\Loc}$ as an argument.

	Secondly, we discuss how to support the trampoline communication between the client and the server. A key pattern is $\ldovoid{send \ v}{loop \ ()}$ where $loop$ is a function waiting for receiving either $Apply \ f \ arg$ to call $f(arg)$ locally and to return its result back to the other  location, or $Ret \ y$ to finish the trampoline communication.
	Both of $\req{V}{W}$ and $\call{V}{W}$ are compiled into a term in this pattern but at one's own location enforced by the {\polycs} type system.
	For $\gen{\Loc}{V}{W}$, the compiled term has a case analysis on a value from the compiled location $\ecomp{\Loc}$ to determine whether $V$ is a remote procedure with an argument $W$.

	Note that our formulation treats this function specially, always placing $loop \ ()$ to follow $send \ v$ immediately. So, $loop \ ()$ can be unfolded  to the function body, $\ldo{x}{receive}{\cdots}$, immediately after sending a value.

	Now we can prove the semantic correctness of the compilation of {\polycs} into {\cs} by proving a generalized lemma (Lemma \ref{theorem:generalizedsemanticcorrectnessoftypeandlocationerasure}) in the next section.

\setcounter{definition}{0}
\renewcommand{\thedefinition}{5.\arabic{definition}}

\setcounter{lemma}{0}
\renewcommand{\thelemma}{5.\arabic{lemma}}

\setcounter{theorem}{0}
\renewcommand{\thetheorem}{5.\arabic{theorem}}

\begin{theorem}[Semantic Correctness of Compilation of {\polycs} into {\cs}]
%\label{theorem:semanticcorrectnessoftypeandlocationerasuretotheend}
If $\confcs{M}{\emptystack} \ \run^* \ \confcs{\lunit V}{\emptystack}$ in {\polycs}
then $\confcs{ \ecomp{M}_\client \ }{ \  loop \ ()} \ \run^* \ \confcs{\ecomp{\lunit V}_\client \ }{\ loop \ ()}$  in {\cs}.
\end{theorem}

\subsection{Semantic correctness of the compilation of the polymorphic CS calculus into the untyped CS calculus}

The definition of $m\subst{v}{x}$ in the untyped CS calculus replacing all occurrences of $x$ in $m$ by $v$ is defined as follows.
	\begin{eqnarray*}
	(y)\subst{v}{x}  & = & \left\{\mbox{\begin{tabular}{l l}
													$v$ &  if  $x=y$ \\
	                                                         				$y$ & otherwise
												\end{tabular}
											} \right.	\\
       (w_1,w_2)\subst{v}{x} &=& (w_1\subst{v}{x} , w_2\subst{v}{x} ) \\
       (Con \ \overline{w}) \subst{v}{x} &=& Con \ \overline{ w\subst{v}{x} } \\
       (\lunit w)\subst{v}{x}  &=& \lunit (w\subst{v}{x} ) \\
       (\ldo{y}{m}{n})\subst{W}{x}  &=&  \left\{\mbox{\begin{tabular}{l l}
													$\ldo{y}{m\subst{v}{x} }{n }$ &  if  $x=y$ \\
	                                                         				$\ldo{y}{m\subst{v}{x} }{n\subst{v}{x} }$ & otherwise
												\end{tabular}
											} \right.	\\
       (\llet{y}{m}{n})\subst{v}{x}  &=&  \left\{\mbox{\begin{tabular}{l l}
													$\llet{y}{m\subst{v}{x} }{ n }$ &  if  $x=y$ \\
	                                                         				$\llet{y}{m\subst{v}{x} }{n\subst{v}{x} }$ & otherwise
												\end{tabular}
											} \right.	\\
       (\pi(w))\subst{v}{x}   &=& \pi(w\subst{v}{x}  ) \\
       (w_f(w_{arg}))\subst{v}{x}   &=& (w_f\subst{v}{x}  (w_{arg}\subst{v}{x}  )) \\
       (p(\overline{w}))\subst{v}{x}  &=& p(\overline{w\subst{W}{x}} ) \\
       (\case{w}{\overline{c\overline{y}\rightarrow m}})\subst{v}{x}   &=&
             \left\{\mbox{\begin{tabular}{l l}
             		$\case{w\subst{v}{x}}{\overline{c \ \overline{y}\rightarrow m}}$ &  if  $\exists y_i.x=y_i$ \\
	              $\case{w\subst{v}{x}}{\overline{c \ \overline{y}\rightarrow m\subst{v}{x} }}$ & otherwise
             	\end{tabular}
	} \right.
	\end{eqnarray*}

\begin{figure}[h]
\begin{tabular}{l l}
\multicolumn{2}{l}{\textbf{Compiling evaluation contexts}} \\[0.1cm]
&
$\ecomp{\ [\ ] \ }_a \ = \ [\ ]$
\\
&
$\ecomp{\  \llet{x}{E_{let}[\ ]}{M} \ }_a \ = \ \llet{x}{ \ecomp{ E_{let}[\ ]}_a }{ \ecomp{ M }_a }$
\\
&
$\ecomp{\  \ldo{x}{E[\ ]}{M} \ }_a \ = \ \ldo{x}{ \ecomp{ E[\ ]}_a }{ \ecomp{ M }_a }$
\\[0.3cm]
\multicolumn{2}{l}{\textbf{Compiling stacks}} \\[0.1cm]
&
  \mbox{
	\begin{prooftree}
		\hypo{  }
		\infer[left label=(C-Stk-Empty-Client)]1{ \ecomp{ \emptystack \ | \ \emptystack }_\client = [\ ] \ | \ loop_{server} \ () }
	\end{prooftree}
  }
\\[0.5cm]
&
  \mbox{
	\begin{prooftree}
		\hypo{ }
		\infer[left label=(C-Stk-Empty-Server)]1{ \ecomp{ \emptystack \ | \ \emptystack }_\server = loop \ ()\ | \  \ldokeyword \ z \leftarrow [\ ];  \ send \ (Ret \ z); \ loop_{server} \ ()}
	\end{prooftree}
  }
\\[0.5cm]
&
  \mbox{
	\begin{prooftree}
		\hypo{  \ecomp{ \ \stackcs \ }_\server = m \ | \ e[\ ] }
		\infer[left label=(C-Stk-Client)]1{ \ \ecomp{ \stackcsWith{\stack_\client }{ E[\ ];\stack_\server} \  }_ \client  =  \ldokeyword \ z \leftarrow [\ ]; \ send \ (Ret \ z); \ m  \ | \ e[ \ \ecomp{ E }_\server[ loop \ () ] \ ]}
	\end{prooftree}
  }
\\[0.5cm]
&
  \mbox{
	\begin{prooftree}
		\hypo{   \ecomp{ \ \stackcs \ }_\client = e[\ ] \ | \ m }
		\infer[left label=(C-Stk-Server)]1{ \ecomp{ \stackcsWith{E[\ ];\stack_\client }{ \stack_\server} }_\server =  e[ \ \ecomp{ E }_\client[ loop \ () ] \ ] \ | \ \ldokeyword \ z \leftarrow [\ ]; \ send \ (Ret \ z); \ m }
	\end{prooftree}
  }
\\[0.5cm]
\multicolumn{2}{l}{\textbf{Compiling configurations}} \\[0.1cm]
&
  \mbox{
	\begin{prooftree}
		\hypo{ \ecomp{  \stackcs }_\client = e[\ ] \ | \ m }
		\infer[left label=(C-Client)]1{ \ecomp{ \confcs{M;\stack_\client}{\stack_\server} } = \confcs{\ e[\ecomp{M}_\client] \ }{ \ m \ } }
	\end{prooftree}
  }
\\[0.5cm]
&
  \mbox{
	\begin{prooftree}
		\hypo{ \ecomp{  \stackcs }_\server = m \ | \ e[\ ] }
		\infer[left label=(C-Server)]1{ \ecomp{ \confcs{\stack_\client}{M;\stack_\server} } = \confcs{\  m \ } {\ e[\ecomp{M}_\server] \ }  }
	\end{prooftree}
  }
\end{tabular}
\caption{Compilation of Stacks and configurations}
\Description[Compilation of Stacks and configurations]{Compilation of Stacks and configurations}
\label{fig:configstackcompilation}
\end{figure}

\begin{lemma}[Evaluation Contexts Under Type and Location Erasure Compilation]
\label{lemma:evaluationcontextsundertypeandlocationerasure}
The structure of evaluation contexts is preserved under the compilation of {\polycs} into {\cs}.
\begin{itemize}
\item $\ecomp{ E[M] }_a \ = \ \ecomp{E}_a[ \ \ecomp{M}_a \ ]$
\end{itemize}
\end{lemma}
\begin{proof} This lemma is proved by induction on the structure of evaluation contexts $E[\ ]$.
	For the base case $[\ ]$, the lemma is true because $\ecomp{[\ ]}_a=[\ ]$.

	For the inductive cases, the lemma is provable as this. When $E[\ ]=\llet{x}{E_0[\ ]}{N}$,

\begin{tabular}{l l l}
&
$\ecomp{ E[M]}_a$
&
\\
$=$
&
$\ecomp{ \llet{x}{E_0[ M]}{N} }_a$
&
by the term compilation
\\
$=$
&
$\llet{x}{\ecomp{E_0[M]}_a}{\ecomp{N}_a}$
& by induction, $\ecomp{E_0[M]}_a=\ecomp{E_0}_a[ \ecomp{M}_a ]$
\\
$=$
&
$\llet{x}{\ecomp{E_0}_a[ \ecomp{M}_a ]}{\ecomp{N}_a}$
& by the evaluation context compilation
\\
$=$
&
$\ecomp{E}_a[ \ \ecomp{M}_a \ ]$
&
\end{tabular}

When $E[\ ]=\ldo{x}{E_0[\ ]}{N}$, the lemma is provable by the same procedure as this.
\end{proof}

\begin{lemma}[Location Invariant Under Type and Location Erasure Compilation]
\label{lemma:locationinvariantundertypeandlocationerasure}
Suppose $V$ is a plain value. Then we have  $\ecomp{V}_\client \ = \ \ecomp{V}_\server$.
\end{lemma}
\begin{proof} This lemma is proved by induction on the structure of plain values. Note that the plain values are those values whose types $A$ are relocatable by $\valty{A}$. For the base case $x$, the lemma is true by the definition of the value compilation as $\ecomp{x}_a=x$.

	For the inductive cases, the lemma is provable by induction. For example, $\ecomp{(V,W)}_\client$ = $\ecomp{(V,W)}_\server$ if $\ecomp{V}_\client=\ecomp{V}_\server$ and $\ecomp{W}_\client=\ecomp{W}_\server$. This condition is true by induction.
	The same way of proving the lemma as this can be applied to the other inductive cases: $\clo{\overline{W}}{F_{name}[\overline{Loc}][\overline{A}]}$ and $\Lambda\alpha.\lunit{V}$ where all subvalues are guaranteed to be plain by the type system for {\polycs}.
\end{proof}

\begin{lemma}[Value Substitution Under Type and Location Erasure Compilation]
\label{lemma:valuesubstundertypeandlocationerasure}
Value substitutions are preserved under the compilation of {\polycs} into {\cs}.
\begin{itemize}
\item $\ecomp{M\subst{V}{x}}_a \ = \ \ecomp{M}_a\subst{ \ecomp{V}_a }{x}$
\item $\ecomp{W\subst{V}{x}}_a \ = \ \ecomp{W}_a\subst{ \ecomp{V}_a }{x}$
\end{itemize}
\end{lemma}
\begin{proof} This lemma is proved by induction on the structure of terms and values.
\end{proof}

\begin{lemma}[Type Substitution Under Type and Location Erasure Compilation]
\label{lemma:typesubstundertypeandlocationerasure}
Type substitutions are preserved under the compilation of {\polycs} into {\cs}.
\begin{itemize}
\item $\ecomp{M\subst{A}{\alpha}}_a \ = \ \ecomp{M}_a$
\item $\ecomp{V\subst{A}{\alpha}}_a \ = \ \ecomp{V}_a$
\end{itemize}
\end{lemma}
\begin{proof} This lemma is proved by induction on the structure of terms and values.
\end{proof}

\begin{lemma}[Location Substitution Under Type and Location Erasure Compilation]
\label{lemma:locationsubstundertypeandlocationerasure}
Location substitutions are preserved under the compilation of {\polycs} into {\cs}.
\begin{itemize}
\item $\ecomp{M\subst{\Loc}{l}}_a \ = \ \ecomp{M}_a\subst{ \ecomp{\Loc}_a }{x_l}$
\item $\ecomp{V\subst{\Loc}{l}}_a \ = \ \ecomp{V}_a\subst{ \ecomp{\Loc}_a }{x_l}$
\end{itemize}
\end{lemma}
\begin{proof} This lemma is proved byinduction on the structure of terms and values.
\end{proof}

\begin{definition}[Structural Equivalence]
\label{def:structuralequivalence}
	Suppose $conf_1$ has one of three terms in the second column as a subterm, and $conf_2$ is a configuration obtained by replacing it by its corresponding term in the third column, or vice versa.
	Then $conf_1$ is said to be structurally equivalent to $conf_2$.
	For notation, we write $conf_1 \ \structeqv \ conf_2$ for it. \\

\begin{tabular} { | l | l | l | } \hline
left identity   & $\ldo{x}{\lunit v}{m}$ & $m\subst{v}{x}$ \\\hline
right identity & $\ldo{x}{m}{\lunit x}$ & $m$ \\\hline
associativity  & $\ldo{y}{ ( \ldo{x}{m_1}{m_2} ) }{m_3}$  & $\ldo{x}{m_1}{ ( \ldo{y}{m_2}{m_3} ) }$ \\\hline
\end{tabular}

\end{definition}

\begin{lemma}[Monad Law]
\label{lemma:associativityofbindings} In {\cs}, unit is a left identity for bind, it is also a right identity for bind, and binds are associative.
\end{lemma}
\begin{proof} This lemma is proved by the following arguments. The left identity is simply supported by the semantic rule (e-do).

	For the right identity, we argue as follows. Suppose $conf_1 \structeqv conf_2$ by the right identity where $conf_2$ is obtained by replacing a subterm $\ldo{x}{m}{\lunit x}$ in $conf_1$.
	Either the $m$ is $\lunit v$ or there exists $conf_1 \run conf_1'$ by the semantic rules for {\cs} such that the $m$ becomes $m'$.
	In the first case, (e-do) lets us have $conf_1 \run \lunit v$, which is $conf_2$.
	In the second case, we also have $conf_2 \run conf_2'$ where $conf_2'$ is the configuration where $m$ is replaced by $m'$, which is $conf_2'$

	For the associativity, we prove the lemma by the similar argument as follows. Either the $m_1$ is $\lunit v$ or there exists $conf_1 \run conf_1'$ by the semantic rules for {\cs} such that the $m_1$ becomes $m_1'$.
	In the first case, the semantic rule (e-do) lets us have $conf_1 = \ldo{y}{ ( \ldo{x}{\lunit v}{m_2} ) }{m_3} \run \ldo{y}{m_2\subst{v}{x}}{m_3}$, which is the configuration that we can get by (e-do) as $\ldo{x}{\lunit v}{ ( \ldo{y}{m_2}{m_3} ) } \run \ldo{y}{m_2\subst{v}{x}}{m_3}$.
	In the second case, we also have $conf_2 \run \ldo{x}{m_1'}{ ( \ldo{y}{m_2}{m_3} ) }$ by the same semantic rule used for $conf_1 \run conf_1'$.
\end{proof}

\begin{lemma}[Generalized Semantic Correctness of Compilation of {\polycs} into {\cs}]
\label{theorem:generalizedsemanticcorrectnessoftypeandlocationerasure} Let us define $\Longrightarrow$ is either $\run$ or $\structeqv$.
If $Conf_1 \ \run \ Conf_2$ then $\ecomp{Conf_1} \ \runequiv* \ \ecomp{Conf_2}$.
\end{lemma}
\begin{proof} This theorem is proved by case analysis on the use of the semantics rules for {\polycs}.

For the cases (E-Let), (E-Do), (E-Proj-i), (E-TApp), (E-App), and (E-LApp), we will prove the theorem by assuming $\conf$ that is either $\confcs{E[\ ];\stack_\client}{\stack_\server}$ or $\confcs{\stack_\client}{E[\ ];\stack_\server}$.
	We have $\confuntyped_0$ as either $\ecomp{\stack_\client \ | \ \stack_\server}_\client = e[ \ ] \ | \ m$ or $\ecomp{\stack_\client \ | \ \stack_\server}_\server = m \ | \ e[ \ ]$.

\ \\
{\bf (E-Let)}
	$Conf_1 = \conf[\ \llet{x}{V}{M} \ ]$ and $Conf_2 = \conf[\ M\subst{V}{x} \ ]$.

\begin{tabular}{l l l}
&
$\ecomp{Conf_1}$
&
\\
&
$\ecomp{\conf[\ \llet{x}{V}{M} \ ]}$
& by (C-Client) or (C-Server)
\\
$=$
&
$\confuntyped_0[ \ \ecomp{ E[\llet{x}{V}{M} ] }_a  \ ]$
& by Lemma \ref{lemma:evaluationcontextsundertypeandlocationerasure}
\\
$=$
&
$\confuntyped_0[ \ \ecomp{ E }_a [ \ecomp{ \llet{x}{V}{M} }_a ] \  ]$
& by replacing $[\ ]$ in $\confuntyped_0$ with $\ecomp{E}_a$
\\
$=$
&
$\confuntyped[\  \ecomp{ \llet{x}{V}{M} }_a  \  ] $
& by the term compilation
\\
$=$
&
$\confuntyped[\  \llet{x}{ \ecomp{V}_a }{ \ecomp{M}_a }  \  ] $
& by (e-let)
\\
$\run$
&
$\confuntyped[\  \ecomp{M}_a  \subst{  \ecomp{V}_a }{x} \  ] $
& by replacing $\ecomp{E}_a$ in $\confuntyped$ with $[\ ]$
\\
$=$
&
$\confuntyped_0[\ \ecomp{ E }_a [\  \ecomp{M}_a  \subst{  \ecomp{V}_a }{x} \ ] \  ] $
& by Lemma \ref{lemma:valuesubstundertypeandlocationerasure}
\\
$=$
&
$\confuntyped_0[\ \ecomp{ E }_a [  \ecomp{ M\subst{V}{x} }_a  ] \  ] $
& by Lemma \ref{lemma:evaluationcontextsundertypeandlocationerasure}
\\
$=$
&
$\confuntyped_0[ \ \ecomp{ E [  M\subst{V}{x}  ] \  ] }_a \ ]$
& by (C-Client) or (C-Server)
\\
$=$
&
$\ecomp{\conf[\ M\subst{V}{x} \ ]}$
&
\\
$=$
&
$\ecomp{Conf_2}$
&
\end{tabular}

\ \\
{\bf (E-Do)}
	$Conf_1 = \conf[\ \ldo{x}{\lunit V}{M} \ ]$ and $Conf_2 = \conf[\ M\subst{V}{x} \ ]$.

\begin{tabular}{l l l}
&
$\ecomp{Conf_1}$
&
\\
&
$\ecomp{\conf[\ \ldo{x}{\lunit V}{M} \ ]}$
& by (C-Client) or (C-Server)
\\
$=$
&
$\confuntyped_0[ \ \ecomp{ E[ \ldo{x}{\lunit V}{M} ] }_a  \ ]$
& by Lemma \ref{lemma:evaluationcontextsundertypeandlocationerasure}
\\
$=$
&
$\confuntyped_0[ \ \ecomp{ E }_a [ \ecomp{ \ldo{x}{\lunit V}{M} }_a ] \  ]$
& by replacing $[\ ]$ in $\confuntyped_0$ with $\ecomp{E}_a$
\\
$=$
&
$\confuntyped[\  \ecomp{ \ldo{x}{\lunit V}{M} }_a  \  ] $
& by the term compilation
\\
$=$
&
$\confuntyped[\  \ldo{x}{ \ecomp{\lunit V}_a }{ \ecomp{M}_a }  \  ] $
& by the value compilation
\\
$=$
&
$\confuntyped[\  \ldo{x}{ \lunit \ecomp{V}_a }{ \ecomp{M}_a }  \  ] $
& by (e-do)
\\
$\run$
&
$\confuntyped[\  \ecomp{M}_a  \subst{  \ecomp{V}_a }{x} \  ] $
& by replacing $\ecomp{E}_a$ in $\confuntyped$ with $[\ ]$
\\
$=$
&
$\confuntyped_0[\ \ecomp{ E }_a [\  \ecomp{M}_a  \subst{  \ecomp{V}_a }{x} \ ] \  ] $
& by Lemma \ref{lemma:valuesubstundertypeandlocationerasure}
\\
$=$
&
$\confuntyped_0[\ \ecomp{ E }_a [  \ecomp{ M\subst{V}{x} }_a  ] \  ] $
& by Lemma \ref{lemma:evaluationcontextsundertypeandlocationerasure}
\\
$=$
&
$\confuntyped_0[ \ \ecomp{ E [  M\subst{V}{x}  ] \  ] }_a \ ]$
& by (C-Client) or (C-Server)
\\
$=$
&
$\ecomp{\conf[\ M\subst{V}{x}}$
&
\\
$=$
&
$\ecomp{Conf_2}$
&
\end{tabular}

\ \\
{\bf (E-Proj-i)}
	$Conf_1 = \conf[\ \pi_i(V_1,V_2) \ ]$ and $Conf_2 = \conf[\ V_i \ ]$.

\begin{tabular}{l l l}
&
$\ecomp{Conf_1}$
&
\\
&
$\ecomp{ \conf[\ \pi_i(V_1,V_2) \ ]}$
& by (C-Client) or (C-Server)
\\
$=$
&
$\confuntyped_0[ \ \ecomp{ E[ \pi_i(V_1,V_2) ] }_a  \ ]$
& by Lemma \ref{lemma:evaluationcontextsundertypeandlocationerasure}
\\
$=$
&
$\confuntyped_0[ \ \ecomp{ E }_a [ \ecomp{ \pi_i(V_1,V_2) }_a ] \  ]$
& by replacing $[\ ]$ in $\confuntyped_0$ with $\ecomp{E}_a$
\\
$=$
&
$\confuntyped[\  \ecomp{ \pi_i(V_1,V_2) }_a  \  ] $
& by the term compilation
\\
$=$
&
$\confuntyped[\  \pi_i \ecomp{ (V_1, V_2) }_a  \  ] $
& by the value compilation
\\
$=$
&
$\confuntyped[\  \pi_i( \ecomp{ V_1}_a,\ecomp{ V_2 }_a)  \  ] $
& by (e-proj-i)
\\
$\run$
&
$\confuntyped[\  \ecomp{ V_i}_a \  ] $
& by replacing $\ecomp{E}_a$ in $\confuntyped$ with $[\ ]$
\\
$=$
&
$\confuntyped_0[\ \ecomp{ E }_a [\  \ecomp{ V_i}_a \ ] \  ] $
& by Lemma \ref{lemma:evaluationcontextsundertypeandlocationerasure}
\\
$=$
&
$\confuntyped_0[ \ \ecomp{ E [  V_i  ] \  ] }_a \ ]$
& by (C-Client) or (C-Server)
\\
$=$
&
$\ecomp{\conf[\ V_i \ ]}$
&
\\
$=$
&
$\ecomp{Conf_2}$
&
\end{tabular}

\ \\
{\bf (E-TApp)}
	$Conf_1 = \conf[\ (\Lambda \alpha.\lunit V)[A] \ ]$ and $Conf_2 = \conf[\ \lunit (V\subst{A}{\alpha}) \ ]$.

\begin{tabular}{l l l}
&
$\ecomp{Conf_1}$
&
\\
&
$\ecomp{\conf[\ (\Lambda \alpha.\lunit V)[A] \ ]}$
& by (C-Client) or (C-Server)
\\
$=$
&
$\confuntyped_0[ \ \ecomp{ E[(\Lambda \alpha.\lunit V)[A] ] }_a  \ ]$
& by Lemma \ref{lemma:evaluationcontextsundertypeandlocationerasure}
\\
$=$
&
$\confuntyped_0[ \ \ecomp{ E }_a [ \ecomp{ (\Lambda \alpha.\lunit V)[A] }_a ] \  ]$
& by replacing $[\ ]$ in $\confuntyped_0$ with $\ecomp{E}_a$
\\
$=$
&
$\confuntyped[\  \ecomp{ (\Lambda \alpha.\lunit V)[A] }_a  \  ] $
& by the term compilation
\\
$=$
&
$\confuntyped[\  \lunit \ecomp{ (\Lambda \alpha.\lunit V)}_a  \  ] $
& by the value compilation
\\
$=$
&
$\confuntyped[\   \lunit \ecomp{V}_a  \  ] $
& by Lemma \ref{lemma:typesubstundertypeandlocationerasure}
\\
$=$
&
$\confuntyped[\   \lunit \ecomp{V\subst{A}{\alpha}}_a  \  ] $
& by replacing $\ecomp{E}_a$ in $\confuntyped$ with $[\ ]$
\\
$=$
&
$\confuntyped_0[\ \ecomp{ E }_a [\  \lunit \ecomp{V\subst{A}{\alpha}}_a \ ] \  ] $
& by Lemma \ref{lemma:valuesubstundertypeandlocationerasure}
\\
$=$
&
$\confuntyped_0[\ \ecomp{ E }_a [  \ecomp{\lunit  (V\subst{A}{\alpha})}_a  ] \  ] $
& by Lemma \ref{lemma:evaluationcontextsundertypeandlocationerasure}
\\
$=$
&
$\confuntyped_0[ \ \ecomp{ E [  \lunit  (V\subst{A}{\alpha})  ] \  ] }_a \ ]$
& by (C-Client) or (C-Server)
\\
$=$
&
$\ecomp{\conf[\ \lunit (V\subst{A}{\alpha}) \ ]}$
&
\\
$=$
&
$\ecomp{Conf_2}$
&
\end{tabular}

\ \\
{\bf (E-App)}
	$Conf_1 = \conf[\ clo(\overline{W},F)(V) \ ]$ and $Conf_2 = \conf[\ M\subst{\overline{W}}{\overline{z}}\subst{V}{x} \ ]$ where $\funcode(F)=\overline{z}.\lambda x.M$.

	Note that $F=F_{name} \, \overline{\Loc} \, \overline{A}$ and $\funcode(F_{name}) = \overline{l} \, \overline{\alpha} \, \overline{z}.\lambda x.M_0$ where $M=M_0\subst{\overline{\Loc}}{\overline{l}}\subst{\overline{A}}{\overline{\alpha}}$.
	By the compilation of function stores, $\funcode(F_{name})=\overline{z_l}\,\overline{z}.\lambda x.\ecomp{M}_a$.

\begin{tabular}{l l l}
&
$\ecomp{Conf_1}$
&
\\
&
$\ecomp{ \conf[\ clo(\overline{W},F)(V) \ ]}$
& by (C-Client) or (C-Server)
\\
$=$
&
$\confuntyped_0[ \ \ecomp{ E[ clo(\overline{W},F)(V) ] }_a  \ ]$
& by Lemma \ref{lemma:evaluationcontextsundertypeandlocationerasure}
\\
$=$
&
$\confuntyped_0[ \ \ecomp{ E }_a [ \ecomp{ clo(\overline{W},F)(V) }_a ] \  ]$
& by replacing $[\ ]$ in $\confuntyped_0$ with $\ecomp{E}_a$
\\
$=$
&
$\confuntyped[\  \ecomp{ clo(\overline{W},F)(V) }_a  \  ] $
& by the term compilation
\\
$=$
&
\multicolumn{2}{l}{
$\confuntyped[\  \case{ \ecomp{clo(\overline{W},F)}_a } {Closure \ \overline{w} \ f \rightarrow \ecomp{M_0}_a \subst{\overline{w}}{\overline{z_l}\,\overline{z}}\subst{ \ecomp{V}_a }{ x }} \  ] $
}
\\
&
& by the value compilation
\\
$=$
&
\multicolumn{2}{l}{
$\confuntyped[\  \case{ Closure (\overline{ \ecomp{\Loc}} \cdot \overline{ \ecomp{W}_a })\,F_{name} } {Closure \ \overline{w} \ f \rightarrow \ecomp{M_0}_a \subst{\overline{w}}{\overline{z_l}\,\overline{z}}\subst{ \ecomp{V}_a }{ x }} \  ] $
}
\\
&
& by (e-case)
\\
$\run$
&
$\confuntyped[\   \ecomp{M_0}_a \subst{\overline{\ecomp{\Loc}}\, \overline{\ecomp{W}_a} }{\overline{z_l}\,\overline{z}}\subst{ \ecomp{V}_a }{ x } \  ] $
& by Lemma \ref{lemma:locationsubstundertypeandlocationerasure}
\\
$=$
&
$\confuntyped[\  \ecomp{M_0\subst{\overline{\Loc}}{\overline{l}}}_a \subst{\overline{\ecomp{W}_a}} {\overline{z}}\subst{ \ecomp{V}_a }{ x } \  ] $
& by Lemma \ref{lemma:typesubstundertypeandlocationerasure}
\\
$=$
&
$\confuntyped[\  \ecomp{M_0
\subst{\overline{\Loc}}{\overline{l}}
\subst{\overline{A}}{\overline{\alpha}}
}_a \subst{\overline{\ecomp{W}_a}}{\overline{z}} \subst{ \ecomp{V}_a }{ x } \  ] $
& by Lemma \ref{lemma:valuesubstundertypeandlocationerasure}
\\
$=$
&
$\confuntyped[\  \ecomp{M_0
\subst{\overline{\Loc}}{\overline{l}}
\subst{\overline{A}}{\overline{\alpha}}
\subst{\overline{W}}{\overline{z}} }_a \subst{ \ecomp{V}_a }{ x } \  ] $
& by Lemma \ref{lemma:valuesubstundertypeandlocationerasure}
\\
$=$
&
$\confuntyped[\  \ecomp{M_0
\subst{\overline{\Loc}}{\overline{l}}
\subst{\overline{A}}{\overline{\alpha}}
\subst{\overline{W}}{\overline{z}}
\subst{ V }{ x } }_a
 \  ] $
& by $M=M_0\subst{\overline{\Loc}}{\overline{l}}\subst{\overline{A}}{\overline{\alpha}}$
\\
$=$
&
$\confuntyped[\  \ecomp{M\subst{\overline{W}}{\overline{z}} \subst{ V }{ x } }_a  \  ] $
& by replacing $\ecomp{E}_a$ in $\confuntyped$ with $[\ ]$
\\
$=$
&
$\confuntyped_0[\ \ecomp{ E }_a [  \ecomp{M\subst{\overline{W}}{\overline{z}} \subst{ V }{ x } }_a  ] \  ] $
& by Lemma \ref{lemma:evaluationcontextsundertypeandlocationerasure}
\\
$=$
&
$\confuntyped_0[ \ \ecomp{ E [  M\subst{\overline{W}}{\overline{z}} \subst{ V }{ x }  ] }_a \ ]$
& by (C-Client) or (C-Server)
\\
$=$
&
$\ecomp{\conf[\ M\subst{\overline{W}}{\overline{z}}\subst{V}{x} \ ]}$
&
\\
$=$
&
$\ecomp{Conf_2}$
&
\end{tabular}

\ \\
{\bf (E-LApp)}
	$Conf_1 = \conf[\ clo(\overline{W},F)[b] \ ]$ and $Conf_2 = \conf[\ V\subst{\overline{W}}{\overline{z}}\subst{b}{l} \ ]$ where $\funcode(F)=\overline{z}.\Lambda l.V$.

	Note that $F=F_{name} \, \overline{\Loc} \, \overline{A}$ and $\funcode(F_{name}) = \overline{l} \, \overline{\alpha} \, \overline{z}.\Lambda l.V_0$ where $V=V_0\subst{\overline{\Loc}}{\overline{l}}\subst{\overline{A}}{\overline{\alpha}}$.
	By the compilation of function stores, $\funcode(F_{name})=\overline{z_l}\,\overline{z}.\lambda x_l.\ecomp{V_0}_a$.

\begin{tabular}{l l l}
&
$\ecomp{Conf_1}$
&
\\
&
$\ecomp{\conf[\ clo(\overline{W},F)[b] \ ]}$
& by (C-Client) or (C-Server)
\\
$=$
&
$\confuntyped_0[ \ \ecomp{ E[ clo(\overline{W},F)[b] ] }_a  \ ]$
& by Lemma \ref{lemma:evaluationcontextsundertypeandlocationerasure}
\\
$=$
&
$\confuntyped_0[ \ \ecomp{ E }_a [ \ecomp{ clo(\overline{W},F)[b] }_a ] \  ]$
& by replacing $[\ ]$ in $\confuntyped_0$ with $\ecomp{E}_a$
\\
$=$
&
$\confuntyped[\  \ecomp{ clo(\overline{W},F)[b] }_a  \  ] $
& by the term compilation
\\
$=$
&
\multicolumn{2}{l}{
$\confuntyped[\  \case{ \ecomp{clo(\overline{W},F)}_a } {Closure \ \overline{w} \ f \rightarrow \ecomp{V_0}_a \subst{\overline{w}}{\overline{z_l}\,\overline{z}}\subst{ \ecomp{b} }{ x_l }} \  ] $
}
\\
&
& by the value compilation
\\
$=$
&
\multicolumn{2}{l}{
$\confuntyped[\  \case{ Closure (\overline{ \ecomp{\Loc}} \cdot \overline{ \ecomp{W}_a })\,F_{name} } {Closure \ \overline{w} \ f \rightarrow \ecomp{V_0}_a \subst{\overline{w}}{\overline{z_l}\,\overline{z}}\subst{ \ecomp{b} }{ x_l }} \  ] $
}
\\
&
& by (e-case)
\\
$\run$
&
$\confuntyped[\   \ecomp{V_0}_a \subst{\overline{\ecomp{\Loc}}\, \overline{\ecomp{W}_a} }{\overline{z_l}\,\overline{z}}\subst{ \ecomp{b} }{ x_l } \  ] $
& by Lemma \ref{lemma:locationsubstundertypeandlocationerasure}
\\
$=$
&
$\confuntyped[\  \ecomp{V_0\subst{\overline{\Loc}}{\overline{l}}}_a \subst{\overline{\ecomp{W}_a}} {\overline{z}}\subst{ \ecomp{b} }{ x_l } \  ] $
& by Lemma \ref{lemma:typesubstundertypeandlocationerasure}
\\
$=$
&
$\confuntyped[\  \ecomp{V_0
\subst{\overline{\Loc}}{\overline{l}}
\subst{\overline{A}}{\overline{\alpha}}
}_a \subst{\overline{\ecomp{W}_a}}{\overline{z}} \subst{ \ecomp{b} }{ x_l } \  ] $
& by Lemma \ref{lemma:valuesubstundertypeandlocationerasure}
\\
$=$
&
$\confuntyped[\  \ecomp{V_0
\subst{\overline{\Loc}}{\overline{l}}
\subst{\overline{A}}{\overline{\alpha}}
\subst{\overline{W}}{\overline{z}} }_a \subst{ \ecomp{b} }{ x_l } \  ] $
& by Lemma \ref{lemma:locationsubstundertypeandlocationerasure}
\\
$=$
&
$\confuntyped[\  \ecomp{V_0
\subst{\overline{\Loc}}{\overline{l}}
\subst{\overline{A}}{\overline{\alpha}}
\subst{\overline{W}}{\overline{z}}
\subst{ b }{ l } }_a
 \  ] $
& by $M=M_0\subst{\overline{\Loc}}{\overline{l}}\subst{\overline{A}}{\overline{\alpha}}$
\\
$=$
&
$\confuntyped[\  \ecomp{V\subst{\overline{W}}{\overline{z}} \subst{ b }{ l } }_a  \  ] $
& by replacing $\ecomp{E}_a$ in $\confuntyped$ with $[\ ]$
\\
$=$
&
$\confuntyped_0[\ \ecomp{ E }_a [  \ecomp{V\subst{\overline{W}}{\overline{z}} \subst{ b }{ l } }_a  ] \  ] $
& by Lemma \ref{lemma:evaluationcontextsundertypeandlocationerasure}
\\
$=$
&
$\confuntyped_0[ \ \ecomp{ E [  V\subst{\overline{W}}{\overline{z}} \subst{ b }{ l }  ] }_a \ ]$
& by (C-Client) or (C-Server)
\\
$=$
&
$\ecomp{\conf[\ V\subst{\overline{W}}{\overline{z}}\subst{b}{l} \ ]}$
&
\\
$=$
&
$\ecomp{Conf_2}$
&
\end{tabular}

\ \\
{\bf (E-Req)}
	$Conf_1=\confcs{E[\req{V}{W}];\stack_\client \ \ }{\ \ \stack_\server}$  and
	$Conf_2=\confcs{E[\ ];\stack_\client\ \ }{\ \ V(W);\stack_\server}$.

{\bf i)} $\stack_\server=\emptystack$: We also have $\stack_\client=\emptystack$ in the compilation of $Conf_1$.
	For compiling $Conf_1$, we need
	$\ecomp{\stack_\client \ | \ \stack_\server}_\client = e[\ ] \ | \ loop_{server} \ ()$.
	For compiling $Conf_2$, we also use
	$\ecomp{ \stack_\client \ | \ \stack_\server }_\server = loop \ ()\ | \  \ldokeyword \ z \leftarrow [\ ];  \ send \ (Ret \ z); \ loop_{server} \ ()$.

\begin{tabular}{l l l}
&
$\ecomp{Conf_1}$
&
\\
$=$
&
$\ecomp{ \confcs{E[\req{V}{W}]; \stack_\client}{\stack_\server} }$
& by (C-Client)
\\
$=$
& $\confcs{ e[\ \ecomp{E[\req{V}{W}]}_\client \ ] }{ loop_{server} \ () }$
& by Lemma \ref{lemma:evaluationcontextsundertypeandlocationerasure}
\\
$=$
& $\confcs{ e[\ \ecomp{E}_\client [ \ \ecomp{\req{V}{W}}_\client \ ] \ ] }{ loop_{server} \ () }$
& by value compilation
\\
$=$
&
\multicolumn{2}{l}{
$\confcs{ e[\ \ecomp{E}_\client [ \ \ldokeyword \  send \ (Apply \ \ecomp{V}_\client  \  \ecomp{N}_\client ); \ loop \ () \ ] \ ] }{ loop_{server} \ () }$
}
\\
&
& by (e-comm-c-s) and by (e-case)
\\
$\run^2$
&
\multicolumn{2}{l}{
$\confcs{ e[\ \ecomp{E}_\client [ \ loop \ () \ ] \ ] }{  \ \ldokeyword \ z \leftarrow \ecomp{V}_\client(\ecomp{W}_\client); \ send \ (Ret \ z);  loop \ () }$
}
\\
&
& by Lemma \ref{lemma:locationinvariantundertypeandlocationerasure}
\\
$=$
&
\multicolumn{2}{l}{
$\confcs{ e[\ \ecomp{E}_\client [ \ loop \ () \ ] \ ] }{  \ \ldokeyword \ z \leftarrow \ecomp{V}_\server(\ecomp{W}_\server); \ send \ (Ret \ z);  loop \ () }$
}
\\
&
& by the term compilation
\\
$=$
&
\multicolumn{2}{l}{
$\confcs{ e[\ \ecomp{E}_\client [ \ loop \ () \ ] \ ] }{  \ \ldokeyword \ z \leftarrow \ecomp{V(W)}_\server; \ send \ (Ret \ z);  loop \ () }$
}
\\
$=$
& $\ecomp{ \confcs{E[\ ];\stack_\client\ }{\ V(W);\stack_\server} }$
& by (C-Server)
\\
$=$
&
$\ecomp{Conf_2}$
&
\end{tabular}

\ \\

{\bf ii)} $\stack_\server=E_0[\ ];\stack_\server'$:
	For compiling $Conf_1$ , we use
	\begin{itemize}
	\item $\ecomp{\stack_\client \ | \ \stack_\server'}_\server = m \ | \ e'[\ ]$
	\item $\ecomp{ \stack_\client \ | \ E_0[\ ];\stack_\server'}_\client = \ldokeyword \ z \leftarrow [\ ]; send \ (Ret \ z); \ m \ | \ e'[\ \ecomp{E_0}_\server[ \ loop \ () \ ] \ ]
	$
	\end{itemize}

	For compiling $Conf_2$, we use $\ecomp{ E[\ ];\stack_\client \ | \ E_0[\ ];\stack_\server'}_\server = $
	\[
	\langle \ldokeyword \ z \leftarrow \ecomp{ E}_\client[loop \ ()]; \ send \ (Ret \ z); m \ \  | \ \ \ldokeyword \ z \leftarrow[\ ]; \ send \ (Ret \ z);  e'[\  loop \ () \ ] \rangle
	\]

\begin{tabular}{l l l}
&
$\ecomp{Conf_1}$
&
\\
$=$
&
$\ecomp{ \confcs{E[\req{V}{W}]; \stack_\client}{\stack_\server} }$
& by (C-Client)
\\
$=$
&
\multicolumn{2}{l}{
$\confcs{\ldokeyword \ z \leftarrow \ecomp{ E[\req{V}{W}]}_\client; \ send \ (Ret \ z); m\ \ }{\ \ e'[\ \ecomp{E_0}_\server[ loop \ ()]]}$
}
\\
&
& by the value compilation
\\
$=$
&
\multicolumn{2}{l}{
$\langle \ldokeyword \ z \leftarrow \ecomp{ E}_\client[\ldokeyword \ send \ (Apply \ \ecomp{V}_\client \ \ecomp{W}_\client); \ loop \ ()]; \ send \ (Ret \ z); m$
}
\\
&
\multicolumn{2}{l}{
$| \ \ e'[\ \ecomp{E_0}_\server[ loop \ ()] \ ] \rangle$
}
\\
&
& by (e-comm-c-s) and (e-case)
\\
$\run^2$
&
\multicolumn{2}{l}{
$\langle \ldokeyword \ z \leftarrow \ecomp{ E}_\client[loop \ ()]; \ send \ (Ret \ z); m$
}
\\
&
\multicolumn{2}{l}{
$| \ \ e'[\  \ldokeyword \ z \leftarrow \ecomp{V}_\client(\ecomp{W}_\client); \ send \ (Ret \ z);  loop \ () \ ] \rangle$
}
\\
&
& by Lemma \ref{lemma:locationinvariantundertypeandlocationerasure}
\\
$=$
&
\multicolumn{2}{l}{
$\langle \ldokeyword \ z \leftarrow \ecomp{ E}_\client[loop \ ()]; \ send \ (Ret \ z); m$
}
\\
&
\multicolumn{2}{l}{
$| \ \ e'[\ \ldokeyword \ z \leftarrow \ecomp{V}_\server(\ecomp{W}_\server); \ send \ (Ret \ z);  loop \ () \ ] \rangle$
}
\\
&
& by the value compilation
\\
$=$
&
\multicolumn{2}{l}{
$\langle \ldokeyword \ z \leftarrow \ecomp{ E}_\client[loop \ ()]; \ send \ (Ret \ z); m$
}
\\
&
\multicolumn{2}{l}{
$| \ \ e'[\ \ldokeyword \ z \leftarrow \ecomp{V(W)}_\server; \ send \ (Ret \ z);  loop \ () \ ] \rangle$
}
\\
&
& by Definition \ref{def:structuralequivalence}
\\
$\structeqv^*$
&
\multicolumn{2}{l}{
$\langle \ldokeyword \ z \leftarrow \ecomp{ E}_\client[loop \ ()]; \ send \ (Ret \ z); m$
}
\\
&
\multicolumn{2}{l}{
$| \ \ \ldokeyword \ z \leftarrow \ecomp{V(W)}_\server; \ send \ (Ret \ z);  e'[\  loop \ () \ ] \rangle$
}
\\
$=$
&
$\ecomp{\confcs{E[\ ];\stack_\client\ \ }{\ \ V(W);\stack_\server}}$
& by (C-Server)
\\
$=$
&
$\ecomp{Conf_2}$
&
\end{tabular}

\ \\
{\bf (E-Call)}
	$Conf_1=\confcs{\stack_\client\ \ }{\ \ E[\call{V}{W}];\stack_\server}$  and
	$Conf_2=\confcs{V(W);\stack_\client\ \ }{\ \ E[\ ];\stack_\server}$.

{\bf i)} $\stack_\server=\emptystack$: We also have $\stack_\client=\emptystack$ in the compilation of $Conf_1$.
	For compiling $Conf_1$, we need $\ecomp{ \stack_\client \ | \ \stack_\server }_\server = loop \ () \ | \ \ldokeyword \ z \leftarrow [\ ];  \ send \ (Ret \ z); \ loop_{server} \ ()$.
	For compiling $Conf_2$, we use $\ecomp{\stack_\client\ | \ E[\ ];\stack_\server}_\client$ that is:

\begin{tabular}{l l l}
&
$\ecomp{Conf_1}$
&
\\
$=$
&
$\ecomp{ \confcs{\stack_\client\ \ }{\ \ E[\call{V}{W}];\stack_\server} }$
& by (C-Server)
\\
$=$
&
\multicolumn{2}{l}{
$\confcs{ loop \ () \ }{ \ \ldokeyword \ z \leftarrow [\ \ecomp{E[\call{V}{W}]}_\server ];  \ send \ (Ret \ z); \ loop_{server} \ ()}$
}
\\
&
& by Lemma \ref{lemma:evaluationcontextsundertypeandlocationerasure}
\\
$=$
&
\multicolumn{2}{l}{
$\confcs{loop \ () \ }{\ \ldokeyword \ z \leftarrow [\ \ecomp{E}_\server[\ecomp{\call{V}{W}}_\server] ];  \ send \ (Ret \ z); \ loop_{server} \ ()}$
}
\\
&
& by the value compilation
\\
$=$
&
\multicolumn{2}{l}{
$\confcs{ loop \ () \ }{ \ \ldokeyword \ z \leftarrow [\ \ecomp{E}_\server[ \ldokeyword \ send \ (Apply \ecomp{V}_\server \ \ecomp{W}_\server); \ loop \ ()] \ ]; send \ (Ret \ z); \ loop_{server} \ ()}$
}
\\
&
&
by (e-comm-s-c) and (e-case)
\\
$\run^2$
&
\multicolumn{2}{l}{
$\langle \ \ldokeyword \ z \leftarrow \ecomp{V}_\client (\ecomp{W}_\client); \ send \ (Ret \ z); \ loop \ ()$
}
\\
&
\multicolumn{2}{l}{
$| \ \ldokeyword \ z \leftarrow [\ \ \ecomp{E}_\server[ loop \ () ] \ ]; send \ (Ret \ z); \ loop_{server} \ () \rangle$
}
\\
&
& by Lemma \ref{lemma:locationinvariantundertypeandlocationerasure}
\\
$=$
&
\multicolumn{2}{l}{
$\langle \ \ldokeyword \ z \leftarrow \ecomp{V}_\server (\ecomp{W}_\server); \ send \ (Ret \ z); \ loop \ ()$
}
\\
&
\multicolumn{2}{l}{
$| \ \ldokeyword \ z \leftarrow [\ \ \ecomp{E}_\server[ loop \ () ] \ ]; send \ (Ret \ z); \ loop_{server} \ () \rangle$
}
\\
&
& by the term compilation
\\
$=$
&
\multicolumn{2}{l}{
$\langle \ \ldokeyword \ z \leftarrow \ecomp{V(W)}_\server; \ send \ (Ret \ z); \ loop \ ()$
}
\\
&
\multicolumn{2}{l}{
$| \ \ldokeyword \ z \leftarrow [\ \ \ecomp{E}_\server[ loop \ () ] \ ]; send \ (Ret \ z); \ loop_{server} \ () \rangle$
}
\\
&
& by the (C-Client)
\\
$=$
&
$\ecomp{ \confcs{V(W);\stack_\client\ \ }{\ \ E[\ ];\stack_\server} }$
&
\\
$=$
&
$\ecomp{Conf_2}$
\end{tabular}

\ \\

{\bf ii)} $\stack_\server=E_1[\ ];\stack_\server'$: We also have $\stack_\client=E_0[\ ];\stack_\client'$ in the compilation of $Conf_1$.
	For compiling $Conf_1$, we need $\ecomp{E_0[\ ];\stack_\client' \ | \ E_1[\ ]; \stack_\server'}_\server$ that is
	$
	e[ \ \ecomp{ E_0 }_\client[ loop \ () ] \ ] \ | \ \ldokeyword \ z \leftarrow [\ ]; \ send \ (Ret \ z); \ m
	$
	where $\ecomp{\stack_\client' \ | \ E_1[\ ]; \stack_\server'}_\client = e[\ ] \ | \ m$

	For compiling $Conf_2$, we need $\ecomp{E_0[\ ];\stack_\client' \ | \ E[\ ]; E_1[\ ]; \stack_\server'}_\client$ that is
	\[
	\ldokeyword \ z \leftarrow [\ ]; \ send \ (Ret \ z); \ e[ \ecomp{E_0}_\client[ loop \ () ]]  \ | \ \ldokeyword \ z \leftarrow [\ \ecomp{E}_\server[ loop \ () ] \ ]; send \ (Ret \ z); \ m \ ()
	\]
	where $\ecomp{E_0[\ ];\stack_\client' \ | \ E_1[\ ]; \stack_\server'}_\server =
	e[ \ecomp{E_0}_\client[ loop \ () ]] \ | \
	\ldokeyword \ z \leftarrow \ [\ ]; \ send \ (Ret \ z); \ m
	$

%	&
%\multicolumn{2}{l}{
%$\langle
%  \ldokeyword \ z \leftarrow \ecomp{V( W)}_\client ;
%  \ send \ (Ret \ z);
%  e[ \ \ecomp{ E_0 }_\client[
%     \ loop \ ()
%  ] \ ]$
%}
%\\
%&
%\multicolumn{2}{l}{
%$| \
%\ldokeyword \ z \leftarrow [ \ecomp{E}_\server[
%  \ loop \ ()
%]; \ send \ (Ret \ z); \ m \rangle$
%}

\begin{tabular}{l l l}
&
$\ecomp{Conf_1}$
&
\\
$=$
&
$\ecomp{ \confcs{\stack_\client\ \ }{\ \ E[\call{V}{W}];\stack_\server} }$
& by (C-Server)
\\
$=$
&
\multicolumn{2}{l}{
$\confcs{ e[ \ \ecomp{ E_0 }_\client[ loop \ () ] \ ]  \ }{ \
\ldokeyword \ z \leftarrow [ \ecomp{E[\call{V}{W}]}_\server ]; \ send \ (Ret \ z); \ m}$
}
\\
&
& by Lemma \ref{lemma:evaluationcontextsundertypeandlocationerasure}
\\
$=$
&
\multicolumn{2}{l}{
$\confcs{ e[ \ \ecomp{ E_0 }_\client[ loop \ () ] \ ]  \ }{ \
\ldokeyword \ z \leftarrow [ \ecomp{E}_\server[\ecomp{ \call{V}{W} }_\server] ]; \ send \ (Ret \ z); \ m}$
}
\\
&
& by the value compilation
\\
$=$
&
\multicolumn{2}{l}{
$\langle  e[ \ \ecomp{ E_0 }_\client[ loop \ () ] \ ]$
}
\\
&
\multicolumn{2}{l}{
$| \
\ldokeyword \ z \leftarrow [ \ecomp{E}_\server[
  \ldokeyword \ send \ ( Apply \ \ecomp{V}_\server \ \ecomp{W}_\server ); \ loop \ ()
]; \ send \ (Ret \ z); \ m \rangle$
}
\\
&
& by (e-comm-s-c) and (e-case)
\\
$\run^2$
&
\multicolumn{2}{l}{
$\langle  e[ \ \ecomp{ E_0 }_\client[
  \ldokeyword \ z \leftarrow \ecomp{V}_\server ( \ecomp{W}_\server );
  \ send \ (Ret \ z);
  \ loop \ ()
] \ ]$
}
\\
&
\multicolumn{2}{l}{
$| \
\ldokeyword \ z \leftarrow [ \ecomp{E}_\server[
  \ loop \ ()
]; \ send \ (Ret \ z); \ m \rangle$
}
\\
&
& by Lemma \ref{lemma:locationinvariantundertypeandlocationerasure}
\\
$=$
&
\multicolumn{2}{l}{
$\langle  e[ \ \ecomp{ E_0 }_\client[
  \ldokeyword \ z \leftarrow \ecomp{V}_\client ( \ecomp{W}_\client );
  \ send \ (Ret \ z);
  \ loop \ ()
] \ ]$
}
\\
&
\multicolumn{2}{l}{
$| \
\ldokeyword \ z \leftarrow [ \ecomp{E}_\server[
  \ loop \ ()
]; \ send \ (Ret \ z); \ m \rangle$
}
\\
&
& by the term compilation
\\
$=$
&
\multicolumn{2}{l}{
$\langle  e[ \ \ecomp{ E_0 }_\client[
  \ldokeyword \ z \leftarrow \ecomp{V( W)}_\client ;
  \ send \ (Ret \ z);
  \ loop \ ()
] \ ]$
}
\\
&
\multicolumn{2}{l}{
$| \
\ldokeyword \ z \leftarrow [ \ecomp{E}_\server[
  \ loop \ ()
]; \ send \ (Ret \ z); \ m \rangle$
}
\\
&
& by Definition \ref{def:structuralequivalence}
\\
$\structeqv^*$
&
\multicolumn{2}{l}{
$\langle
  \ldokeyword \ z \leftarrow \ecomp{V( W)}_\client ;
  \ send \ (Ret \ z);
  e[ \ \ecomp{ E_0 }_\client[
     \ loop \ ()
  ] \ ]$
}
\\
&
\multicolumn{2}{l}{
$| \
\ldokeyword \ z \leftarrow [ \ecomp{E}_\server[
  \ loop \ ()
]; \ send \ (Ret \ z); \ m \rangle$
}
\\
&
& by (C-Server)
\\
$=$
&
$\ecomp{ \confcs{V(W);\stack_\client\ \ }{\ \ E[\ ];\stack_\server}  } $
&
\\
$=$
&
$\ecomp{Conf_2}$
&
\end{tabular}

\ \\
{\bf (E-Unit-C)}
	$Conf_1 = \confcs{\lunit{V};\stack_\client\ \ }{\ \ E[\ ];\stack_\server}$ and $Conf_2 = \confcs{\stack_\client\ \ }{\ \ E[\lunit{V}];\stack_\server}$.

	In the compilation of $Conf_1$, we have
	\begin{itemize}
	\item $\ecomp{\stack_\client \ | \ \stack_\server}_\server = m \ | \ e[ \ ]$
	\item $\ecomp{\stack_\client \ | \ E[\ ]; \ \stack _\server}_\client =
	   \ldokeyword \ z \leftarrow [ \ ]; \ send \ (Ret \ z); \ m
	   \ | \ e[ \ecomp{E}_\server[ loop \ ()] ]$
	\end{itemize}

\begin{tabular}{l l l}
&
$\ecomp{Conf_1}$
&
\\
$=$
&
$\ecomp{ \confcs{\lunit{V};\stack_\client\ \ }{\ \ E[\ ];\stack_\server }}$
& by (C-Server)
\\
$=$
&
$\confcs{\ldokeyword \ z \leftarrow \ecomp{\lunit V}_\client; \ send \ (Ret \ z); \ m \ }
	  { \ e[ \ecomp{E}_\server[ loop \ ()] ]}$
& by the value compilation
\\
$=$
&
$\confcs{\ldokeyword \ z \leftarrow \lunit \ecomp{ V}_\client; \ send \ (Ret \ z); \ m \ }
	  { \ e[ \ecomp{E}_\server[ loop \ ()] ]}$
& by (e-do)
\\
$=$
&
$\confcs{\ldokeyword \ \ send \ (Ret \  \ecomp{ V}_\client); \ m \ }
	  { \ e[ \ecomp{E}_\server[ loop \ ()] ]}$
& by (e-comm-c-s) and (e-case)
\\
$=$
&
$\confcs{\ m \ }
	  { \ e[ \ecomp{E}_\server[ \lunit  \ecomp{ V}_\client] ]}$
& by Lemma \ref{lemma:locationinvariantundertypeandlocationerasure}
\\
$=$
&
$\confcs{\ m \ }
	  { \ e[ \ecomp{E}_\server[ \lunit  \ecomp{ V}_\server] ]}$
& by the value compilation
\\
$=$
&
$\confcs{\ m \ }
	  { \ e[ \ecomp{E}_\server[  \ecomp{\lunit  V}_\server] ]}$
& by (C-Server)
\\
$=$
&
$\ecomp{  \confcs{\stack_\client\ \ }{\ \ E[\lunit{V}];\stack_\server} }$
&
\\
$=$
&
$\ecomp{  Conf_2 }$
&
\end{tabular}

\ \\
{\bf (E-Unit-S)}
	$Conf_1=\confcs{E[\ ];\stack_\client\ \ }{\ \ \lunit{V};\stack_\server}$ and $Conf_2=\confcs{E[\lunit{V}];\stack_\client\ \ }{\ \ \stack_\server}$.

	In the compilation of $Conf_1$, we have
	\begin{itemize}
	\item $\ecomp{\stack_\client \ | \ \stack_\server}_\client = e[\ ] \ | \ m$
	\item $\ecomp{E[\ ]; \stack_\client \ | \ \stack_\server}_\server = e[ \ \ecomp{E}_\client[loop \ ()] \ ] \ | \ \ldokeyword \ z \leftarrow [\ ]; \ send \ (Ret \ z); \ m$
	\end{itemize}

\begin{tabular}{l l l}
&
$\ecomp{Conf_1}$
&
\\
$=$
&
$\ecomp{  \confcs{E[\ ];\stack_\client\ \ }{\ \ \lunit{V};\stack_\server} }$
& by (C-Server)
\\
$=$
&
$e[ \ \ecomp{E}_\client[loop \ ()] \ ] \ | \ \ldokeyword \ z \leftarrow \ecomp{\lunit V}_\server; \ send \ (Ret \ z); \ m$
& by the value compilation
\\
$=$
&
$e[ \ \ecomp{E}_\client[loop \ ()] \ ] \ | \ \ldokeyword \ z \leftarrow \lunit \ecomp{V}_\server; \ send \ (Ret \ z); \ m$
& by (e-do) (e-comm-s-c)
\\
$=$
&
$e[ \ \ecomp{E}_\client[ \lunit \ecomp{V}_\client ] \ ] \ | \  \ m$
& by  Lemma \ref{lemma:locationinvariantundertypeandlocationerasure}
\\
$=$
&
$e[ \ \ecomp{E}_\client[ \lunit \ecomp{V}_\server ] \ ] \ | \  \ m$
& by   the value compilation
\\
$=$
&
$e[ \ \ecomp{E}_\client[  \ecomp{\lunit V}_\server ] \ ] \ | \  \ m$
& by (C-Client)
\\
$=$
&
$\ecomp{ \confcs{E[\lunit{V}];\stack_\client\ \ }{\ \ \stack_\server} }$
&
\\
$=$
&
$\ecomp{ Conf_2 }$
&
\end{tabular}

\ \\
{\bf (E-Unit-S-E)}
	$Conf_1=\confcs{\emptystack\ \ }{\ \ \lunit{V}}$ and $Conf_2=\confcs{\lunit{V}\ \ }{\ \ \emptystack}$.

	In the compilation of $Conf_1$, we have
	$
		\ecomp{\emptystack \ | \ \emptystack}_\server =
		loop \ () \ | \ \ldokeyword \ z \leftarrow [\ ]; \ send \ (Ret \ z); \ loop_{server} \ ()
	$.

\begin{tabular}{l l l}
&
$\ecomp{Conf_1}$
&
\\
$=$
&
$\ecomp{  \confcs{\emptystack\ \ }{\ \ \lunit{V}} }$
& by (C-Server)
\\
$=$
&
$loop \ () \ | \ \ldokeyword \ z \leftarrow \ecomp{\lunit V}_\server; \ send \ (Ret \ z); loop_{server} \ ()$
& by the value compilation
\\
$=$
&
$loop \ () \ | \ \ldokeyword \ z \leftarrow \lunit \ecomp{V}_\server; \ send \ (Ret \ z); loop_{server} \ ()$
& by (e-do) and (e-comm-s-c)
\\
$=$
&
$\lunit \ecomp{V}_\server \ | \ loop_{server} \ ()$
& by Lemma \ref{lemma:locationinvariantundertypeandlocationerasure}
\\
$=$
&
$\lunit \ecomp{V}_\client \ | \ loop_{server} \ ()$
& by the value compilation
\\
$=$
&
$\ecomp{\lunit  V}_\client \ | \ loop_{server} \ ()$
& by (C-Client)
\\
$=$
&
$\ecomp{ \confcs{\lunit{V}\ \ }{\ \ \emptystack} }$
& \\
$=$
&
$\ecomp{Conf_2}$
&
\end{tabular}

\ \\
{\bf (E-Gen-C-C)}
	$Conf_1=\confcs{E[\gen{\client}{V}{W}];\stack_\client\ \ }{\ \ \stack_\server}$  and
	$Conf_2=\confcs{E[V(W)];\stack_\client\ \ }{\ \ \stack_\server}$. Let us have $\ecomp{\stack_\client \ | \ \stack_\server}_\client = e[\ ] \ | \ m$.

\begin{tabular}{l l l}
&
$\ecomp{Conf_1}$
&
\\
$=$
&
$\ecomp{  \confcs{E[\gen{\client}{V}{W}];\stack_\client\ \ }{\ \ \stack_\server} }$
& by (C-Client)
\\
$=$
&
$\confcs{\ e[ \ecomp{ E[\gen{\client}{V}{W}] }_\client] \ }{ \ m \ }$
& by Lemma \ref{lemma:evaluationcontextsundertypeandlocationerasure}
\\
$=$
&
$\confcs{\ e[ \ecomp{ E }_\client [ \ecomp{ \gen{\client}{V}{W}  }_\client ] \ ] \ }{ \ m \ }$
& by the value compilation
\\
$=$
&
$\confcs{\ e[ \ecomp{ E }_\client [ \ldo{x}{\case{ \ecomp{\client} }{ \cdots}}{\lunit{x}} ] \ ] \ }{ \ m \ }$
& by $\ecomp{\client}=Client$ and (e-case)
\\
$=$
&
$\confcs{\ e[ \ecomp{ E }_\client [ \ldo{x}{ \ecomp{V(W)}_\client }{\lunit{x}} ] \ ] \ }{ \ m \ }$
& by Lemma \ref{def:structuralequivalence}
\\
$\structeqv$
&
$\confcs{\ e[ \ecomp{ E }_\client [ \ecomp{V(W)}_\client  ] \ ] \ }{ \ m \ }$
& by Lemma \ref{lemma:evaluationcontextsundertypeandlocationerasure}
\\
$=$
&
$\confcs{\ e[ \ecomp{ E [ V(W)  ] \ ] }_\client \ }{ \ m \ }$
& by (C-Client)
\\
$=$
&
$\ecomp{ \confcs{E[V(W)];\stack_\client\ \ }{\ \ \stack_\server} }$
&
\\
$=$
&
$\ecomp{ Conf_2 }$
&
\end{tabular}

\ \\
{\bf (E-Gen-S-C)}
	$Conf_1=\confcs{E[\gen{\server}{V}{W}];\stack_\client\ \ }{\ \ \stack_\server}$  and
	$Conf_2=\confcs{E[\req{V}{W}];\stack_\client\ \ }{\ \ \stack_\server}$.
	Let us have $\ecomp{\stack_\client \ | \ \stack_\server}_\client = e[\ ] \ | \ m$.

\begin{tabular}{l l l}
&
$\ecomp{Conf_1}$
&
\\
$=$
&
$\ecomp{  \confcs{E[\gen{\server}{V}{W}];\stack_\client\ \ }{\ \ \stack_\server} }$
& by (C-Client)
\\
$=$
&
$\confcs{\ e[ \ecomp{ E[\gen{\server}{V}{W}] }_\client] \ }{ \ m \ }$
& by Lemma \ref{lemma:evaluationcontextsundertypeandlocationerasure}
\\
$=$
&
$\confcs{\ e[ \ecomp{ E }_\client [ \ecomp{ \gen{\server}{V}{W}  }_\client ] \ ] \ }{ \ m \ }$
& by the value compilation
\\
$=$
&
$\confcs{\ e[ \ecomp{ E }_\client [ \ldo{x}{\case{ \ecomp{\server} }{ \cdots}}{\lunit{x}} ] \ ] \ }{ \ m \ }$
& by $\ecomp{\server}=Server$ and (e-case)
\\
$=$
&
$\confcs{\ e[ \ecomp{ E }_\client [ \ldo{x}{ \ecomp{\req{V}{W}}_\client }{\lunit{x}} ] \ ] \ }{ \ m \ }$
& by Def. \ref{def:structuralequivalence}
\\
$\structeqv$
&
$\confcs{\ e[ \ecomp{ E }_\client [ \ecomp{\req{V}{W}}_\client  ] \ ] \ }{ \ m \ }$
& by Lemma \ref{lemma:evaluationcontextsundertypeandlocationerasure}
\\
$=$
&
$\confcs{\ e[ \ecomp{ E [ \req{V}{W}  ] \ ] }_\client \ }{ \ m \ }$
& by (C-Client)
\\
$=$
&
$\ecomp{ \confcs{E[\req{V}{W}];\stack_\client\ \ }{\ \ \stack_\server} }$
&
\\
$=$
&
$\ecomp{ Conf_2 }$
&
\end{tabular}

\ \\
{\bf (E-Gen-C-S)}
	$Conf_1=\confcs{\stack_\client\ \ }{\ \ E[\gen{\client}{V}{W}];\stack_\server}$ and
	$Conf_2=\confcs{\stack_\client\ \ }{\ \ E[\call{V}{W}];\stack_\server}$.
	Let us have $\ecomp{\stack_\client \ | \ \stack_\server}_\server = m \ | \ e[\ ]$.

\begin{tabular}{l l l}
&
$\ecomp{Conf_1}$
&
\\
$=$
&
$\ecomp{  \confcs{\stack_\client\ \ }{\ \ E[\gen{\client}{V}{W}];\stack_\server} }$
& by (C-Server)
\\
$=$
&
$ \confcs{m\ \ }{\ \ e[ \ecomp{E [ \gen{\client}{V}{W}] _\server  } ] } $
& by Lemma \ref{lemma:evaluationcontextsundertypeandlocationerasure}
\\
$=$
&
$ \confcs{m\ \ }{\ \ e[ \ecomp{E}_\server [ \ecomp{ \gen{\client}{V}{W} }_\server ]] } $
& by the value compilation
\\
$=$
&
$ \confcs{m\ \ }{\ \ e[ \ecomp{E}_\server [ \ldo{x}{\case{\ecomp{\client} }{\cdots}}{\lunit x} ]] } $
& by $\ecomp{\client}=Client$ and (e-case)
\\
$=$
&
$ \confcs{m\ \ }{\ \ e[ \ecomp{E}_\server [ \ldo{x}{\ecomp{\call{V}{W}}_\server}{\lunit x} ]] } $
& by  Def. \ref{def:structuralequivalence}
\\
$\structeqv$
&
$ \confcs{m\ \ }{\ \ e[ \ecomp{E}_\server [ \ecomp{\call{V}{W}}_\server ]] } $
& by Lemma \ref{lemma:evaluationcontextsundertypeandlocationerasure}
\\
$=$
&
$ \confcs{m\ \ }{\ \ e[ \ecomp{E [\call{V}{W} ]}_\server] } $
& by (C-Server)
\\
$=$
&
$\ecomp{ \confcs{\stack_\client\ \ }{\ \ E[\call{V}{W}];\stack_\server} } $
&
\\
$=$
&
$\ecomp{ Conf_2 } $
&
\end{tabular}

\ \\
{\bf (E-Gen-S-S)}
	$Conf_1=\confcs{\stack_\client\ \ }{\ \ E[\gen{\server}{V}{W}];\stack_\server}$ and
	$Conf_2=\confcs{\stack_\client\ \ }{\ \ E[V(W)];\stack_\server}$.
	Let us have $\ecomp{\stack_\client \ | \ \stack_\server}_\server = m \ | \ e[\ ]$.

\begin{tabular}{l l l}
&
$\ecomp{Conf_1}$
&
\\
$=$
&
$\ecomp{  \confcs{\stack_\client\ \ }{\ \ E[\gen{\server}{V}{W}];\stack_\server} }$
& by (C-Server)
\\
$=$
&
$ \confcs{m\ \ }{\ \ e[ \ecomp{E [ \gen{\server}{V}{W}] _\server  } ] } $
& by Lemma \ref{lemma:evaluationcontextsundertypeandlocationerasure}
\\
$=$
&
$ \confcs{m\ \ }{\ \ e[ \ecomp{E}_\server [ \ecomp{ \gen{\server}{V}{W} }_\server ]] } $
& by the value compilation
\\
$=$
&
$ \confcs{m\ \ }{\ \ e[ \ecomp{E}_\server [ \ldo{x}{\case{\ecomp{\server} }{\cdots}}{\lunit x} ]] } $
& by $\ecomp{V(W)}=Server$ and (e-case)
\\
$=$
&
$ \confcs{m\ \ }{\ \ e[ \ecomp{E}_\server [ \ldo{x}{\ecomp{V(W)}_\server}{\lunit x} ]] } $
& by  Def. \ref{def:structuralequivalence}
\\
$\structeqv$
&
$ \confcs{m\ \ }{\ \ e[ \ecomp{E}_\server [ \ecomp{V(W)}_\server ]] } $
& by Lemma \ref{lemma:evaluationcontextsundertypeandlocationerasure}
\\
$=$
&
$ \confcs{m\ \ }{\ \ e[ \ecomp{E [V(W) ]}_\server] } $
& by (C-Server)
\\
$=$
&
$\ecomp{ \confcs{\stack_\client\ \ }{\ \ E[V(W)];\stack_\server} } $
&
\\
$=$
&
$\ecomp{ Conf_2 } $
&
\end{tabular}

\end{proof}

\section{A running example}
\label{sec:polyrpc:runningexample}

Here is a running example in \polyrpc.
\[
(\Lambda l. \lamL{l}{g}{g \ 1}) [\server] \ (\lamL{\client}{x}{x})
\]

Evaluation starting at client goes to server by $(\lamL{\server}{g}{g
  \ 1}) \ (\lamL{\client}{x}{x})$ and then to the client by
$(\lamL{\client}{x}{x}) \ 1$ resulting in $1$ there. The result comes
back to the server and then to the client, ending the evaluation.

\ \\

A \polycs program that is compiled from the running RPC example is as
follows:
        \[
          main \ = \ \ldo{ \ h \ } {\lunit{ (\clo{\emptyset}{f_1}[\server]) } } {\req{\ h}{\ \clo{\emptyset}{f_3} \ }}
        \]
where $(\funstore_\client,\funstore_\server)$ is 
\begin{center}
\begin{tabular}{l l l l l}
  $f_1 : \emptyset.\emptyset.\forall l.T \ \cloty{\cloty{Int\funL{\client}T \ Int} \funL{l} T \ Int} $ & $=$ & $ \emptyset.\emptyset. \ \Lambda l. \ \lunit{ \ (\clo{\emptyset}{f_2[l]}) \ }$
  &
  $\in$ & $ \funstore_\client, \funstore_\server$
  \\
  $f_2 : l.\emptyset.\cloty{Int\funL{\client}T \ Int} \funL{l} T \ Int  $ & $=$ & $ l.\emptyset. \ \lambda g. \ \gen{ \ \client}{ \ g}{ \ 1 \ }$
  &
  $\in$ & $ \funstore_\client, \funstore_\server$
  \\
  $f_3 : \emptyset.\emptyset.Int \funL{\client} T\ Int  $ & $=$ & $ \emptyset.\emptyset. \ \lambda x. \ \lunit{ x }$
  &
  $\in$ & $ \funstore_\client$
\end{tabular}
\end{center}

\ \\

Let us evaluate the \polycs program.

\begin{tabular}{l l}
  & $\confcs{main \ }{\ \epsilon}$
\\  
$\longrightarrow^2
$  & $\confcs{\ \req{ \ \clo{\emptyset}{f_2[\server]} \ }{ \ \clo{\emptyset}{f_3} \ } \ }{\ \epsilon \ }$
\\  
$\longrightarrow$  & $\confcs{\ [\ ] \ }{\ \clo{\emptyset}{f_2[\server]} \ ( \ \clo{\emptyset}{f_3} \ ) \ }$
\\  
$\longrightarrow$  & $\confcs{\ [\ ] \ }{\ \gen{\client}{ \ \clo{\emptyset}{f_3} \ }{1} \ }$
\\  
$\longrightarrow$  & $\confcs{\ [\ ] \ }{\ \call{ \ \clo{\emptyset}{f_3} \ }{1} \ }$
\\  
$\longrightarrow$  & $\confcs{\ \clo{\emptyset}{f_3} \ (1);[\ ] \ }{\ [\ ] \ }$
\\  
$\longrightarrow$  & $\confcs{\ \lunit{1};[\ ] \ }{\ [\ ] \ }$
\\  
$\longrightarrow$  & $\confcs{\ [\ ] \ }{\ \lunit{1} \ }$
\\  
$\longrightarrow$  & $\confcs{\ \lunit{1} \ }{\ \epsilon \ }$
\end{tabular}

\ \\

An untyped \cs program that is compiled from the \polycs example is as
follows:
\[
main \ = \ \ldo{ \ h \ } {\lunit{ \ (Closure \ Server \ f_2)}}{ \ \ldokeyword \ \{ \ send \ (Apply \ h \ (Closure \ \emptyset \ f_3)); \ loop \ () \ \}}
  %\textsfCase \ Closure \ \emptyset \ f_1 \ \textsfOf \ Closure \ \overline{w} \ f \ \rightarrow
  %\  m_{f}\subst{\overline{w}}{\overline{z_f}}\subst{Server}{x_{f}}
\]

where

\begin{tabular}{l l}
  $f_1 \mapsto \emptyset. \ \lambda x_l. \ \lunit{ \ Closure \ x_l \ f_2 \ }$
  &
  $\in \funstore_\client, \funstore_\server$
  \\
  $f_2 \mapsto z_l.\ \lambda g. \ \textsfCase \ g \ \textsfOf \ Closure \ w \ f \rightarrow m_f \subst{w}{z_l}\subst{1}{x_f}$
  &
  $\in \funstore_\client$
  \\
  $f_2 \mapsto z_l.\ \lambda g. \ \ldokeyword \ \{ \ send \ (Apply \ g \ 1); \ loop \ () \ \}$
  &
  $\in \funstore_\server$
  \\
  $f_3 \mapsto \emptyset.\ \lambda x. \ \lunit{x}$
  &
  $\in \funstore_\client$
\end{tabular}

\ \\

Let us see how the example \cs program is running.

%% \begin{tabular}{l l}
%%   & $\confcs{\ main \ \ }{\ \ loop \ () \ }$
%% \\  
%% $\longrightarrow$
%% & $\confcs{\ send \ (Apply \ (Closure \ Server \ f_2) \ (Closure \ \emptyset \ f_3)); \ loop \ () \ \ }{\ \ loop \ () \ }$
%% \\  
%% $\longrightarrow^2$
%% & $\confcs{\ loop \ () \ \ }{\ \ e[\textsfCase \ (Closure \ Server \ f_2) \ \textsfOf \ Closure \ w \ f \rightarrow m_{f_2} \subst{w}{z_l}\subst{(Closure \ \emptyset \ f_3)}{g}] \ }$
%% \\
%% &
%% where $e[ \ ] = \ldokeyword \ z \leftarrow [ \ ]; send \ (Ret \ z); \ loop \ ()$
%% \\  
%% $\longrightarrow$
%% & $\confcs{\ loop \ () \ \ }{\ \ e[m_{f_2} \subst{Server}{z_l}\subst{(Closure \ \emptyset \ f_3)}{g}] \ }$
%% \\
%% $=$
%% & $\confcs{\ loop \ () \ \ }{\ \ e[\ldokeyword \ { \ send \ (Apply \ (Closure \ \emptyset \ f_3) \ 1); \ loop \ () \ }]  \ }$
%% \\  
%% $\longrightarrow^2$
%% & $\confcs{\ e[\textsfCase \ (Closure \ \emptyset \ f_3) \ \textsfOf \ Closure \ \emptyset \ f \rightarrow m_{f_3} \subst{1}{x}] \ \ }{\ \ e[ \ loop \ () ] \ }$
%% \\  
%% $\longrightarrow$
%% & $\confcs{\ e[m_{f_3} \subst{1}{x}] \ \ }{\ \ e[ \ loop \ () ] \ }$
%% \\  
%% $=$
%% & $\confcs{\ e[\lunit{1}] \ \ }{\ \ e[ \ loop \ () ] \ }$
%% \\  
%% $\longrightarrow$
%% & $\confcs{\ send \ (Ret \ 1); \ loop \ () \ \ }{\ \ e[ \ loop \ () ] \ }$
%% \\  
%% $\longrightarrow^2$
%% & $\confcs{\ loop \ () \ \ }{\ \ e[ \ \lunit{1} ] \ }$
%% \\  
%% $\longrightarrow$
%% & $\confcs{\ loop \ () \ \ }{\ \ send \ (Ret \ 1); \ loop \ () \ }$
%% \\  
%% $\longrightarrow^2$
%% & $\confcs{\ \lunit{1} \ \ }{\ \ loop \ () \ }$
%% \end{tabular}

\begin{tabular}{l l}
  & $\confcs{\ main \ \ }{\ \ loop_{body} \ }$ \ \ where $loop_{body} = \ldokeyword \ x \leftarrow receive; \ \textsfCase \ x \ \textsfOf \ \{ \ Apply \ f \ arg \rightarrow \cdots; \ Ret \ y \rightarrow \lunit{y} \ \}$
\\
$=$
&
$\confcs{\ldo{h}{ \lunit{(Closure \ Server \ f_2)} }{\ \ldokeyword \ \{ \ send \ (Apply \ h \ (Closure \ \emptyset \ f_3)); \ loop \ () \ \}} \ \ }{\ \ loop_{body} \ }$
\\
$\longrightarrow$
& $\confcs{\ \ldokeyword \ \{ \ send \ (Apply \ (Closure \ Server \ f_2) \ (Closure \ \emptyset \ f_3)); \ loop \ () \ \} \ }{\ \ loop_{body} \ }$
\\  
$\longrightarrow^2$
& $\confcs{\ loop_{body} \ \ }{\ \ e[(Closure \ Server \ f_2) \  (Closure \ \emptyset \ f_3)] \ }$
%& $\confcs{\ loop_{body} \ \ }{\ \ e[\textsfCase \ (Closure \ Server \ f_2) \ \textsfOf \ Closure \ w \ f \rightarrow m_{f_2} \subst%{w}{z_l}\subst{(Closure \ \emptyset \ f_3)}{g}] \ }$
\ \ where $e[ \ ] = \ldokeyword \ z \leftarrow [ \ ]; send \ (Ret \ z); \ loop \ ()$
\\  
$\longrightarrow$
& $\confcs{\ loop_{body} \ \ }{\ \ e[m_{f_2} \subst{Server}{z_l}\subst{(Closure \ \emptyset \ f_3)}{g}] \ }$ \ \ where $m_{f_2}$ is the body of $f_2$
\\
$=$
& $\confcs{\ loop_{body} \ \ }{\ \ e[\ldokeyword \ { \ send \ (Apply \ (Closure \ \emptyset \ f_3) \ 1); \ loop \ () \ }]  \ }$
\\  
$\longrightarrow^2$
& $\confcs{\ e[ (Closure \ \emptyset \ f_3) \ (1)] \ \ }{\ \ e[ \ loop_{body} ] \ }$
%& $\confcs{\ e[\textsfCase \ (Closure \ \emptyset \ f_3) \ \textsfOf \ Closure \ \emptyset \ f \rightarrow m_{f_3} \subst{1}{x}] \ \ }{\ \ e[ \ loop_{body} ] \ }$
\\  
$\longrightarrow$
& $\confcs{\ e[m_{f_3} \subst{1}{x}] \ \ }{\ \ e[ \ loop_{body} ] \ }$ \ \ where $m_{f_3}$ is the body of $f_3$
\\  
$=$
& $\confcs{\ e[\lunit{1}] \ \ }{\ \ e[ \ loop_{body} ] \ }$
\\  
$\longrightarrow$
& $\confcs{\ send \ (Ret \ 1); \ loop \ () \ \ }{\ \ e[ \ loop_{body} ] \ }$
\\  
$\longrightarrow^2$
& $\confcs{\ loop_{body} \ \ }{\ \ e[ \ \lunit{1} ] \ }$
\\  
$\longrightarrow$
& $\confcs{\ loop_{body} \ \ }{\ \ send \ (Ret \ 1); \ loop \ () \ }$
\\  
$\longrightarrow^2$
& $\confcs{\ \lunit{1} \ \ }{\ \ loop_{body} \ }$
\end{tabular}

\end{document}